\documentclass[twocolumn]{aastex63}
\usepackage{color}

\begin{document}

\newcommand       \msun        	{$M_{\odot}$}
\newcommand       \lsun      	{$L_{\odot}$} 
\newcommand       \zsun      	{$Z_{\odot}$} 
\newcommand       \hub {{\rm km~s}$^{-1}$~{\rm Mpc}$^{-1}$}
\newcommand	     \myr              {$M_{\odot}$~yr$^{-1}$}
\newcommand	     \myre              {{\rm M}_{\odot}~{\rm yr}^{-1}}
\newcommand       \kms        {km~s$^{-1}$}
\newcommand       \ergs        {erg~s$^{-1}$}
\newcommand       \mstar      {M$_{\star}$}
\newcommand 	  \nwats            {nW m$^{-2}$ sr$^{-1}$}
\newcommand	      \mpc              {Mpc$^{-3}$}
\newcommand	      \cc             {cm$^{-3}$}
\newcommand	      \yr              {yr$^{-1}$}
\newcommand	      \sfr              {$M_{\odot}$~yr$^{-1}$}
\newcommand        \mic        	 {$\mu$m}
\newcommand        \nun		{$\nu$}
\newcommand         \lya		{Ly$\alpha$}
\newcommand         \ha		{H$\alpha$}
\newcommand	      \zmin      	{$z_{min}$}
\newcommand       \zmax      	{$z_{max}$}
\newcommand       \gray       {$\gamma$-ray}
\newcommand       \lcdm       {$\Lambda$CDM}
\newcommand    \cii   {[C~{\footnotesize II}]}
\newcommand   \chisq     {$\chi^2$}
\newcommand      \spitz  {{\it Spitzer}}   
\newcommand      \hersh  {{\it Herschel}}
\newcommand      \hst   {{\it HST}}
\newcommand \rosat		{{\it ROSAT}}
\newcommand \galex		{{\it GALEX}}
\newcommand \chandra	{{\it Chandra}}


\title{THE INFRARED ECHO OF SN2010${\rm jl}$ AND ITS IMPLICATIONS FOR \\SHOCK BREAKOUT CHARACTERISTICS}

\author{Eli Dwek}
\affiliation{Observational Cosmology Lab, NASA Goddard Space Flight Center, Mail Code 665, Greenbelt, MD 20771, USA}
\email{eli.dwek@nasa.gov}

\author{Arkaprabha Sarangi}
\affiliation{Observational Cosmology Lab, NASA Goddard Space Flight Center, Mail Code 665, Greenbelt, MD 20771, USA}
\affiliation{CRESST/CUA/GSFC, USA}
\affiliation{Dark Cosmology Center, Niels Bohr Institute for Astronomy, University of Copenhagen
VIbenshuset, Lyngbyvej 2, 4. sal, 2100  Copenhagen, Denmark}

\author{Richard G. Arendt}
\affiliation{Observational Cosmology Lab, NASA Goddard Space Flight Center, Mail Code 665, Greenbelt, MD 20771, USA}
\affiliation{CRESST/UMBC/GSFC, USA}

\author{Timothy Kallman}
\affiliation{X-Ray Astrophysics Lab, NASA Goddard Space Flight Center, Mail Code 662, Greenbelt, MD 20771, USA}

\author{Demos Kazanas}
\affiliation{Gravitational Astrophysics Lab, NASA Goddard Space Flight Center, Mail Code 663, Greenbelt, MD 20771, USA}

\author{Ori D. Fox}
\affiliation{Space Telescope Science Institute, 3700 San Martin Drive, Baltimore, MD 21218, USA}

\received{receipt date}
\revised{revision date}
\accepted{acceptance date}
\published{published date}
\submitjournal{The Astrophysical Journal}
\begin{abstract}
SN~2010jl is a Type~IIn core collapse supernova whose radiative output is powered by the interaction of the SN shock wave with its surrounding dense circumstellar medium (CSM). After day $\sim 60$, its light curve developed a NIR excess emission from dust. 
This excess could be a thermal IR echo from pre-existing CSM dust, or emission from newly-formed  dust either in the cooling postshock region of the CSM, or  in the cooling SN ejecta.  Recent analysis has shown that dust formation in the CSM can commence only after day $\sim 380$, and has also ruled out newly-formed ejecta dust as the source of the NIR emission. 
The early ($ < 380$~d) NIR emission can therefore only be attributed to an IR echo. The $H$-$K$ color temperature of the echo is about 1250~K. The best fitting model requires the presence of about $1.6\times 10^{-4}$~\msun\ of amorphous carbon dust at a distance of $2.2\times 10^{16}$~cm from the explosion. The CSM-powered luminosity is preceded by an intense burst of hard radiation generated by the breakout of the SN shock through the stellar surface. The peak burst luminosity seen by the CSM dust is significantly reduced by Thomson scattering in the CSM, but still has the potential of evaporating the dust needed to produce the echo. We show that the survival of the echo-producing dust provides important constraints on the intensity, effective temperature, and duration of the burst.
\end{abstract}
\keywords{Circumstellar shells(242), Circumstellar grains(239), Circumstellar dust(236), Light curves(918), Shocks(2086), Core-collapse supernovae(304)}

\section{Introduction}
SN~2010jl is a Type~IIn core collapse supernova (CCSN) whose combined X-ray, UV-optical (UVO), and near-infrared (NIR) luminosities greatly exceeds that generated by the energy releases from radioactive elements in the ejecta, and must therefore be powered by the interaction of the SN shock wave with its dense, $n \gtrsim 10^8$~cm$^{-3}$, circumstellar medium (CSM). The CSM was created by mass loss from the progenitor star with estimated mass loss rates ranging from $10^{-4}$ to 0.1~\myr \citep[see][and references therein]{fox17}.  
    The natures of  progenitor stars with such mass loss rates range from red supergiants to luminous blue variables \citep{smith14}. 
    
    Searches for the progenitor of SN~2010jl, and the observed X-ray spectra and  UVO-NIR photometric light curves from the SN have provided important information and constraints on the nature of its progenitor star and its surrounding CSM. These observations and their conclusions can be briefly summarized as follows: 

 (1) Pre-explosion {\it Hubble}/WFPC2 and {\it Spitzer}/IRAC images of the region around SN~2010jl yielded upper limits at 0.29, 0.82, 3.6, 4.5, 5.8, and 8.0~\mic\ for the emission from any progenitor star \citep{fox17}.
   These upper limits require a minimum amount of extinction from pre-existing dust to be present in the CSM in order to hide any luminous hot  progenitor. However, the extinction from this pre-existing dust is limited by the requirement that the absorbed and reradiated IR emission not exceed the upper limits set by the IRAC observations \citep{dwek17}. 

%

(2) Post explosion observations of the SN give an upper limit of $A_V \leq 0.15\pm0.07$ on the amount of extinction through the host galaxy  \citep[summarized in][]{dwek17}. 
This upper limit suggests that only a moderate amount of dust, obscuring a faint cool progenitor, was present in the CSM before the explosion. Alternatively, a large amount of dust, enough to extinguish a luminous hot progenitor, may have been present in the CSM and subsequently mostly evaporated by the  initial burst of radiation generated by the shock breakout through the stellar surface.

(3) {\it Chandra} observations of SN~2010jl show that the X-ray spectrum from the SN was generated by a blast wave propagating at a velocity of $\sim 3000$~\kms\ through the CSM. The intervening H-column density needed to fit the X-ray spectrum decreased with time, and was used to derive the density profile of the CSM \citep{ofek14,chandra15, sarangi18}. 

(4) The photometric SN light curve shows the presence of two distinct emission components:  A UVO component with a blackbody temperature of $\sim 7000$~K, and a NIR dust emission component with a blackbody temperature of about 1800~K \citep{gall14,fransson14,sarangi18}. The IR emission component rose above the UVO continuum around day 90 after the explosion. The NIR light curve remained fairly flat until day $\sim 300$, increased sharply between days $\sim$ 300 and 400, and steeply declined thereafter (see Figure~\ref{uvoir} below, \cite[][Figure 7]{gall14}, and \cite[][Figure 4]{sarangi18}.
     
(5) Observations with {\it Hubble} and the VLT/X-shooter spectrograph showed a trend of increased wavelength-dependent absorption of the red wings of the hydrogen and oxygen line profiles in the spectrum of SN~2010jl \citep{smith12, gall14}. 
    The blueing and wavelength dependent extinction of the emission lines have been presented  as evidence for the rapid formation of dust in the CSM \citep{smith12, gall14}, and as evidence for dust formation in the SN ejecta at later ($\gtrsim 380$~d) times \citep{gall14}.
    
    The  suggestion that the evolution of the line emission was caused by the early formation of dust in the CSM was challenged by \cite{chugai04,dessart15, fransson14} and \cite{jencson16}. They argued that the blueing of the lines could be produced by electron scattering of the lines in the dense CSM, since receding lines have to traverse larger Thomson optical depths. 
 
Any evolving NIR excess around CCSNe can be caused by one or more of the following: (1) the development of an IR echo, consisting of the absorption and reradiation of the SN luminosity by pre-existing circimstellar of interstellar dust \citep{bode79,dwek83b}; (2) the formation of dust in the expanding SN ejecta \citep[e.g.][]{clayton79,dwek92c,wooden93}; or (3) the formation of dust in the postshock region of the circumstellar medium around the SN \citep{smith12,gall14,sarangi18}. In the following we eliminate the latter two possibilities for the origin of the early NIR emission around SN2010jl.

To elucidate the origin of the NIR emission from SN~2010jl, \cite{sarangi18} developed a detailed model for the shock-induced formation of dust in the CSM. 
    Their study showed that the formation of dust is impeded  by the downstream propagation of the radiation from the shocked gas. This effect prevented the cooling of the shocked gas to temperatures below the dust condensation temperature for the first $\sim 300$~days.  The early NIR emission therefore cannot be attributed to the formation of dust in the CSM.  Dust formation commences only after that epoch, when the shock has sufficiently weakened.  
 
    The early NIR emission also cannot be attributed to the formation of dust in the SN ejecta. Expanding at a typical velocity of 3000~\kms, the slow-moving metal-rich ejecta will only reach a radius of $\sim 2\times10^{15}$~cm around day 90, when the dust emission component first rises above the SN ``photosphere". However, the blackbody radius of the NIR emitting region is $\sim 1.5\times 10^{16}$~cm, setting a firm lower limit on the distance of the emitting dust from the SN \citep{sarangi18}. This leaves an IR echo, the reradiated emission from pre-existing CSM dust heated by the SN luminosity that produces the observed light curve, as the only viable source of the NIR emission during the early epoch of the evolution of SN~2010jl.
    
An IR echo model for SN~2010jl was developed by \citep{andrews11} and later updated by \cite{bevan20}. 
In their model, the CSM consists of an inner clumpy and dust-free shell, and an outer dusty torus that is inclined by 60\arcdeg\ with respect to the observer in order to avoid the obscuration of the UVO light from the SN by the dust. A dusty spherical CSM would have provided an excessive amount of extinction to the SN. Based on the analog SN~2006tf, which showed no reddening, \cite{andrews11} assumed little or no reddening towards SN~2010jl as well, and adopted a CSM morphology of an inclined torus allowing for a low optical depth along the line of sight to the SN. The same toroidal configuration was maintained by \cite{bevan20}.

In their model, the echo arises from ACAR dust in the torus that is heated by a flash of light from the SN. This flash of light presumably evaporated all the dust in the inner shell, so the only remaining pre-existing dust resides in the torus. The torus is at a distance of $\sim 1.0$~ly or $\sim 9.4\times10^{17}$~cm. For the flash to heat the dust to the temperatures required to fit the spectrum of the echo, \cite{bevan20} adopted a flash luminosity and temperature of $8\times 10^9$~\lsun, and $1.75\times10^5$~K, respectively. In order to fit the intensity of the echo the flash had to be sustained for 100~d, the light crossing time across the radius of the toroidal tube. A shorter flash, would only have illuminated a fraction of the toroidal dust at any given time. With these parameters, the total radiative energy generated by the flash is about $3\times 10^{50}$~erg. The flash is presumably generated by the shock breakout through the stellar surface. This energy reprises a significant fraction of the total explosive energy of CCSN which is nominally $\sim 10^{51}$~erg. For comparison, the energy generated by the shock breakout in SN1987A was only $(1-2)\times 10^{47}$~erg. Theoretical models predict peak luminosities of $\sim 10^{46}$~erg lasting for $\sim 10^3$~s, giving a total energy of $\sim 10^{49}$~erg \cite{fryer20}. The high shock breakout energy in the \cite{bevan20} model is driven by the toroidal geometry adopted in their model. 

    
 More recent observations and studies of SN~2010jl, presented by \cite{fox17} and \cite{dwek17},  depict a different picture of the CSM. We find that the $H$-$K$ color temperature of the dust is $\sim 1250$~K, significantly higher  than the dust temperature of $\sim 750$~K derived by \cite{andrews11}, yielding a much lower blackbody radius of $(1-2)\times 10^{16}$~cm, than that derived in their model. This high color temperature is an approximate representation of the physical dust temperature. Consequently, less dust is required to generate the observed IR echo. Furthermore, the searches for a progenitor suggest that a dusty CSM may be needed to hide a hot luminous progenitor, and the low but non-zero post explosion extinction alleviates the need to resort to an extreme toroidal configuration for the CSM.
      
    Based on the recent observations and studies of SN~2010jl we re-examine the toroidal echo model for the early IR emission from the CSM, taking the effects of the delayed emission from different parts of the CSM into account. In this paper we concentrate on reproducing the NIR photometry during the  $\sim 27 - 230$ time period, before the UV-optical luminosity of the SN exhibited a precipitous drop. We will show that ignoring the later data is justified, since, deprived of its heating source, the IR echo makes a negligible contribution to the IR emission at later times. Furthermore, the later emission has been shown to be consistent with dust formation in the swept up CSM \citep{andrews11,fransson14,gall14,sarangi19,bevan20}.  
    
  We also introduce a new element in the analysis of IR echoes from CCSNe, namely the constraints these echoes provide on the intense burst of radiation generated by the breakout of the shock through the stellar surface. The burst luminosity undergoes significant processing by Thomson scattering and nebular absorption/reemission, but it still capable of evaporating the dust. The requirement for the survival of the CSM dust that generates the IR echo provides important constraints on the luminosity, the effective temperature, and the duration of the burst.
    
    We first present the UV-optical and IR observations of the SN, and derive the echo contribution to the near-IR fluxes (Section 2). In Section~3, we present the mathematical formalism for the evolution of the IR echo, and in Section~4 the model input parameters: the SN light curve, the CSM density profile, and the dust properties. In Section~5 we  calculate the evolution of the echo for a grid of dust compositions, grain radii, and distances of the dust from the center of explosion. CSM and dust characteristics are determined by the best fitting echo to the data, and the model results are summarized in this section. Section~6 presents the constraints on the characteristics of the shock breakout imposed by the need to preserve the dust needed to generate the observed echo.   The results of our paper and their broader astrophysical implications are summarized in Section~7.

The distance to UGC~5189A, the host galaxy of SN2010jl is $49 \pm 4$~Mpc \citep{smith11, fransson14}. We adopted  a distance of  50~Mpc in all our calculations.
\section{Data}                                             
\begin{figure*}[t]
\begin{center}
\includegraphics[width=3.3in]{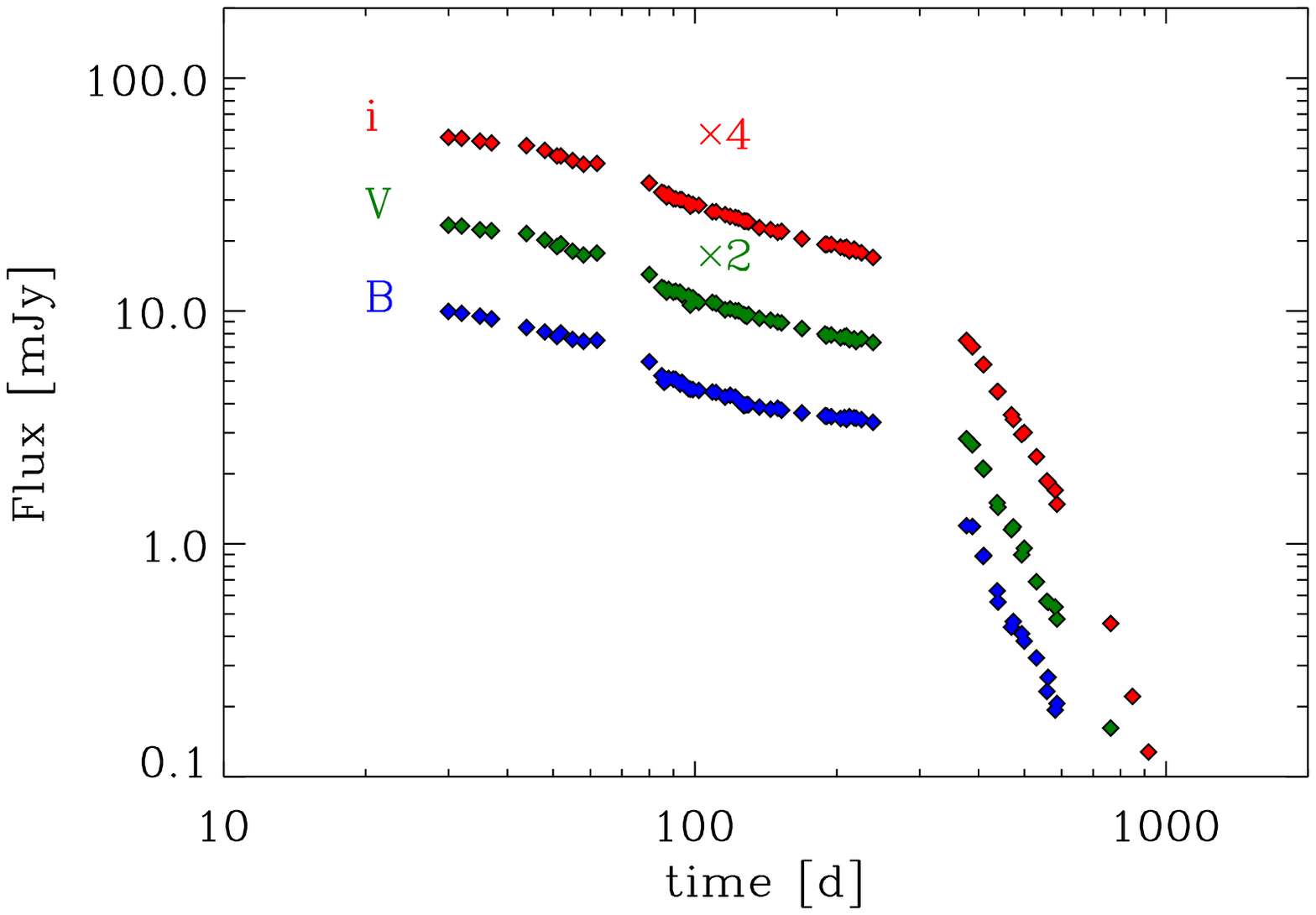}
\includegraphics[width=3.3in]{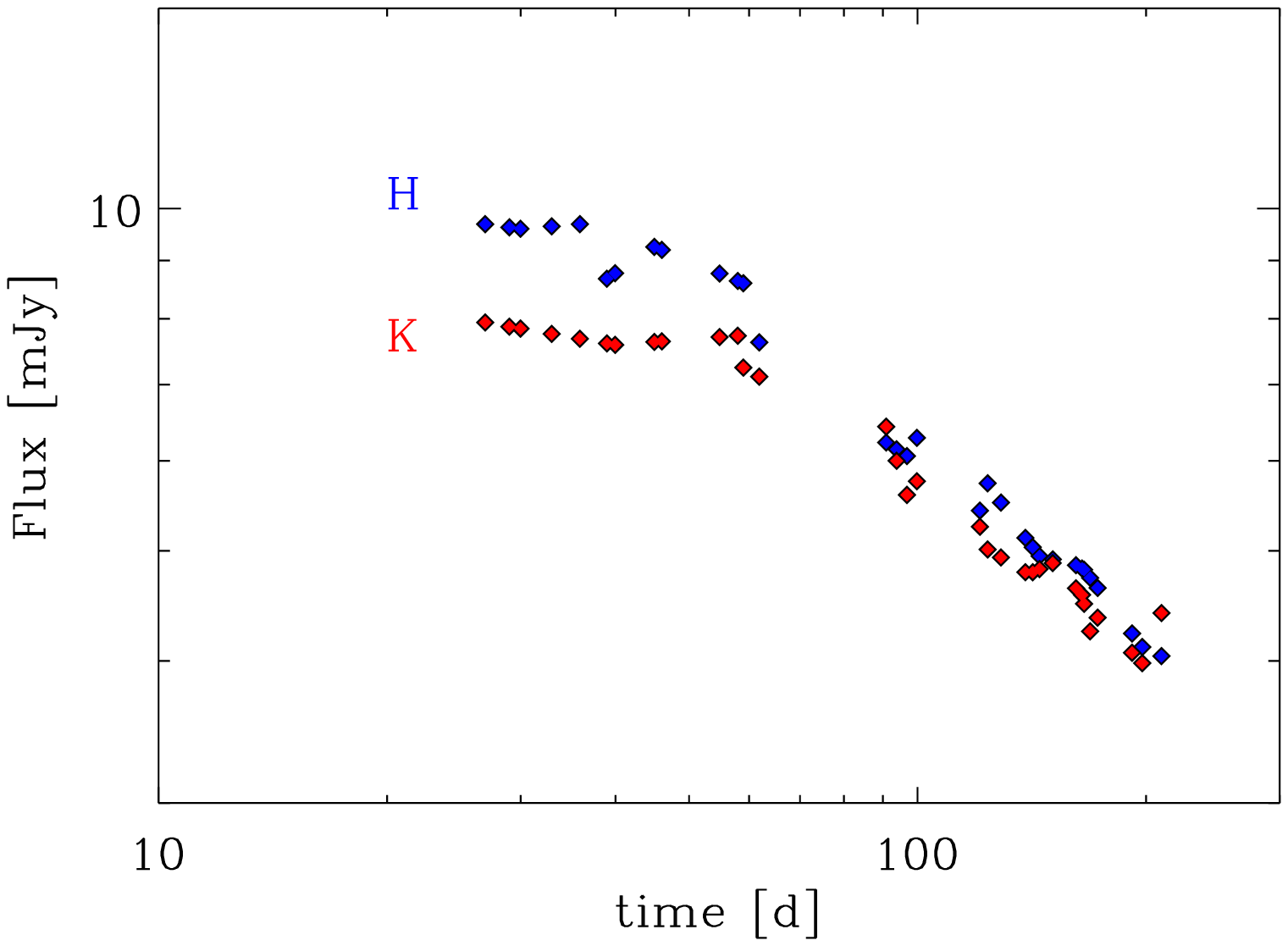}
\caption{\label{uvoir}  Left panel: The light curves of SN~2010jl in the $i'$, $B$, and $V$ bands.   Right panel: The light curves of SN~2010jl in the $H$, and $K$ bands. Data taken from \cite{fransson14}.}
\end{center}
\end{figure*}

 We used the available photometric UVO to NIR data obtained during the $\sim 26$ - 230 day of observations to model the echo from the SN. 
The photometric data in the $u'$(0.359\mic)  $B$(0.433\mic),  $V$(0.550\mic),  $i'$(0.763\mic),  $J$(1.235\mic),  $H$(1.662\mic), and  $K$(2.159\mic) bands were taken from \cite{fransson14}. The tabulated magnitudes were converted to fluxes using the zero magnitude fluxes of Vega\footnote{\url{http://www.gemini.edu, https://cassis.sirtf.com/herschel} }. 
 The 3.6 and 4.5~\mic\ fluxes obtained by the Infrared Array Camera (IRAC) on board the {\it Spitzer} satellite were taken from \cite{andrews11}, \cite{fox13b}, and {\it Spitzer} archival data. 
Figure~\ref{uvoir} presents the SN light curve in the $B$, $V$, and $i'$ bands (left panel) and in the $H$ and $K$ bands (right panel).

\begin{figure*}[t]
\begin{center}
\includegraphics[width=3.3in]{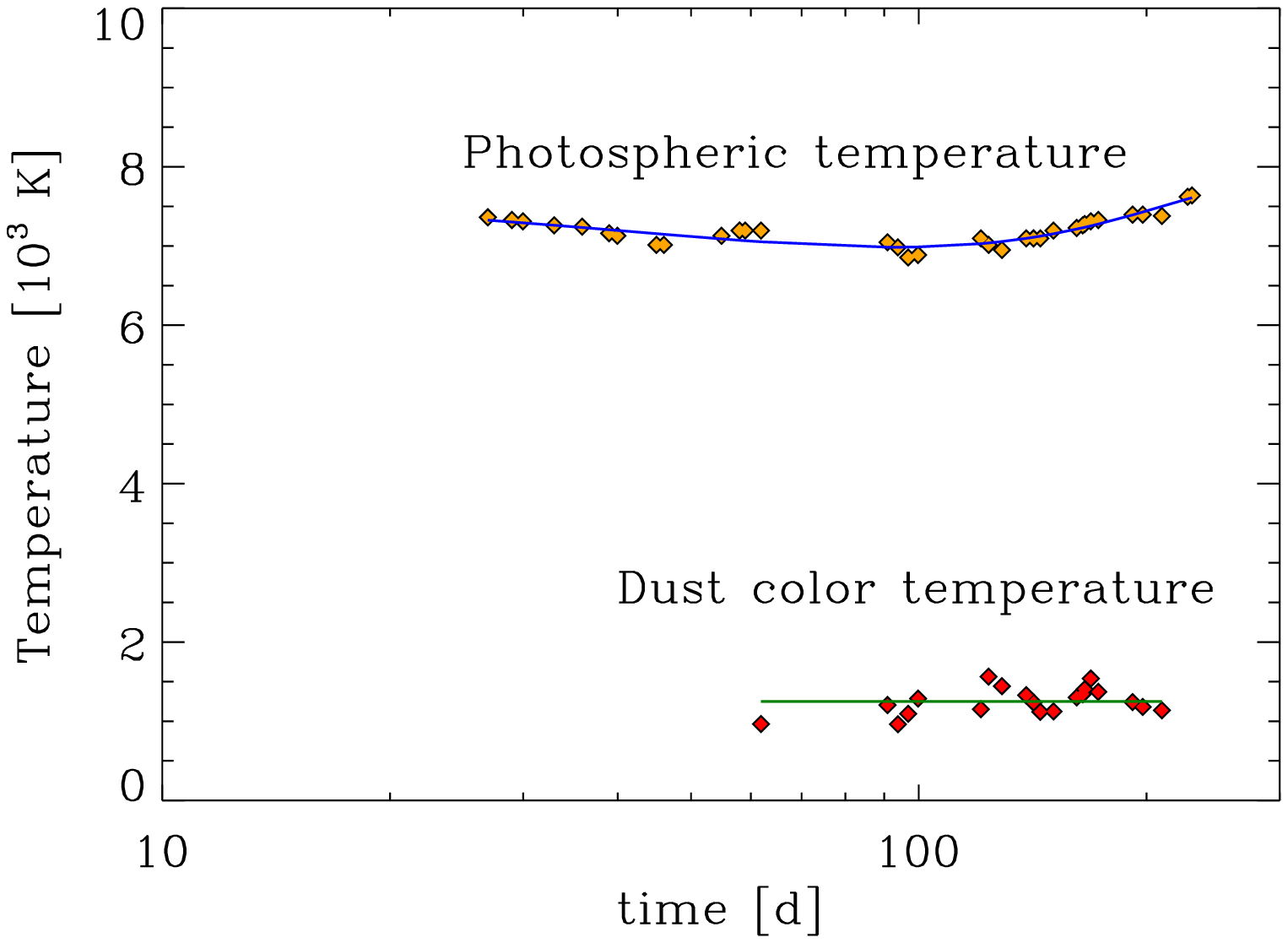}
\includegraphics[width=3.3in]{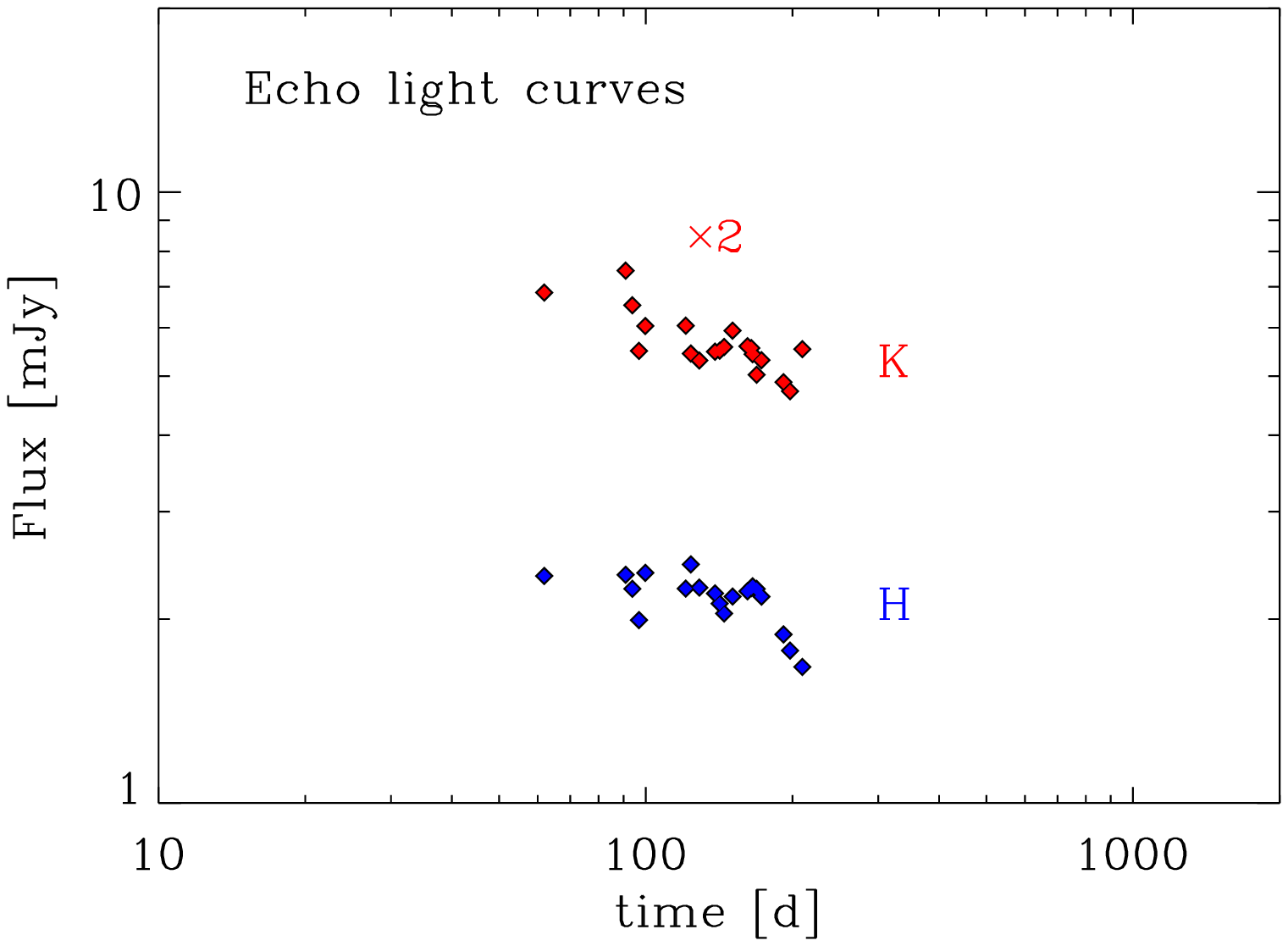}
\caption{\label{echo}   Left panel: The blackbody temperature fits to the  $B$, $V$, and $i'$ band fluxes (orange diamonds) and the color temperature of the $H$ and $K$ bands flux ratio assuming a blackbody with a $\lambda^{-1.5}$ emissivity law. The blue curve is a third order polynomial fit to the photospheric temperature, and the green line represents the average dust color temperature at 1250~K. Right panel: The temporal behavior of the echo in the $H$ and $K$ bands. The echo component was derived by subtracting the photospheric contribution from the observed $H$ and $K$ band fluxes shown in Figs~\ref{uvoir} and \ref{2comp}.}
\end{center}
\end{figure*}

\begin{figure*}[t]
\begin{center}
\includegraphics[width=2.2in]{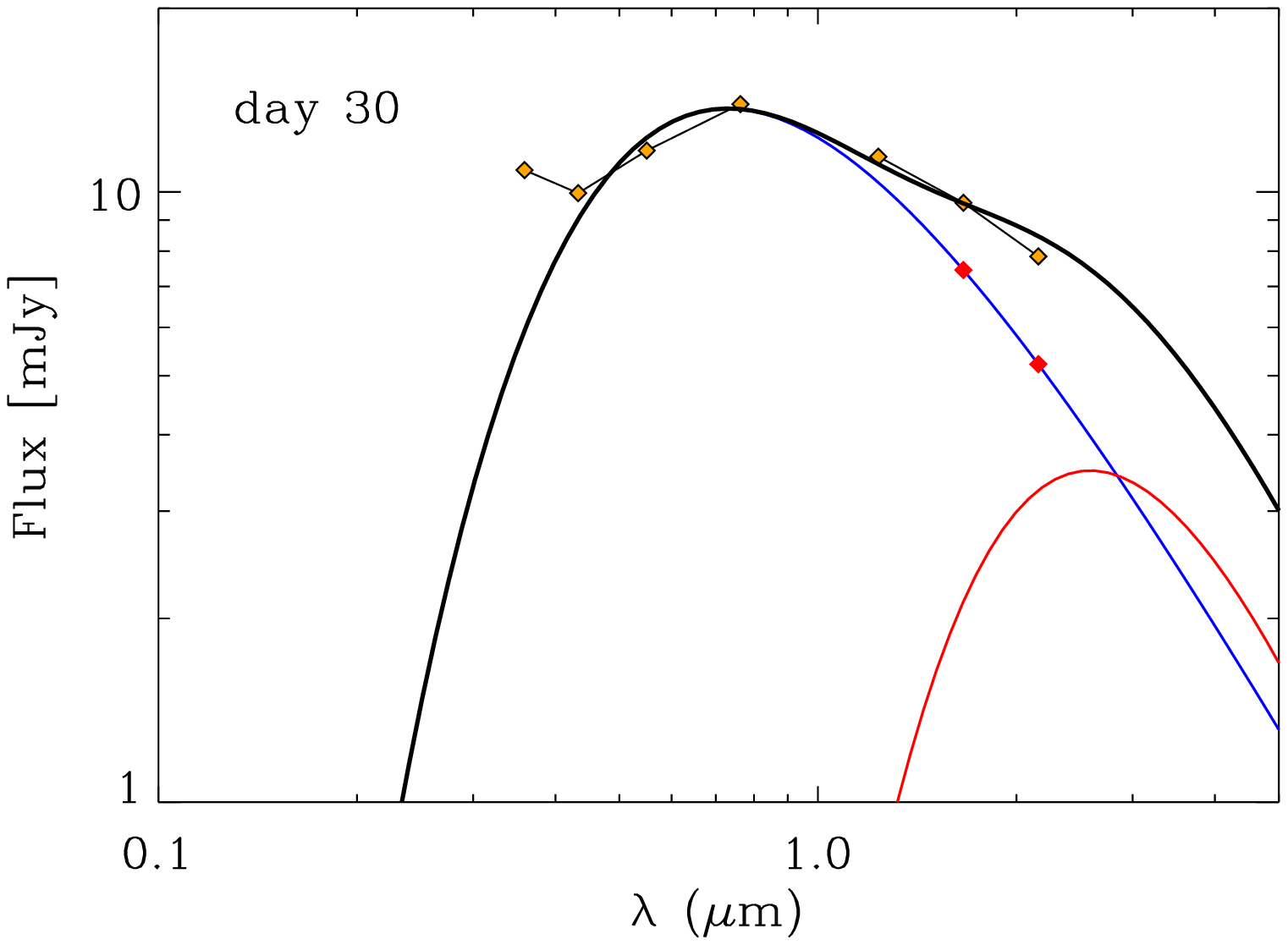}
\includegraphics[width=2.2in]{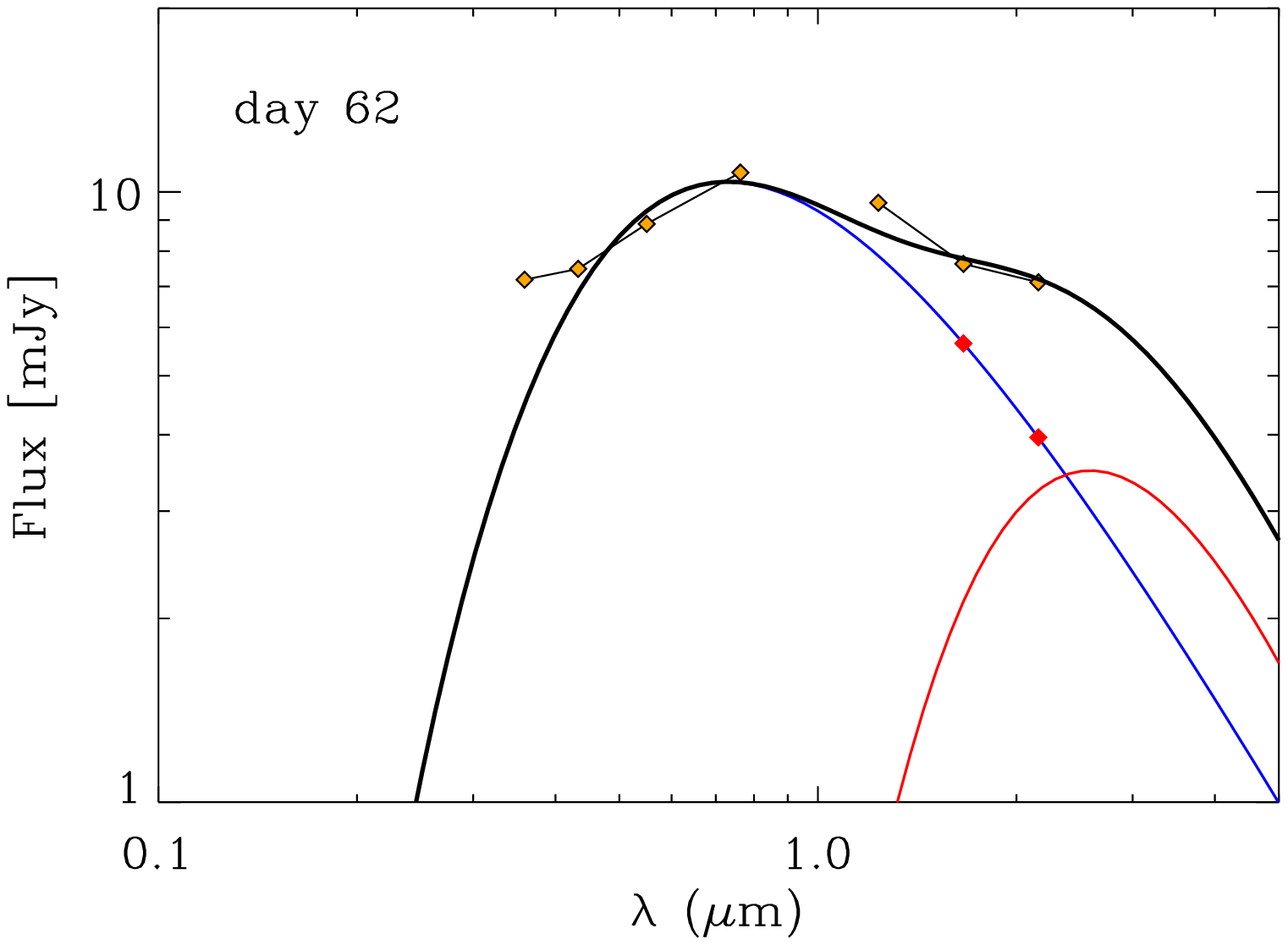}
\includegraphics[width=2.2in]{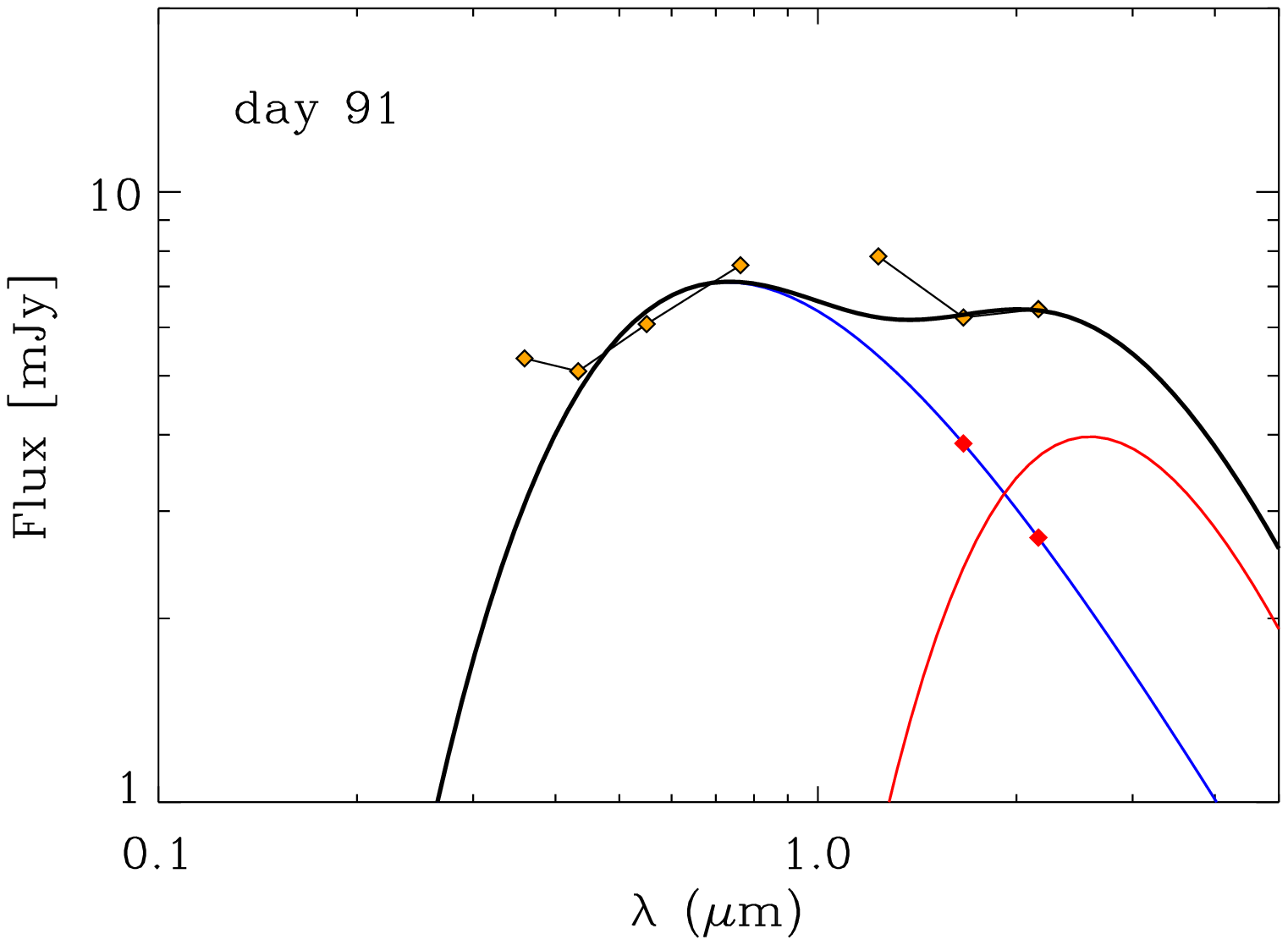}\\
\includegraphics[width=2.2in]{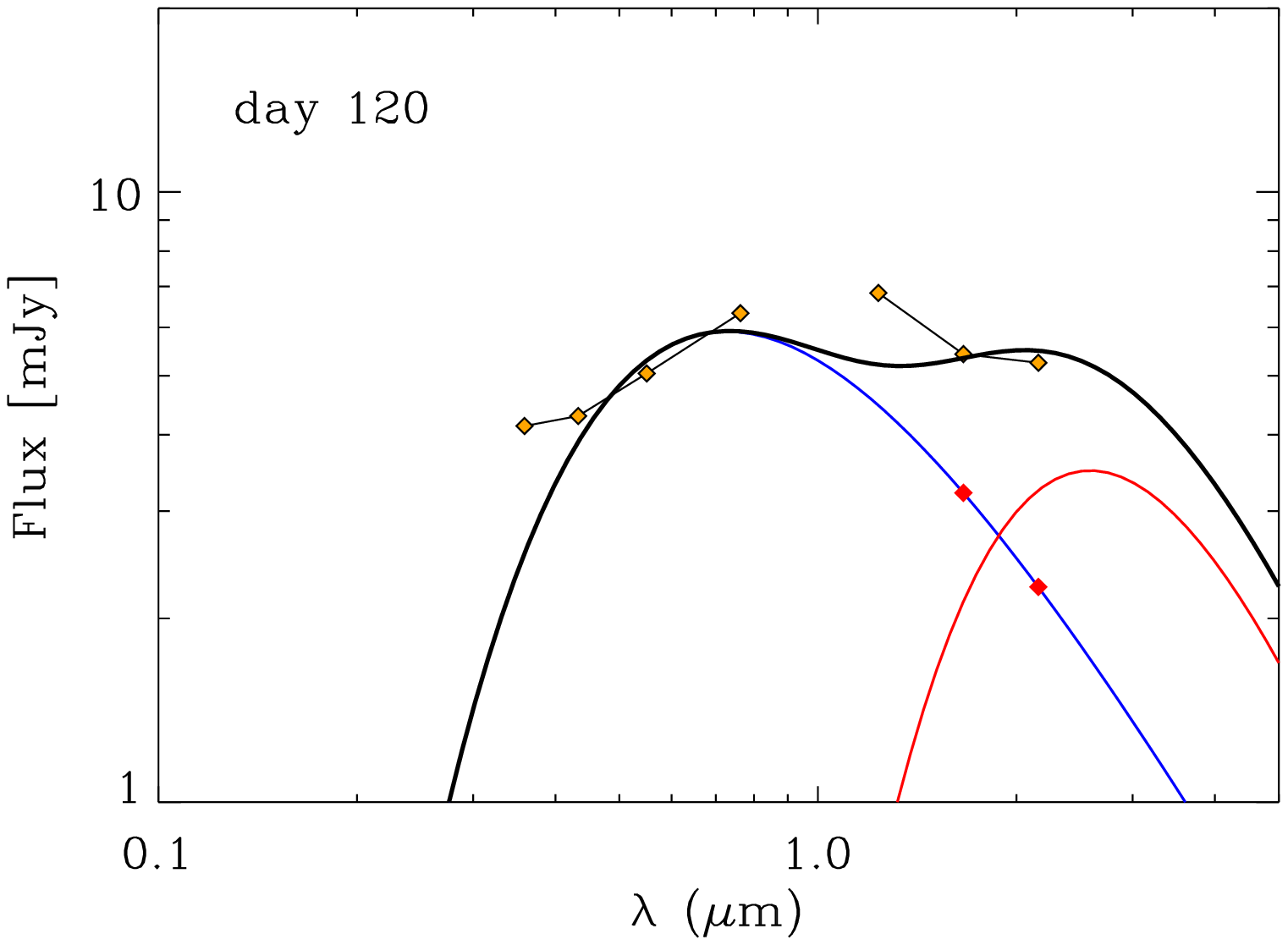}
\includegraphics[width=2.2in]{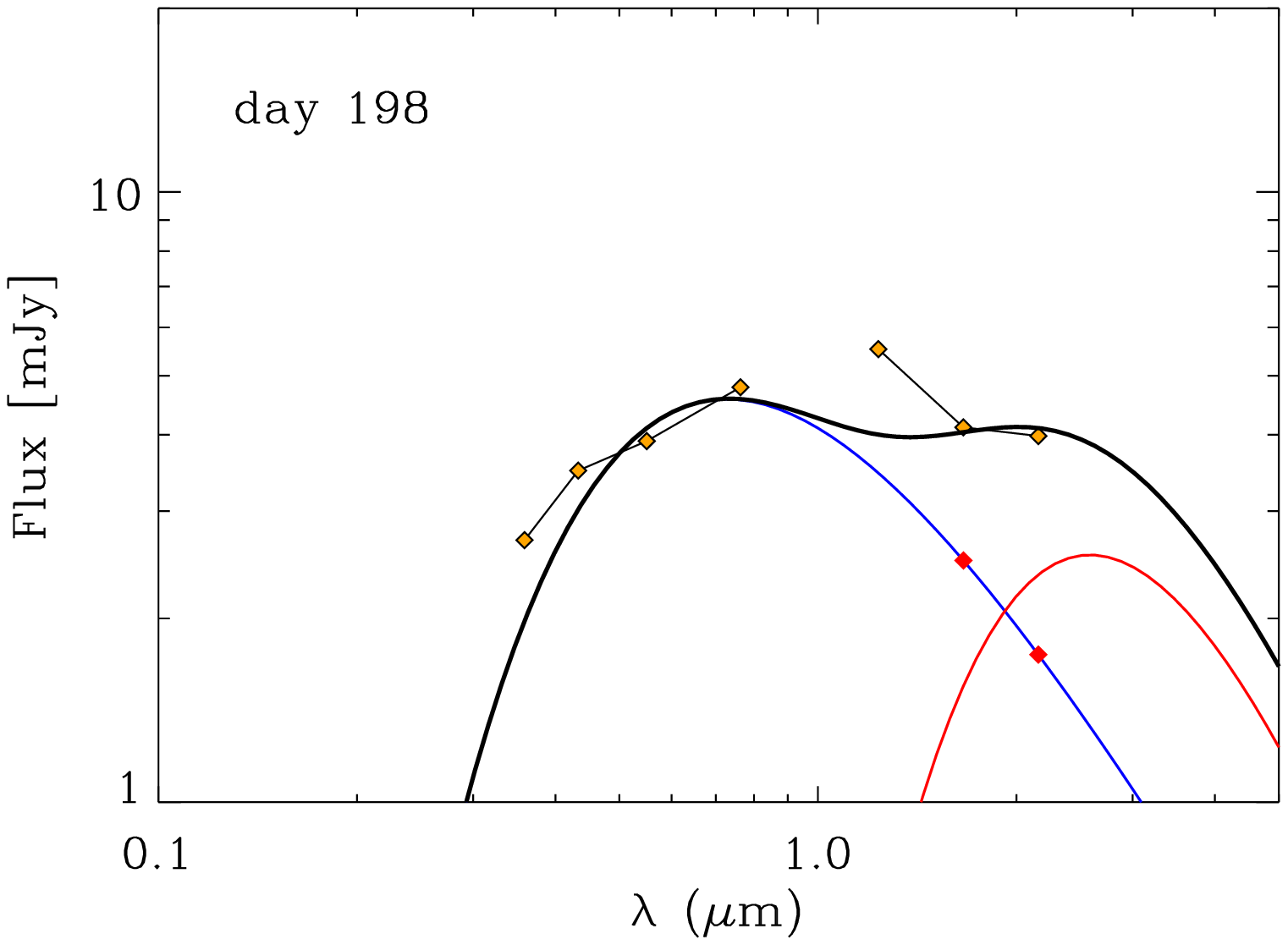}
\includegraphics[width=2.2in]{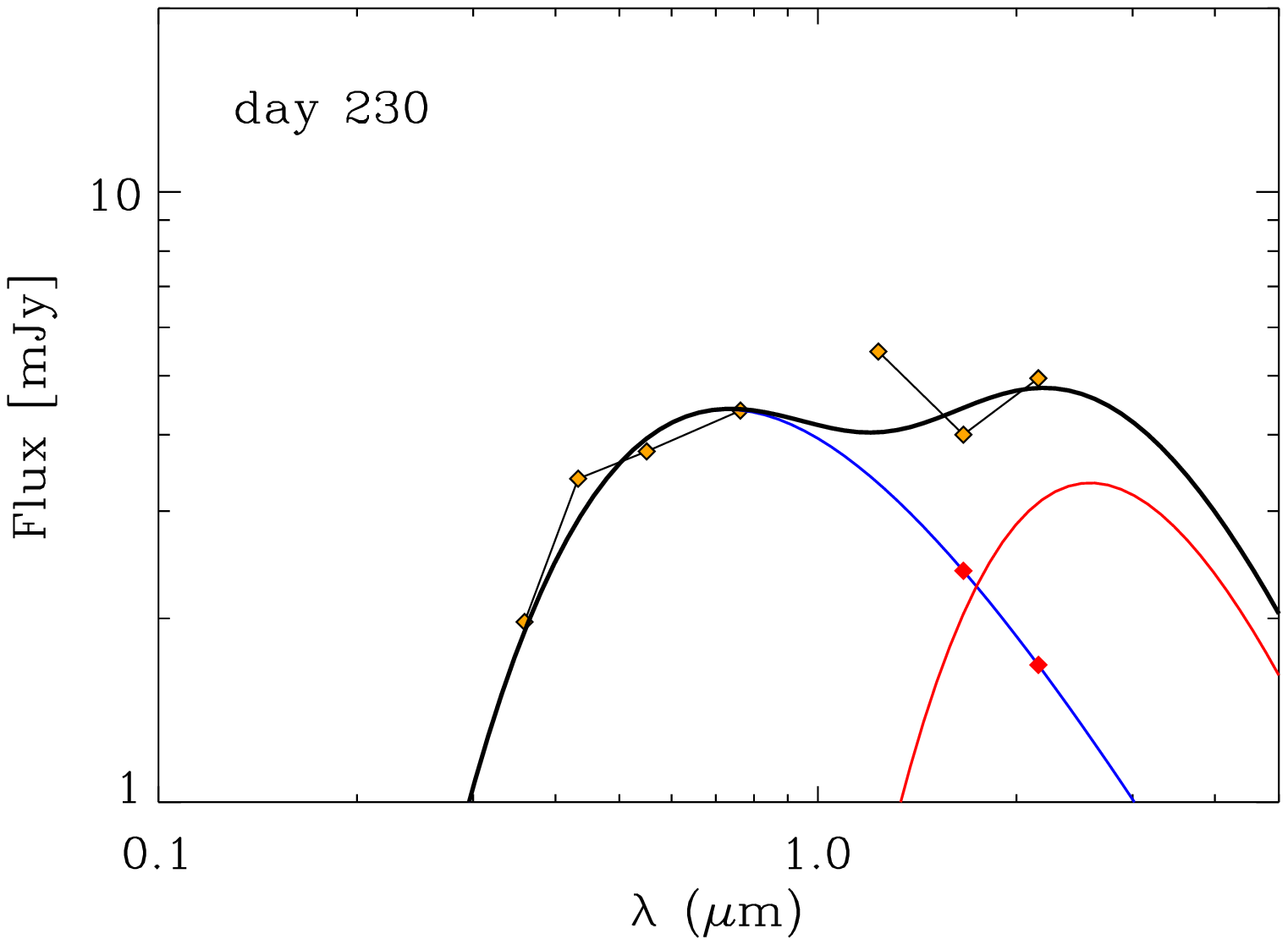}
\caption{\label{2comp} Results of 2 components fit to the SN light curves for select days during the evolution of the echo. The SN photosphere is represented by a 7000~K blackbody, and the dust emission component is represented by a 1250~K blackbody modified by a $\lambda^{-1.5}$ emissivity law. The emission from the echo is the difference between the observed $H$ and $K$ fluxes and the extrapolated flux of the photosphere in these bands, represented by the red diamonds.}
\end{center}
\end{figure*}

 The $H$ and $K$ bands include contributions from the photosphere and the echo. To derive the contribution of the echo to the  NIR emission we fitted a blackbody curve to the $V$, $B$, and $i'$ fluxes at the same epochs of the $H$ and $K$ observations. 
Figure~\ref{echo} (orange diamonds in left panel) shows the blackbody temperature of the photosphere obtained from the best fit to the UVO data alone. The range of temperatures is in good agreement with values of $\sim 7300$~K and 6900-7450~K, derived by \cite{gall14} and \cite{fransson14}, respectively. Figure~\ref{2comp} shows a 2-component fit to the UVO-NIR fluxes for select days during the 26-230 observing period. In these fits, the photospheric temperature was best fit by a blackbody with a constant temperature of 7000~K. The dust spectrum was characterized by a modified blackbody with a $\lambda^{-1.5}$ emissivity law.  A constant dust temperature, equal to the dust color temperature of  of 1250~K, provided a good fit to the observed fluxes. We regard this as an effective temperature giving rise to the $H$ and $K$ band light curves. Longer wavelength observations are required to determine the presence of any colder dust in the CSM. Also shown in figure~\ref{2comp} are the extrapolated photospheric contributions to the $H$ and $K$ bands (red diamonds). The contribution to the IR echo in these bands is obtained by subtracting these fluxes from the observations. The right panel in figure~\ref{echo} depicts the  residual $H$ and $K$ band fluxes that form the echo. 
The red diamonds in the left panel of figure~\ref{echo} shows the color temperature of the residual $H$ and $K$ bands. The average $H-K$ color temperature is 1250~K, consistent with the dust temperature used in the two component fits.

The derived dust temperature is significantly higher than the value of about 750~K derived by \cite{andrews11}, and lower than the average values of $\sim 1700$~K derived by \cite{gall14} and \cite{fransson14}. The dust temperature and the temporal behavior of the NIR echo provide important constraints on the distribution and composition of the echoing dust in the CSM.

\begin{figure}
 \includegraphics[width=3.3in]{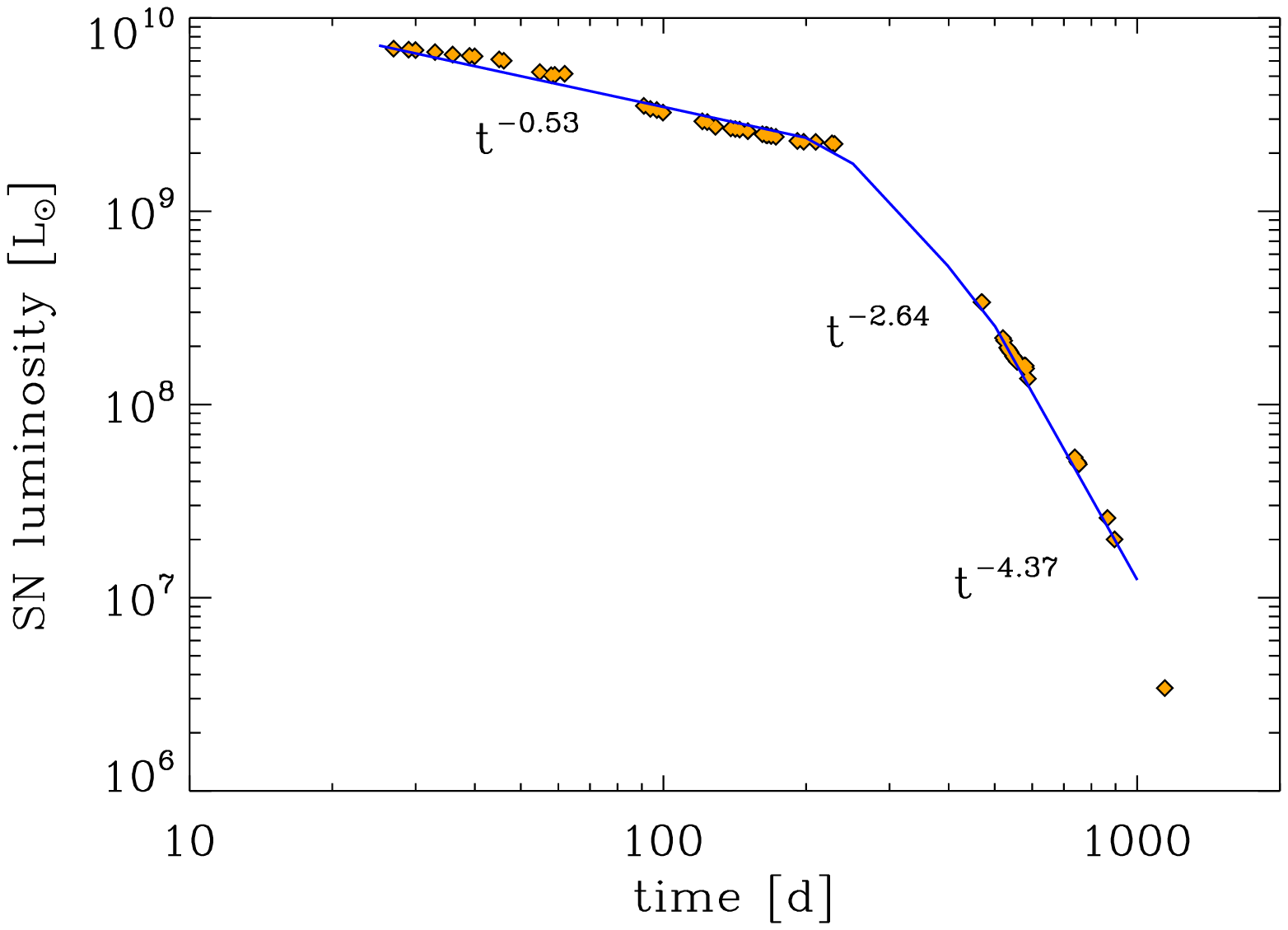}
\caption{\label{Lsn} The supernova optical luminosity as a function of time. The orange diamonds represent the luminosity calculated from the blackbody and temperature fits to the $V$, $B$, and $i'$ band fluxes at the epochs of the NIR observations. The blue curve is a broken power law fit to the luminosity. A constant luminosity of $6.9\times 10^9$~\lsun was adopted for $t \leq 20$~d.}
\end{figure}

Figure~\ref{Lsn} shows the evolution of the SN luminosity as a function of time. The orange diamonds represent the luminosity calculated by blackbody fits to the $V$, $B$, and $i'$ band fluxes at the epochs of the $H$ and $K$ band observations. The blackbody temperatures were presented in Figure~\ref{echo}. The blue curve is a broken power law fit to the luminosity given by
\begin{eqnarray}
\label{Lsneq}
L_{SN}(t) & = & L_0\ \qquad \qquad \qquad \qquad t \leq 27 \nonumber \\
L_{SN}(t) & = & L_0\times (t/t_0)^ {-0.54} \qquad 27 \leq t \leq 230 \nonumber \\
 & = & L_1\times (t/t_1)^{-2.64} \qquad  230 \leq t \leq 470   \\
  & = & L_2\times (t/t_2)^{-4.37} \qquad t >  470 \nonumber 
\end{eqnarray}
where  \{$L_0, L_1, L_2$\}= \{6.9, 2.2, 0.34\} $\times 10^9$~\lsun, and \{$t_0, t_1, t_2$\}= \{27, 230, 470\} $d$. A constant photospheric temperature of 7000~K was adopted for the SN spectrum.

 \section{Mathematical Formalism}      

\begin{figure}
 \includegraphics[width=3.3in]{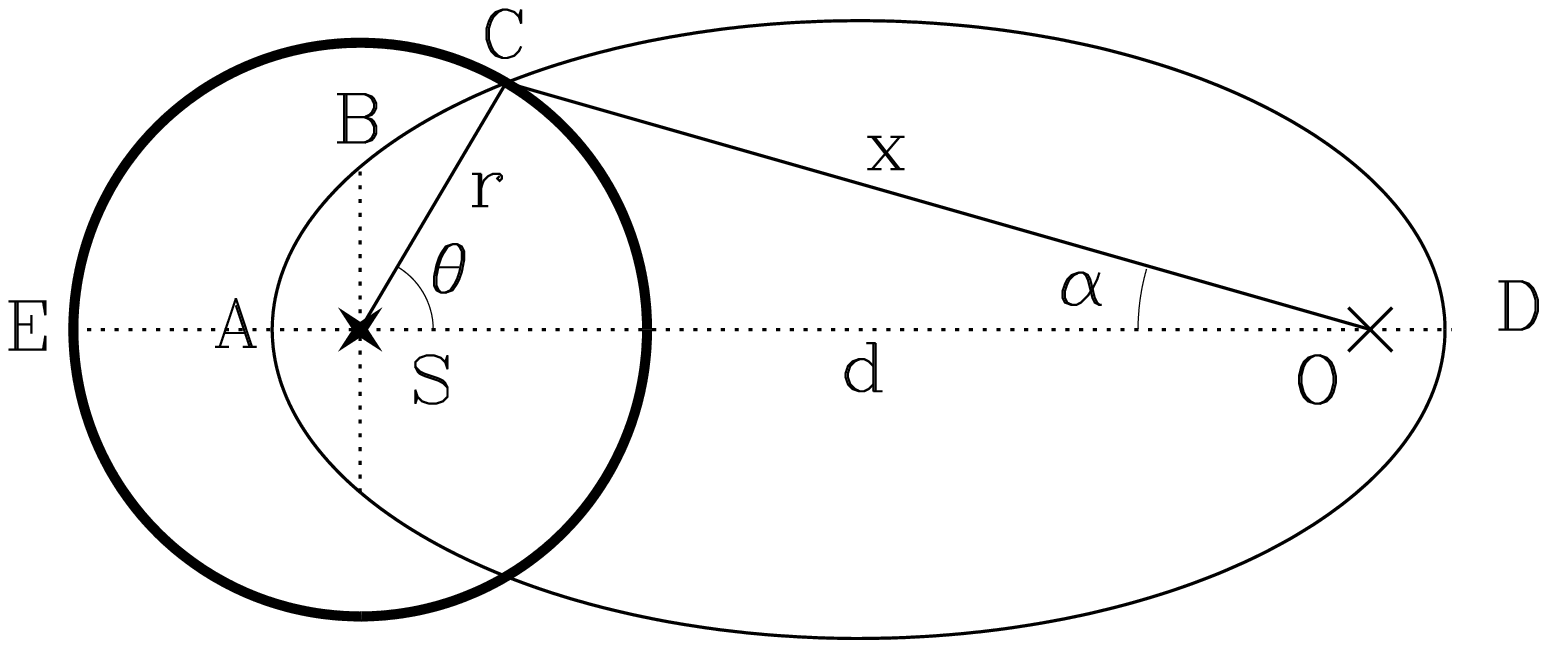}
\caption{\label{ellipse} The geometry of an echo. The radiation arriving at the observer at any given time $t$ is emitted by dust located at the intersection of an ellipsoid of revolution and the CSM. The ellipsoid is characterized by the distance, $d=\overline{SO}$ between the SN (S) and the observer (O), both located at the focal points of the ellipse, and $2\overline{AS}/c$ which is the delay time between the direct and echoed radiation.}
\end{figure}

The evolution of an echo is the results of a convolution of a function that represents the morphology of the CSM with the delayed response of the medium to the SN luminosity.
Figure~\ref{ellipse} shows the projection of an ellipsoid of revolution with focal points at the supernova  $S$ and the observer $O$.  The ellipsoid is the locus of all points for which the sum of distances to the focal points is equal to the length of the major axis, $\overline{AD}$.  For point $C$ on the ellipsoid, $r+x =  \overline{AD} = 2\overline{AS}+d$, where $d\equiv \overline{SO}$ is the distance of the SN to the observer. Any emission from the SN arriving at the observer via any point on the ellipsoid will be delayed with respect to the direct signal from the SN by a time $\tau=(r+x-d)/c= 2\overline{AS}/c$. Using the law of cosines: $x^2=r^2+d^2-2rd\cos\theta$, the delay time between the arrival of the  direct signal from the SN and that received through point $C$ can be written as
\begin{eqnarray}
\label{taud}
\tau &= &  \left({d\over c}\right) \left\{\left[\left(1-{r\over d}\right)^2 + {2r\over d} [1-\cos\theta]\right]^{1/2}  + {r\over d} - 1\right\}  \nonumber \\
 & = & {r\over c}\, [1-\cos\theta] \qquad {\rm for} \ \  r/d << 1 
\end{eqnarray}
or inversely
\begin{equation}
\label{cost}
\cos\theta = 1-{c\tau\over r}\, \left[1+ {c\tau\over 2d} -{r\over d}\right] \ .
\end{equation}

The specific IR flux, $F_{\nu}(\lambda, t)$, from all dust grains on the circumstellar shell of radius $r$ that can give rise to the emission at time $t$ is given by the convolution integral \citep{dwek92b} over the mass of emitting dust, 
\begin{equation}
\label{flux1}
F_{\nu}(\lambda, r, t) = {1\over4 \pi d^2}\,\int_1^{-1}\, \epsilon_{\nu}[\lambda, t-\tau(\cos\theta)]\, {dm_d\over d\cos\theta}\, d\cos\theta \ \ \  , 
\end{equation}
where  $\epsilon_{\nu}(\lambda, t-\tau)$ is the specific luminosity of the dust per unit dust mass, and 
\begin{equation}
{dm_d \over d\cos\theta}  =  -2 \pi \rho_d\, r^2\, \Delta r  \qquad ,
\end{equation}
where the minus reflects the increase in dust mass with decreasing $\cos\theta$.
 
 By changing variables from $\cos\theta$ to $\tau$ we can express the IR flux in eq.~(\ref{flux1}) as an integral over delay times using the following transformation
 \begin{equation}
   dm_d(\tau) =  {dm_d\over d\tau}\, d\tau={dm_d\over d\cos\theta}\ {d\cos\theta\over d\tau}\, d\tau
\end{equation}   
where
\begin{equation}
\label{dcost}
{d\cos\theta\over d\tau}   =   - {c\over r} \left[1+{c\tau\over d} -{r\over d}\right] \, \Pi(\tau)
\end{equation}
The $\Pi$ function given by 
\begin{eqnarray}
\Pi(\tau) & = & 1  \ \ {\rm for} \ \ 0<\tau<2r/c  \nonumber \\
 & = & 0 \qquad \qquad \qquad \qquad {\rm otherwise} .
 \end{eqnarray}
The expression for $dm_d(\tau)$ becomes
\begin{eqnarray}
 dm_d(\tau) & = &  2 \pi \rho_d\,r\, \Delta r\,  \left[1+{c\tau\over d} -{r\over d}\right]\, \Pi(\tau)\, c\, d\tau \\
   & = & 2 \pi \rho_d\,r\, \Delta r\, \Pi(\tau)\, c\, d\tau \qquad {\rm for} \  r/d, c\tau/d\ \ll 1  \, .  \nonumber
\end{eqnarray}

\noindent
Equation~(\ref{flux1}) can now be rewritten as
\begin{equation}
\label{flux2}
F_{\nu}(\lambda, r, t) =   {2 \pi \rho_d\,r\, \Delta r\over 4 \pi d^2}\,  \int_0^{t}\  \Pi(\tau)\, \epsilon_{\nu}(\lambda, r, t-\tau)\, c\, d\tau \ .
\end{equation}
For an optically-thin shell extending from an inner radius (cavity) $R_0$ to an outer radius $R_{max}$, the echo is given by:
\begin{equation}
\label{flux3}
F_{\nu}(\lambda, t) =  \int_{R_{min}}^{R_{max}}\  {2 \pi \rho_d(r)\,r\, dr\over 4 \pi d^2}\,  \int_0^{t}\  \Pi(\tau)\, \epsilon_{\nu}(\lambda, r, t-\tau)\, c\, d\tau \ .
\end{equation}
where $R_{min}$=max$\{R_0, R_{evap}, R_d\}$, where $R_d$ is the inner radius of the pre-existing dust shell, and $R_{evap}$ is the  radius at which the dust temperature is equal to its vaporization temperature. No dust can exist at smaller radii.

 The specific luminosity of the dust, in units of erg~s$^{-1}$~Hz$^{-1}$~g$^{-1}$, is given by an integral 
\begin{eqnarray}
\label{eps}
& & \epsilon_{\nu}(\lambda, r, t)  =  \\
& & {4\over \left<m_{gr}\right>}  \int_{a_1}^{a_2} m_{gr}(a)\, f(a)\, \pi B_{\nu}[\lambda, T_d(a,r,t)]\, \kappa(\lambda, a) da \nonumber
\end{eqnarray}
where  $\kappa(\lambda, a)$ is the dust mass absorption coefficient in cm$^2$~g$^{-1}$,  $\pi B_{\nu}(\lambda, T_d)$ is the Planck function, $f(a)$ is the grain size distribution normalized to unity in the \{$a_1$,$a_2$\} radius interval, $m_{gr}(a)$ is the mass of a dust grain of radius $a$, and $\left<m_{gr}\right>$ is the dust mass averaged over the grain size distribution.  

Assuming that the CSM is optically thin at UVO wavelengths, the dust temperature, $T_d$,  is given by the energy balance equation,
\begin{eqnarray}
\label{tdust}
 & & {L_{SN}(t)\over 4\pi r^2}\ \int_0^{\infty}\ {\pi B_{\nu}(\lambda, T_{SN})\over \sigma T_{SN}^4}\, \kappa(\lambda, a)\, d\nu = \\ \nonumber
& &  \int_0^{\infty}\ 4\pi B_{\nu}[\lambda, T_d(a,r,t)]\, \kappa(\lambda, a)\, d\nu \qquad ,
\end{eqnarray} 
where $L_{SN}(t)$ is the SN luminosity at time $t$, and $T_{SN}$ is the effective temperature of its assumed blackbody spectrum. 

 With the dust temperature and emissivity determined from eqs.~(\ref{tdust}) and (\ref{eps}), the only free parameters of the model, given by eq.~(\ref{flux3}), are $R_{min}$, and the mass density, $\rho_d(r)$, of the dust at each radius. The latter can be expressed in terms of the dust-to-hydrogen mass ratio in the shell, $Z_{dH}$, and the hydrogen number density, $n_H$,  as
\begin{eqnarray}
\label{rho}
\rho_d(r) & \equiv & n_d(r)\,\int_{a_1}^{a_2}\, m_{gr}(a) f(a) da   \nonumber \\
 & = & n_d(r) \left<m_{gr}\right> = m_H\, n_H(r)\, Z_{dH} \qquad ,
\end{eqnarray}
where $n_d(r)$ is the total number density of dust grains at radius $r$, and  $m_H$ is the mass of an H atom.

The total optical depth through the shell, $\tau(\lambda, a)$, produced by a population of grains of radius $a$  at wavelength $\lambda$ is given by,
\begin{equation}
\label{tauW}
\tau(\lambda,a) = \int_{R_{min}}^{R_{max}} \rho_d(r)\, dr\, \int_{a_1}^{a_2} m_{gr}(a) f(a)\,  \kappa(\lambda, a)\, da \qquad.
\end{equation} 

\section{Model input parametes} 

\subsection{CSM density profile}

\begin{figure*}[t]
\begin{center}
\includegraphics[width=3.3in]{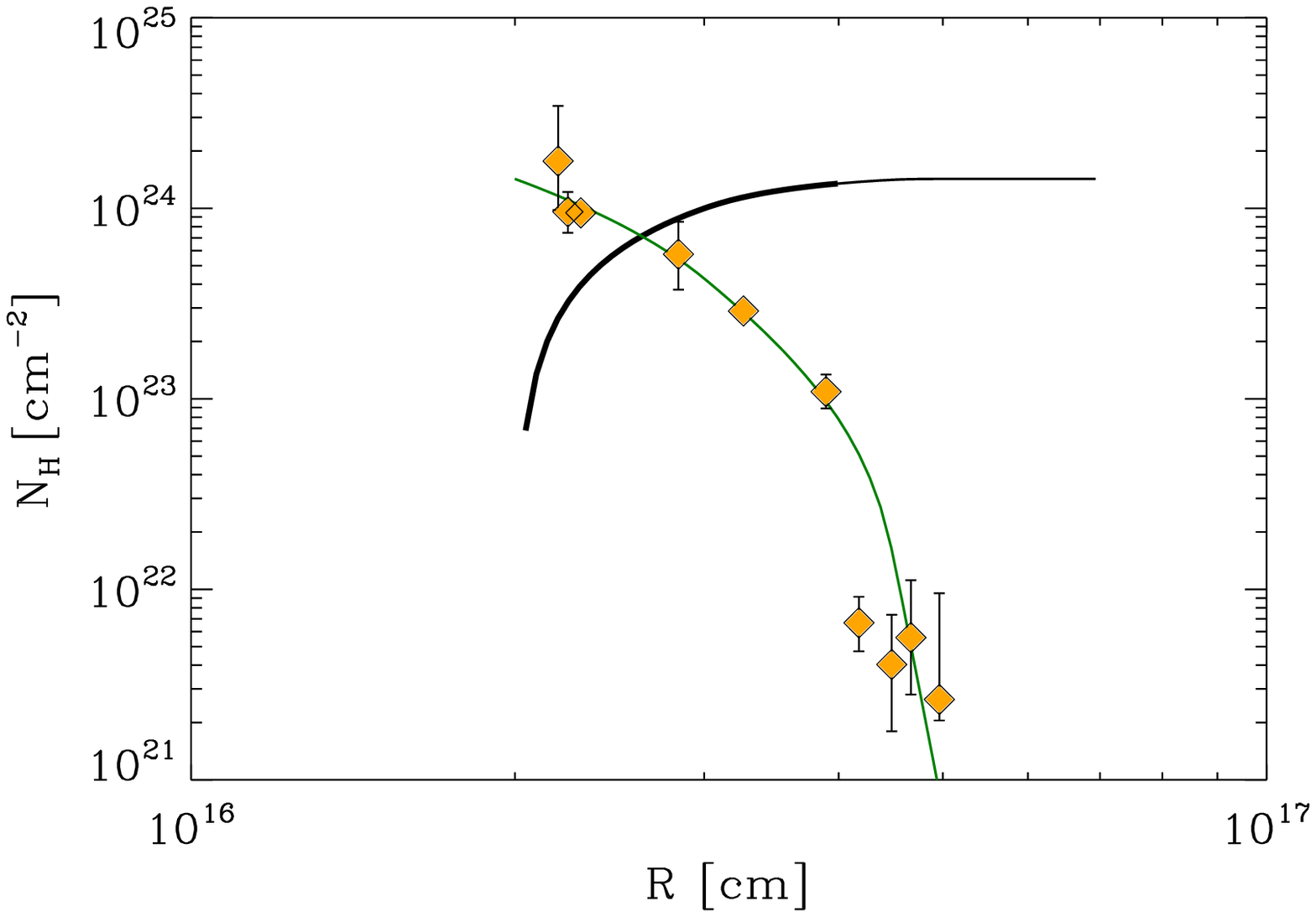}
\includegraphics[width=3.3in]{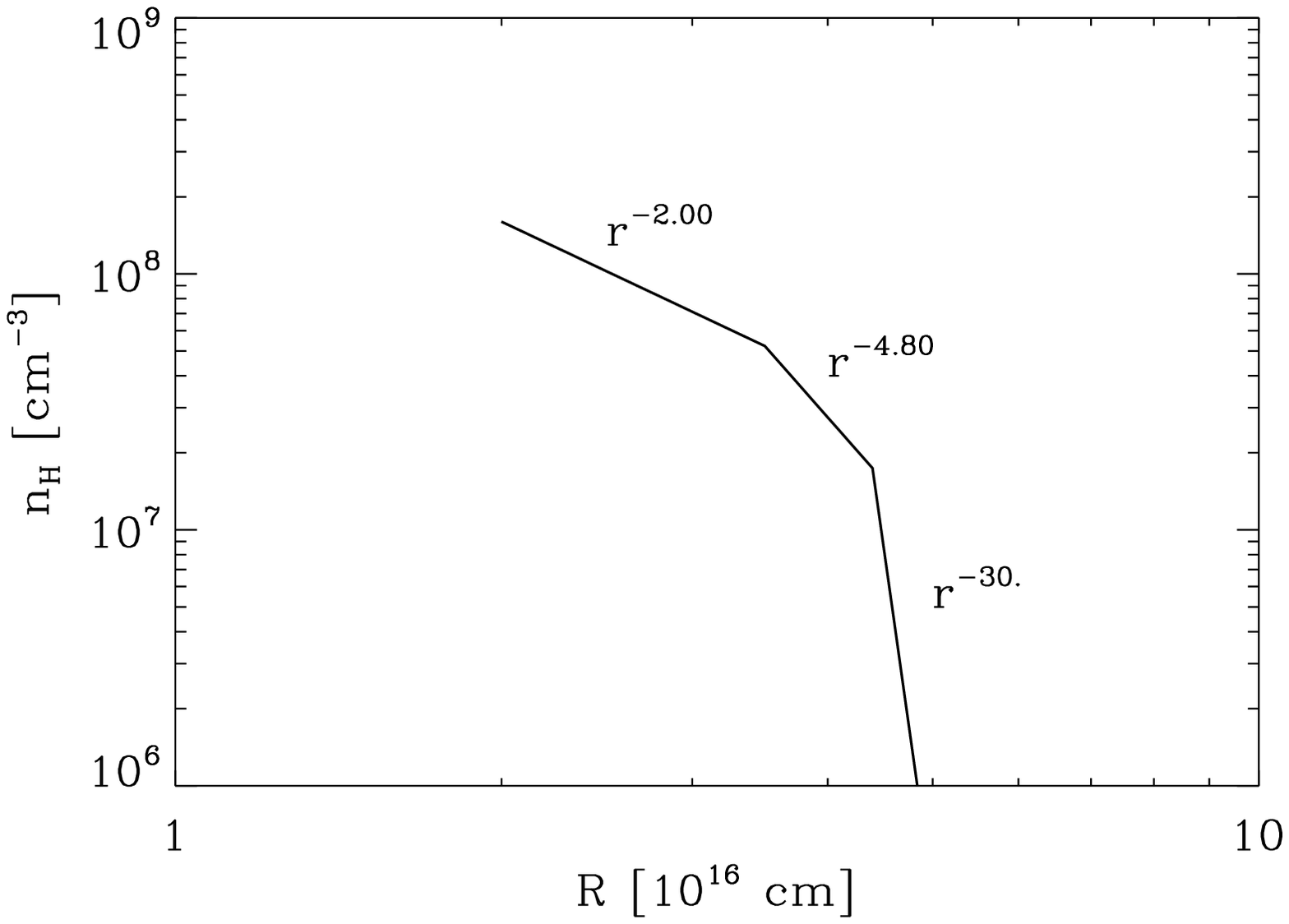}
\caption{\label{csm} Left panel: Intervening hydrogen column densities to the SN versus shell radius as derived from X-ray observations are presented by yellow diamonds. The green curve is a fit to the data derived from the CSM density profile shown in the right panel. 
The black curve in the left panel is the cumulative column density of the CSM. 
The total H-column density through the CSM is $1.7\times 10^{24}$~cm$^{-2}$. 
Right panel: The  density profile of the CSM giving rise to the H column density profile in the left panel. The CSM has a peak density of $1.60\times 10^8$~\cc\ at the inner radius, $R_0=2\times 10^{16}$~cm. Its density structure is characterized by a broken power law presented in eq.~(\ref{nHeq}).}
\end{center}
\end{figure*}

The hydrogen density profile of the CSM is constrained by the Chandra X-ray observations of the SN \citep{chandra15}. Figure~\ref{csm} depicts the intervening H column densities to the X-ray emitting shell in the CSM as a function of radius which is determined  from the shock dynamics through the CSM. The green curve in the panel is a fit through the Chandra data, calculated for the density profile shown in the right panel of the figure. Also shown in the left panel is the cumulative value of $N_H$ as seen from the center of the explosion (black curve). 
The CSM density profile, derived from the \cite{chandra15} data, is shown in the right panel. It is characterized by a continuous broken power law given by:

\begin{eqnarray}
\label{nHeq}
n_{H}(R) & = & n_0\times (R/R_0)^ {-2.0} \qquad R_0 \leq R \leq R_1 \nonumber \\
 & = & n_1\times (R/R_1)^{-4.80} \qquad  R_1 \leq R \leq R_2   \\
  & = & n_2\times (R/R_2)^{-30} \qquad R >  R_2 \nonumber 
\end{eqnarray}
where  \{$n_0, n_1, n_2$\}= \{1.60, 0.52, 0.17\} $\times 10^8$~cm$^{-3}$, and \{$R_0, R_1, R_2$\}= \{2.0, 3.5, 4.4\}  $\times 10^{16}$~cm.
 The total H-mass of the CSM is 15~\msun. 
The inner radius of the CSM was determined by the fit to the X-ray observations \citep[see also][]{sarangi18}. A smaller inner radius with an $r^{-2}$ density profile would not have been able to fit the evolution of the H-column density. To give rise to the echo, the SN blast wave must traverse the cavity at large enough speeds to reach the inner radius by day $\sim 26$. This requires the shock velocity to be about 10,000~\kms. After reaching the dense CSM, the shock must slow down to a velocity of $\sim 3,000$~\kms\ in order to produce the X-ray spectra. The shock would have penetrated to a distance of $\sim 5\times 10^{15}$ into the CSM by day 230, the last period of the echo observations.  
 
\subsection{Dust properties}
\begin{figure}[t]
\includegraphics[width=3.3in]{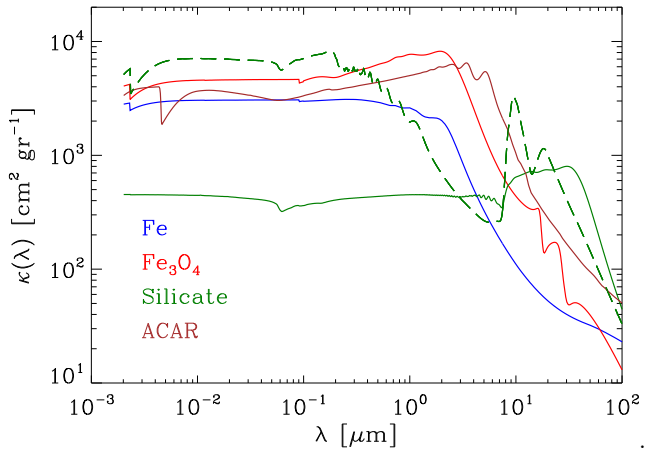}. 
\caption{\label{kappa} The mass absorption coefficients of the dust different dust compositions. The solid curves are the values of $\kappa(\lambda)$ plotted for grain sizes that provided the best fit to the IR echo, given in Table~1. The long-dashed green curve represents the value of $\kappa(\lambda)$ for 0.30~\mic\ silicate grains.}
\end{figure}

The IR echo from the CSM was calculated for a 3-parameter grid consisting of dust composition, grain radius, and the location of the dust in the CSM. Dust compositions considered were metallic iron (Fe), Fe$_3$O$_4$, astronomical silicates, and amorphous carbon (ACAR). 
Optical constants for the various dust grains were taken from the Jena data base \citep{semenov03,jaeger94,dorschner95,jaeger03}, from \cite[][article by Jaeger et al.]{palik91}, and Kozasa (2006, private communications). Optical constants for astronomical silicates and ACAR dust were taken from \cite{draine07a} and  \cite{rouleau91}, respectively. The IR emission was calculated for single-sized grains with radii of 0.01, 0.03, 0.20, 0.30, 1.0, and 5.0~\mic.
Figure~\ref{kappa} shows examples of the mass absorption coefficient of the dust compositions considered in this paper at the radii that best fit the echo (Table~1).

The CSM dust is heated by the SN luminosity,  whose light curve is presented in the right panel of Figure~\ref{Lsn}. The source of the emission was assumed to be at the center of the explosion. Motivated by the uniformity of the $B$, $V$, and $i'$ observations, a constant blackbody temperature of 7000~K was adopted for the SN spectrum (see Fig.~\ref{2comp}). 
 
The radiation generated by the shock-CSM interaction will evaporate the CSM dust out to a distance that is determined by the dust composition and radius. The lower limit on the integral is then given by max\{$R_0, R_{evap}$\}.
Dust evaporation is a time dependent process which has been studied in detail by \citep{voit91}. Exposed to an intense radiation field a dust grain can reach a sufficiently high temperature for atoms in the grain to acquire enough energy to overcome their binding and surface barrier energies to escape the grain. The evaporation process cools the grain, which additionally cools by radiative emission. The study of \cite{voit91} shows that sublimation is substantial even after the absorption of a single energetic photon. Continuous exposure to the same radiation field, will accelerate the evaporation process, since the grain's temperature will increase with decreasing grain radius. For the purpose of our calculations, we adopted sublimation temperatures of 2000 and 1500~K for, respectively, carbon and silicate grains \citep{kruegel03,voit91,dullemond10, pollack94, nagel13}. Similar values were used in previous analyses of IR echoes \citep[e.g.][]{dwek85, gall14, fransson14}.

\begin{figure}[t]
\includegraphics[width=3.3in]{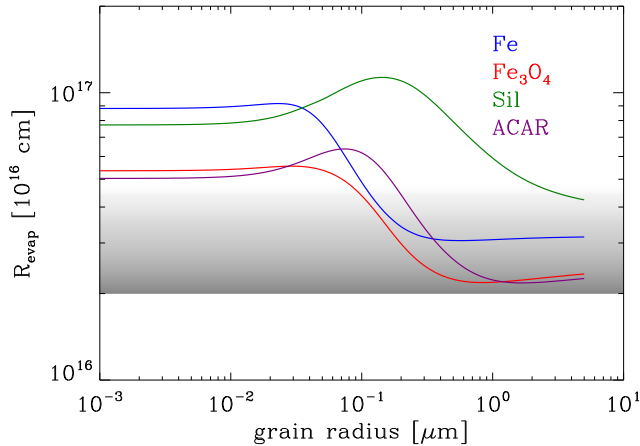}
\caption{\label{Revap} The evaporation radius of the dust as a function of grain radius for the different dust compositions. The shaded region represents the density profile of the CSM cavity.}
\end{figure}

 Figure~\ref{Revap} shows the dust evaporation radius as a function of grain radius for the different dust compositions. The grey shading of the picture represents the density profile of the CSM, with the darkest region corresponding to the highest density. Grains with radii $\lesssim 0.1$~\mic\ are evaporated out to distances $> 5\times 10^{16}$~cm, and will make a negligible contribution to the IR echo. 
Only very large,  $\sim 5$~\mic, silicate grains will survive evaporation at distances $\gtrsim 4.5\times 10^{16}$~cm. However, their contribution to the echo would require an excessive mass of silicates because of their very low mass absorption coefficient (see Fig.~\ref{kappa}). Furthermore, the rapid falloff of the H-density at these distances would require a dust-to-H mass ratio of $\sim 2\times 10^{-2}$ that exceeds solar abundances.

\begin{figure*}[t]
\begin{center}
\includegraphics[width=3.3in]{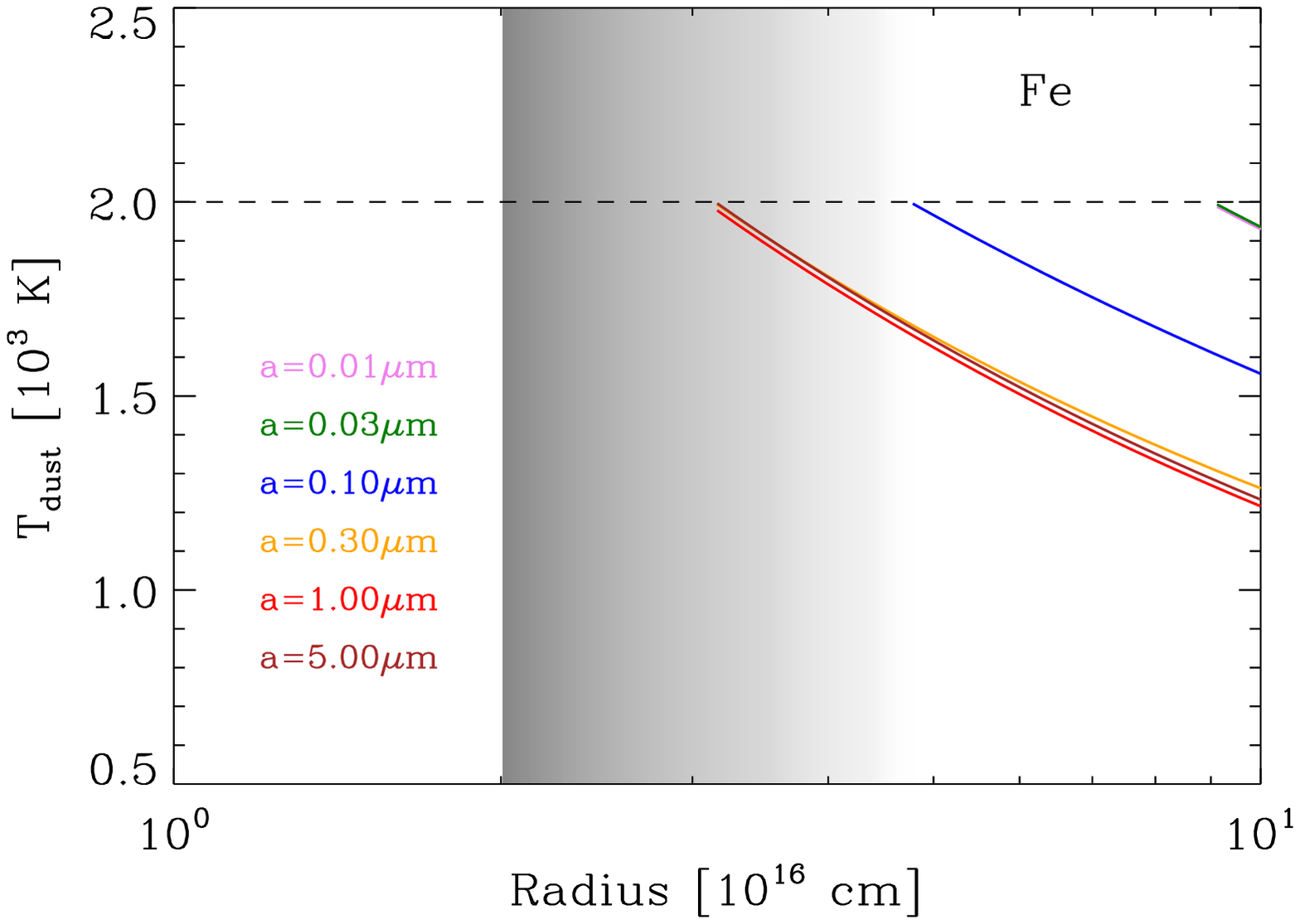}
\includegraphics[width=3.3in]{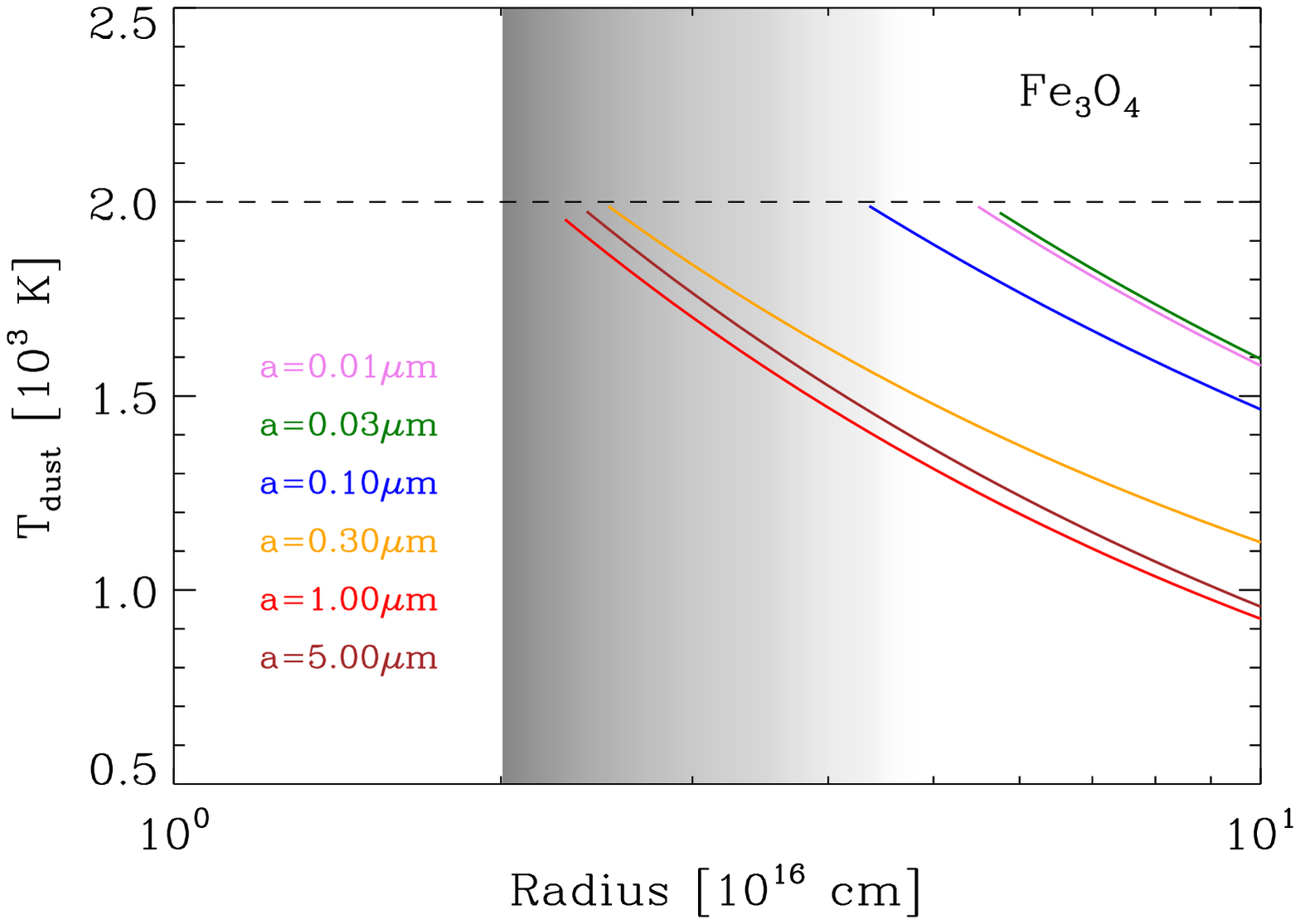}\\
\includegraphics[width=3.3in]{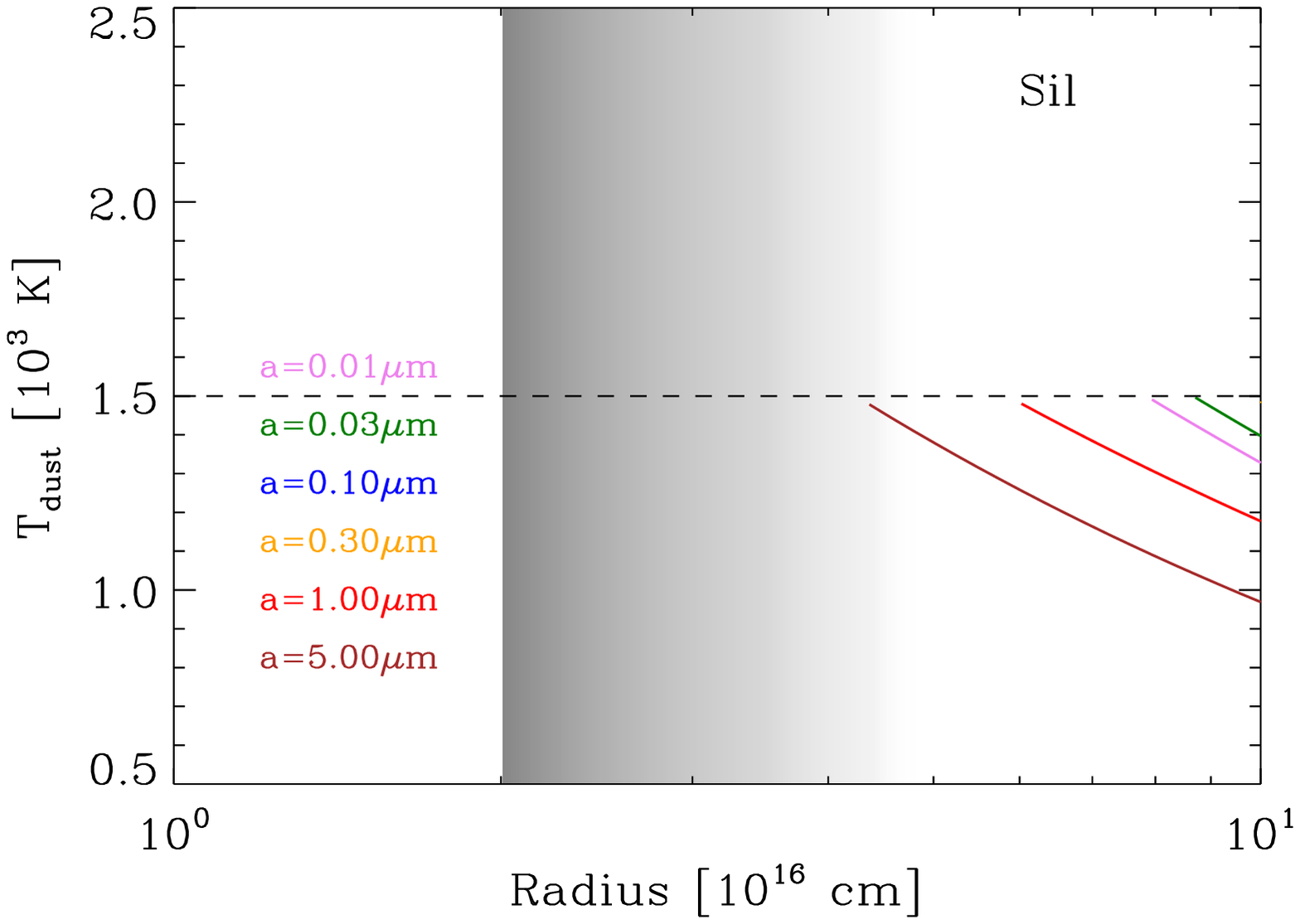}
\includegraphics[width=3.3in]{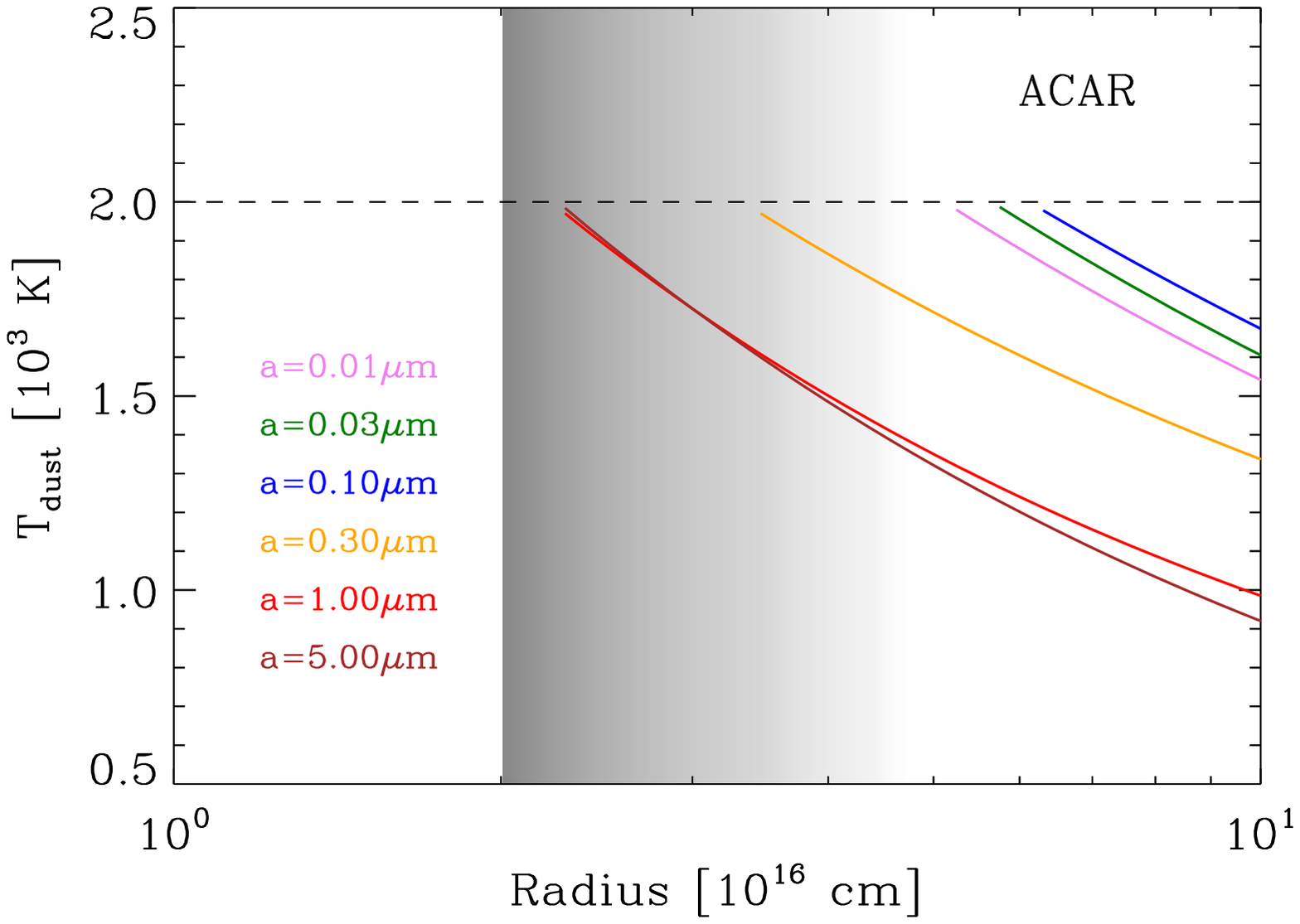}
\caption{\label{Trad} The dust temperature profile at $t=0$ as a function of radius. The dashed horizontal line corresponds to the dust evaporation temperature, and the shaded region represents the density profile of the CSM cavity.}
\end{center}
\end{figure*}

Figure~\ref{Trad} shows the dust temperature profile through the shell at time $t=0$, when it is first exposed to the SN radiation. The different panels and curves correspond to the different dust compositions and grain radii, respectively. The dashed horizontal red line indicates the dust evaporation temperature.  Each curve intersects the evaporation temperature at the location of the evaporation radius. In conjunction with Figure~\ref{csm}, the figure shows that most of the echo from Fe$_3$O$_4$ and ACAR grains arises from the innermost regions of the CSM where its density falls of as $r^{-2}$, whereas most of the echo from Fe and silicate dust arises from the region where the CSM density declines more steeply as $r^{-4.8}$. The figure also shows that most of the echo arises from dust radiating over a narrow range of temperatures below the evaporation value. 
 
\section{The Thermal Echo from SN2010jl }          

Infrared echoes represent dust radiating at a range of physical temperatures. As seen by the central source, each shell of the CSM is characterized by a single dust temperature, which evolves as the SN luminosity declines. From the observer point of view, ellipsoids of equal time delay sample the emission from many shells with different dust temperatures. For a source with declining luminosity, leading ellipsoids  represent dust with higher temperatures than trailing ones, causing a change in the evolution of the echo colors and intensity. The temporal behavior of an echo contains therefore important information on the morphology of the ambient dusty medium, and the composition and size distribution of the CSM dust. 

\begin{figure*}[t]
\begin{center}
\includegraphics[width=3.3in]{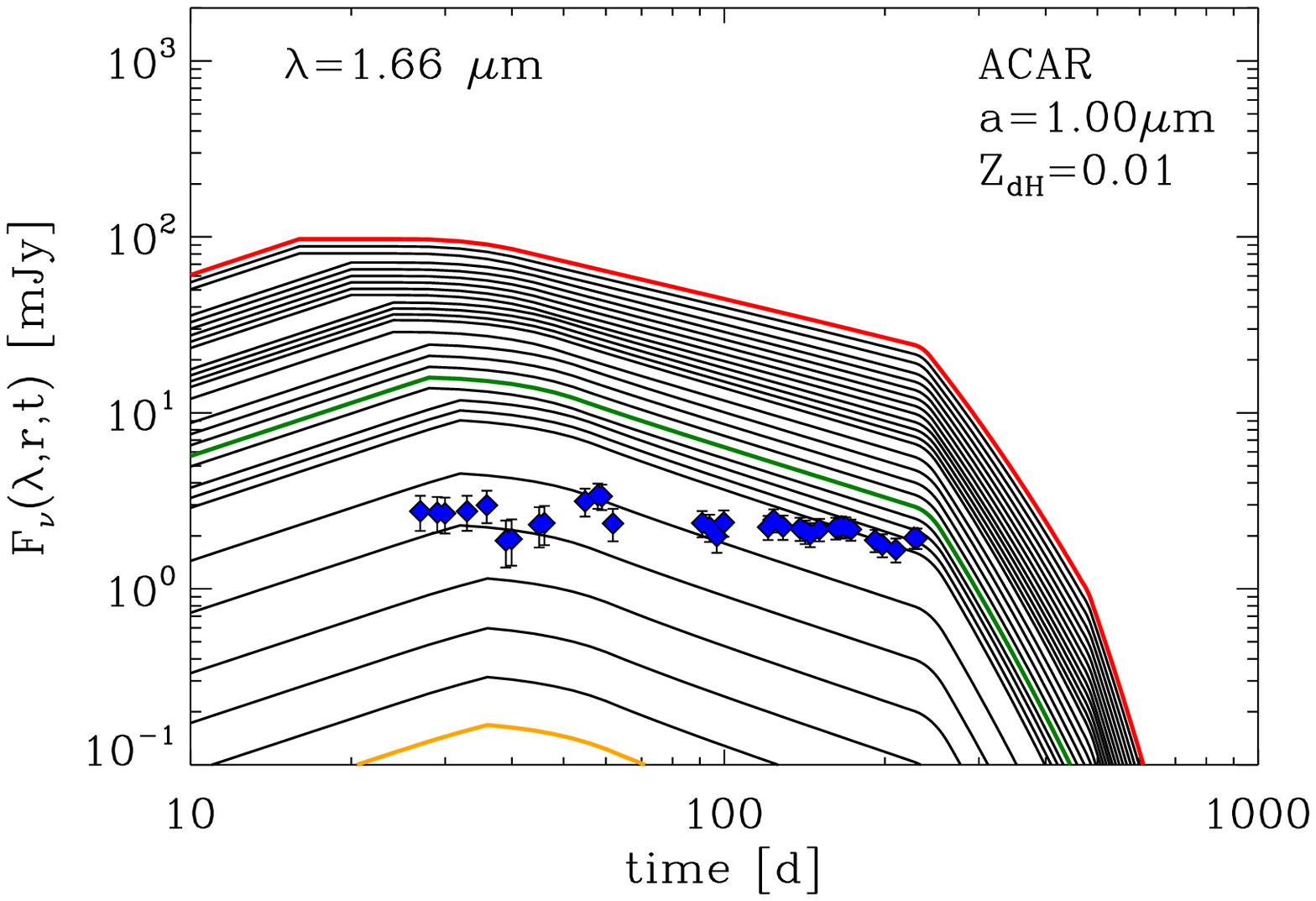}
\includegraphics[width=3.3in]{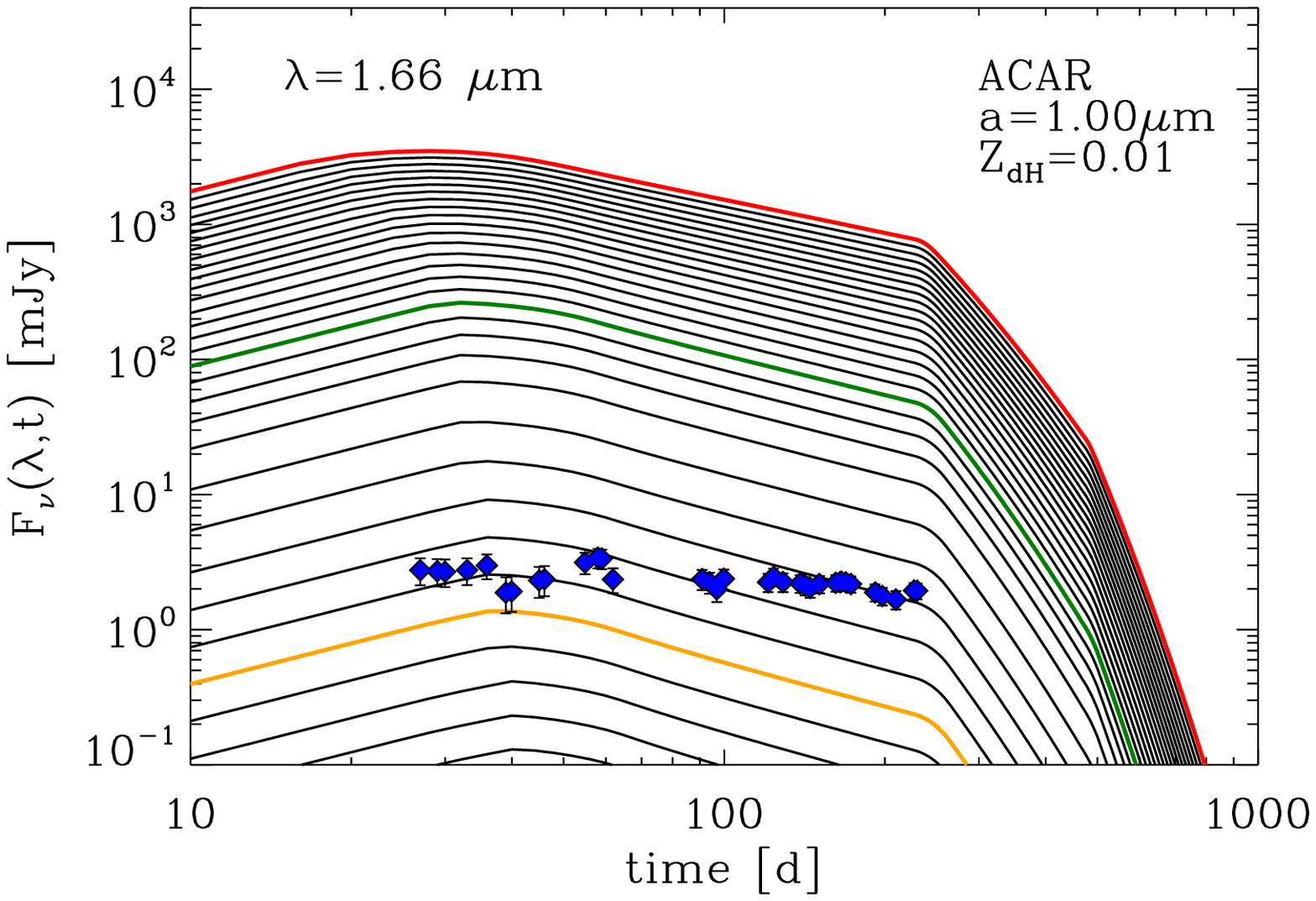} \\
\includegraphics[width=3.3in]{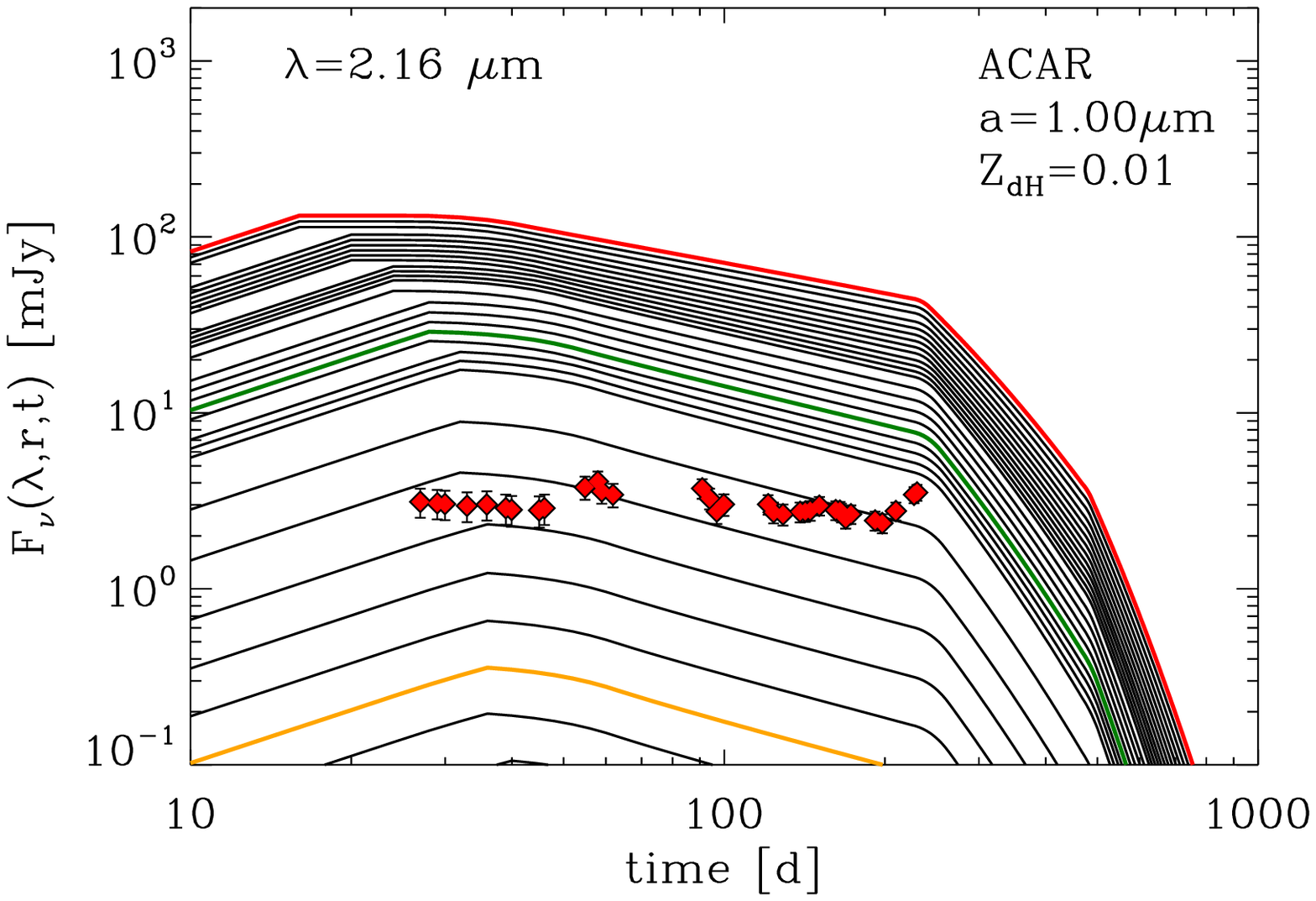}
\includegraphics[width=3.3in]{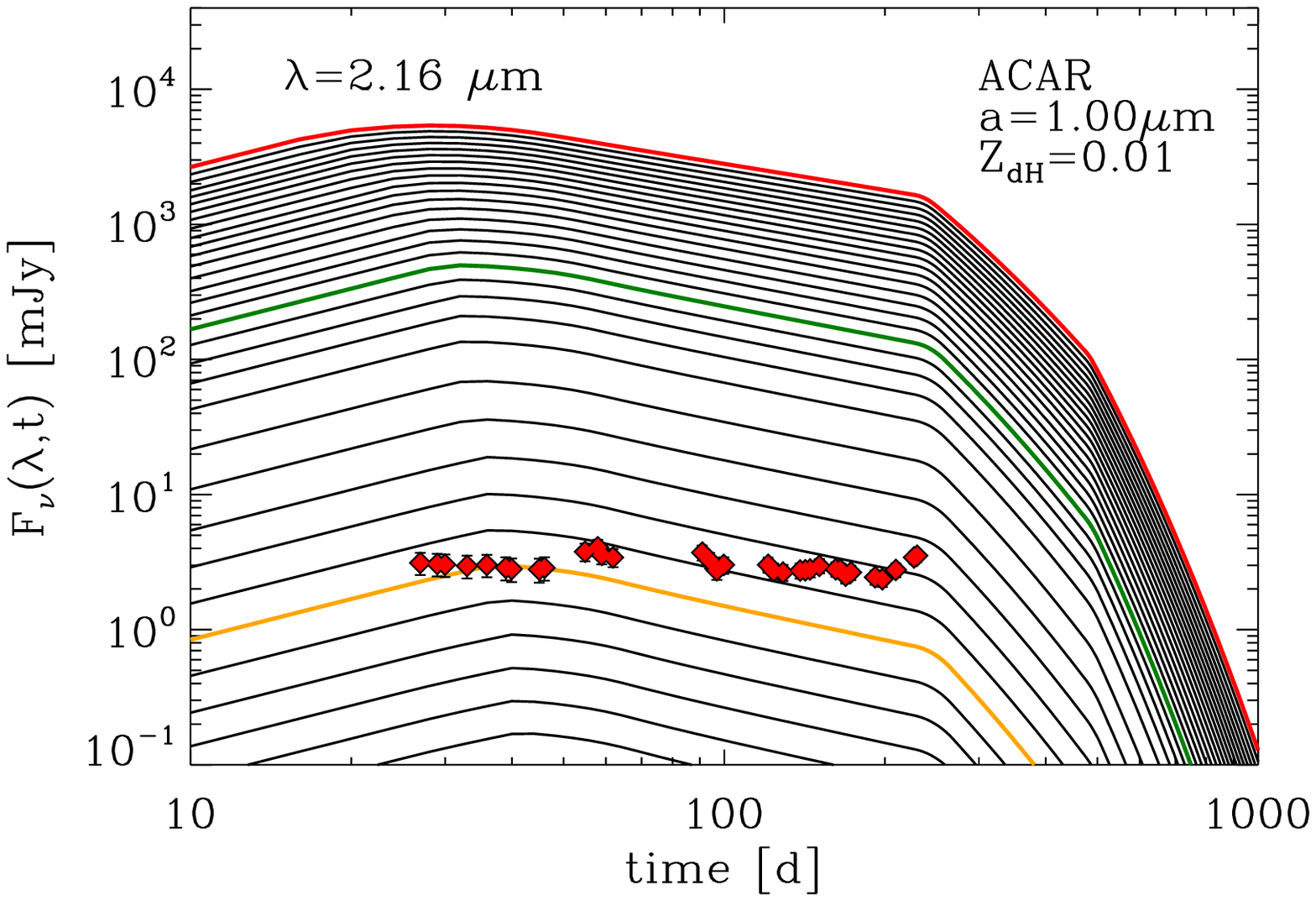}
\caption{\label{shells}  Left panel: The contribution of individual shells to the IR echo in the $H$ and $K$ bands. Right panel: Their corresponding cumulative contribution to the IR echo given by eq.~(\ref{flux3}), for different values of $R_{min}$. The colored curves correspond to values of $R_{min}=5\times 10^{16}$ (orange), $4\times 10^{16}$ (green), and $2.2\times 10^{16}$~cm (red). Results are plotted for amorphous carbon grains with radii of 1.0~\mic. For the sake of presentation, all calculations were done for a dust-to-hydrogen mass ratio of 0.01.}
\end{center}
\end{figure*}

We assume that before the SN event the preexisting dust was present throughout the CSM down to the radius of the inner cavity, so that initially, $R_d = R_0$ in eq.~(\ref{flux3}). 
Figure~\ref{shells} (left column) shows the individual contribution of each shell, consisting of 0.30~\mic\ radii ACAR grains, to the IR echo in the $H$ and $K$ bands for $t=0$. The fluxes, given by $F_{\nu}(\lambda,r,t)$ in eq.~(\ref{flux2}), were calculated for a value of $Z_{dH}=0.01$. The orange, green, and red curves show the individual contribution of shells with radii of $5\times 10^{16}$, $4\times 10^{16}$, and $3.2\times 10^{16}$~cm, respectively.
The evolution of the echo from each shell is determined by the convolution of the dust emissivity with the kernel $\Pi$ (eq.~\ref{flux2}). Initially, all calculated light curves rise as the ellipsoid of equal delay time sweeps over the shell, reaching a maximum value when the entire shell is swept up. The length of this period is determined by the shell radius and is given by $\sim 2R/c$. It is shortest for the innermost shell, which corresponds to the dust vaporization temperature (red curve).
The light crossing time across each shell is shorter than the decline time of the SN luminosity. Consequently, the subsequent evolution of the NIR echo is characterized by an initial slow decline until day 230, reflecting the $t^{-0.53}$ decline in the UVO luminosity of the SN, followed by steeper declines thereafter.

The IR echo consists of the contribution of the emission from all shells with radii larger than $R_{evap}$.
The right column of figure~~\ref{shells} shows the cumulative contribution of all shells, as given by eq.~(\ref{flux3}), for different values of $R_{min}$. The orange, green, and red curves correspond to the integrated flux from the CSM shells for values of $R_{min}=5\times 10^{16}$, $4\times 10^{16}$, and $3.2\times 10^{16}$~cm, respectively, where the latter radius is equal to the vaporization radius of the dust.

\begin{figure*}[t]
\begin{center}
\includegraphics[width=3.3in]{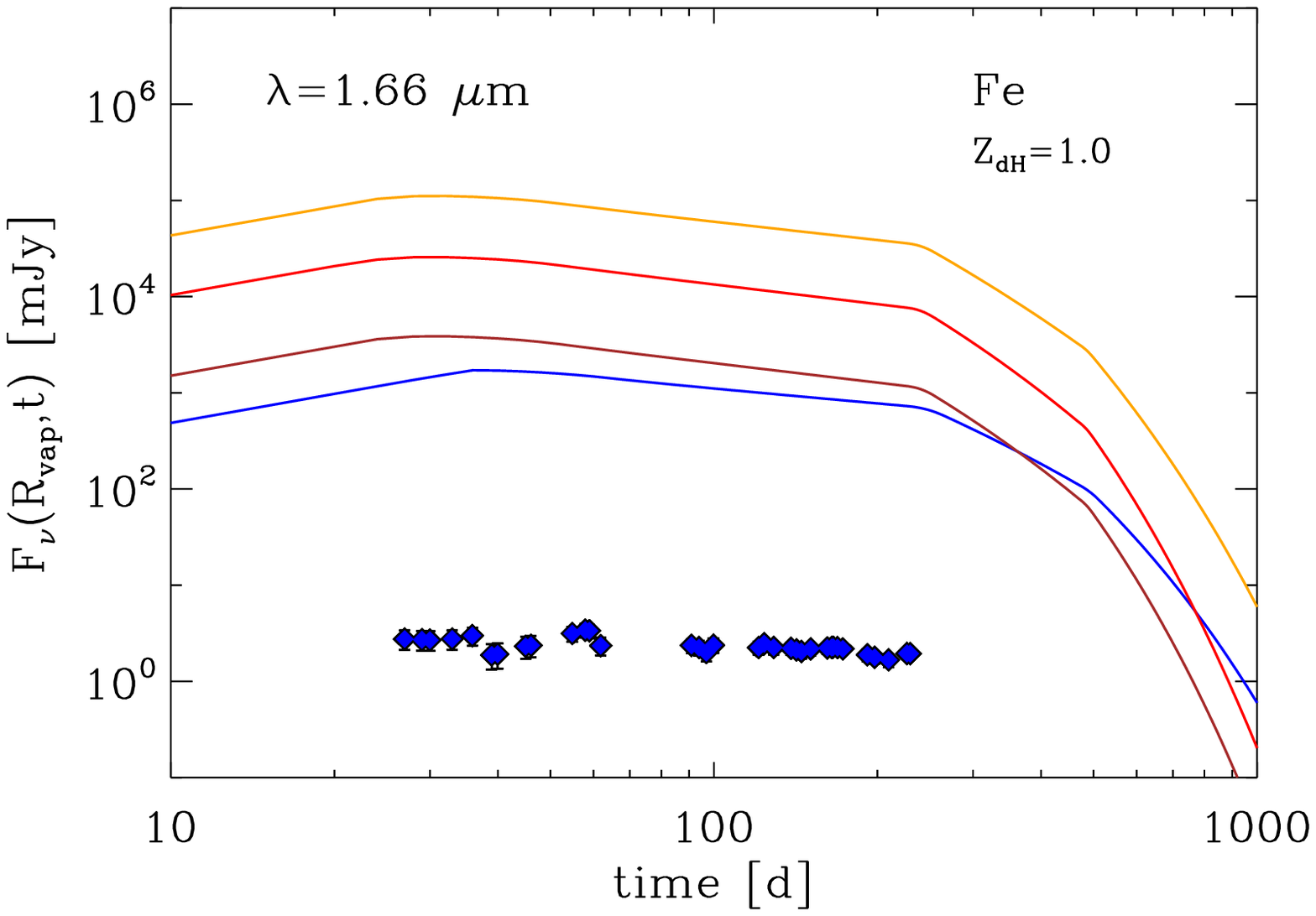}
\includegraphics[width=3.3in]{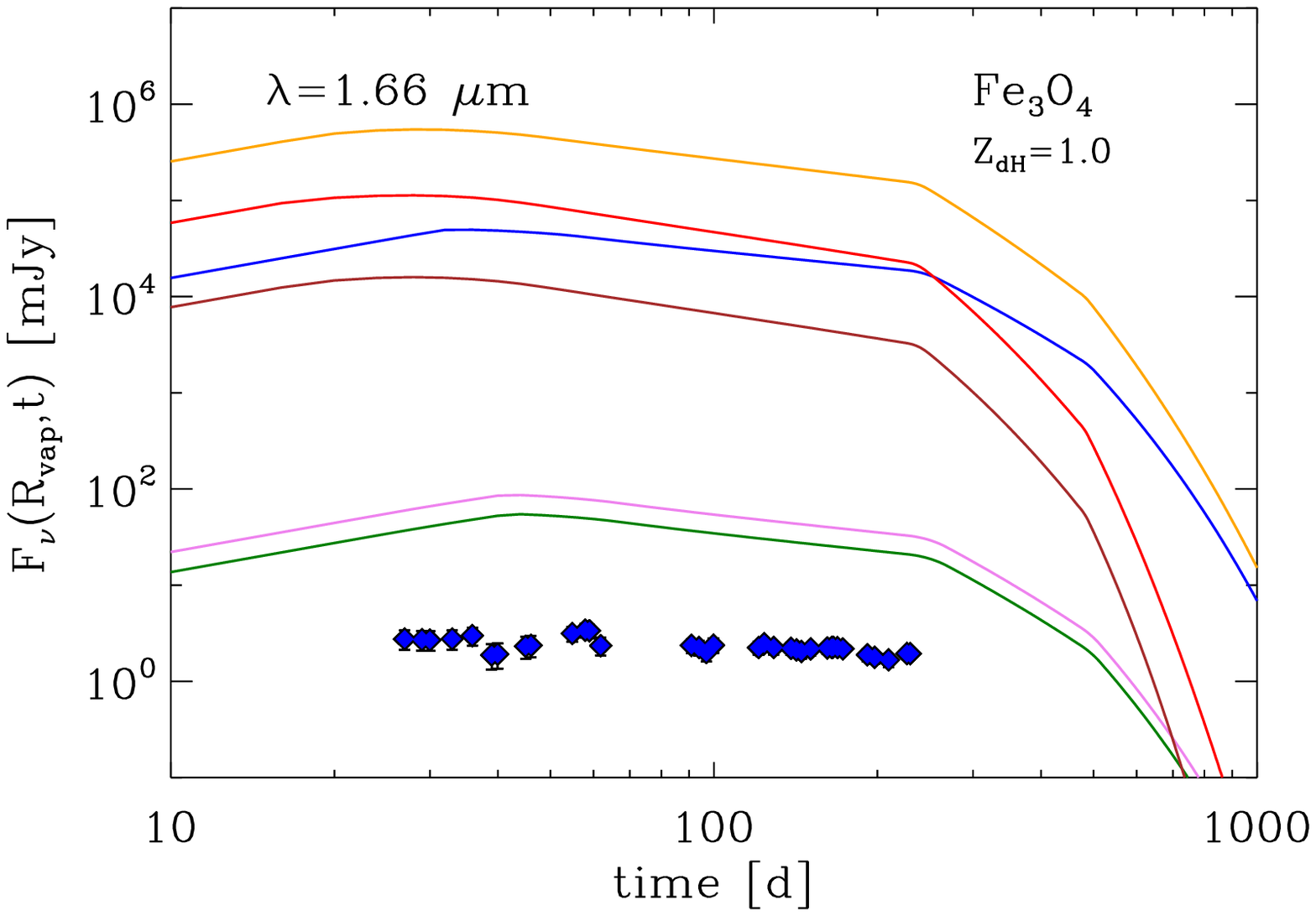}\\
\includegraphics[width=3.3in]{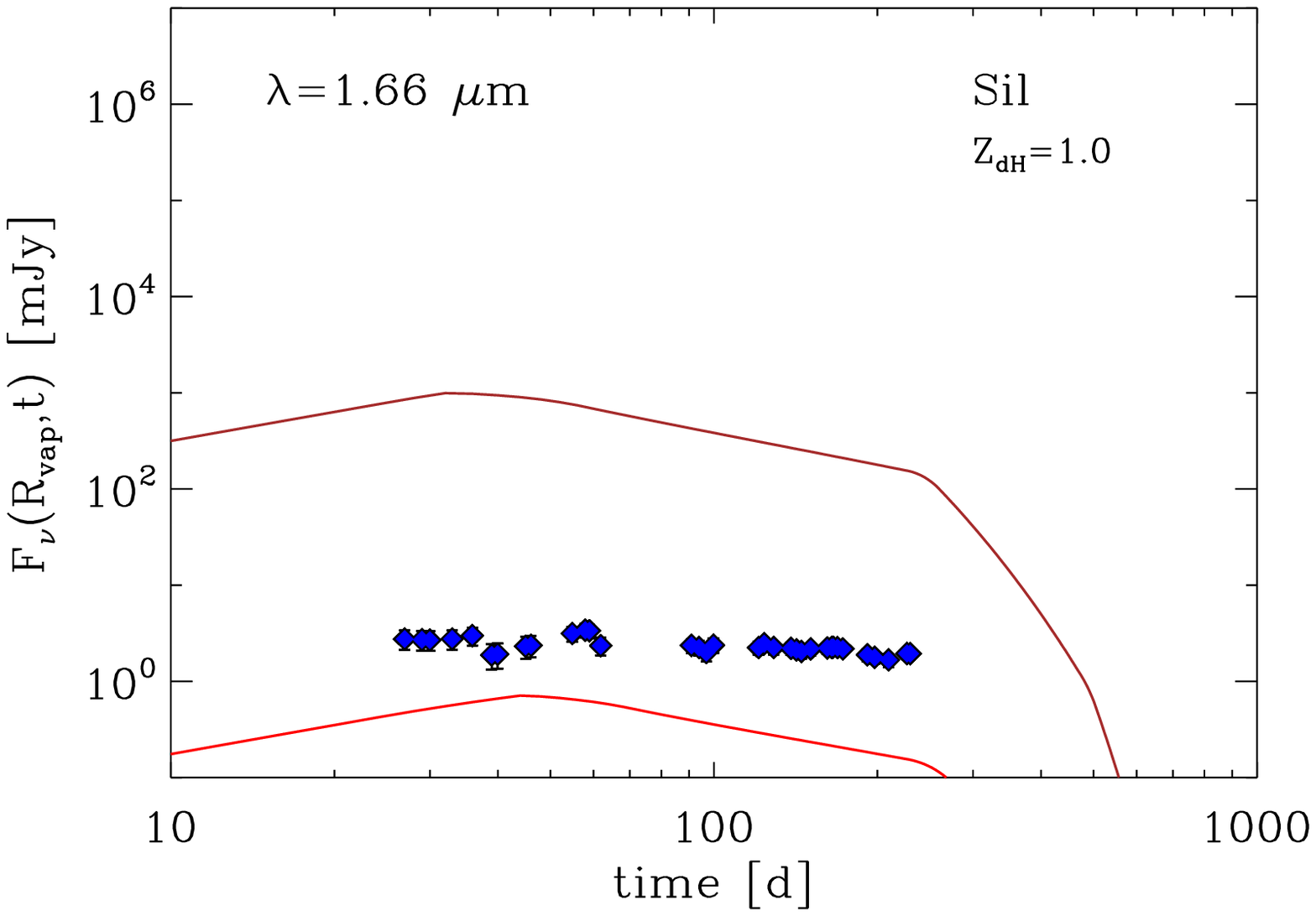}
\includegraphics[width=3.3in]{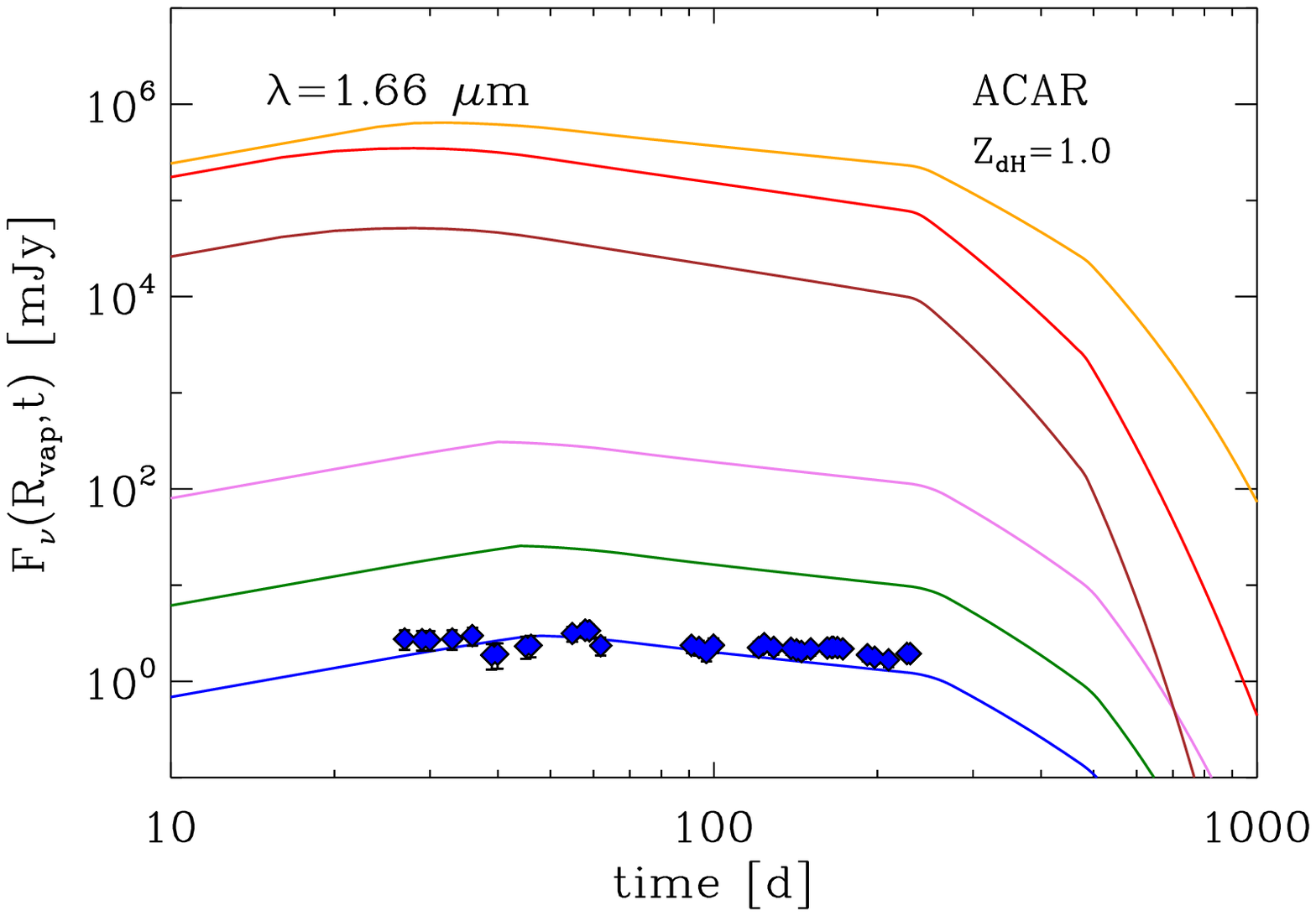}
\caption{\label{FshellsH} The echo in the $H$, band for the different dust compositions and grain radii. The different colored curves correspond to grain radii of 0.01 (magenta), 0.03 (green), 0.1 (blue), 0.3 (orange), 1.0 (red), and 5.0~\mic\ (brown). The inner radius of the dust shell is given by the evaporation radius for each dust composition and grain radius. For sake of the discussion, the results are presented for a dust-to-H mass ratio of 1.0.}
\end{center}
\end{figure*}

\begin{figure*}[t]
\begin{center}
\includegraphics[width=3.3in]{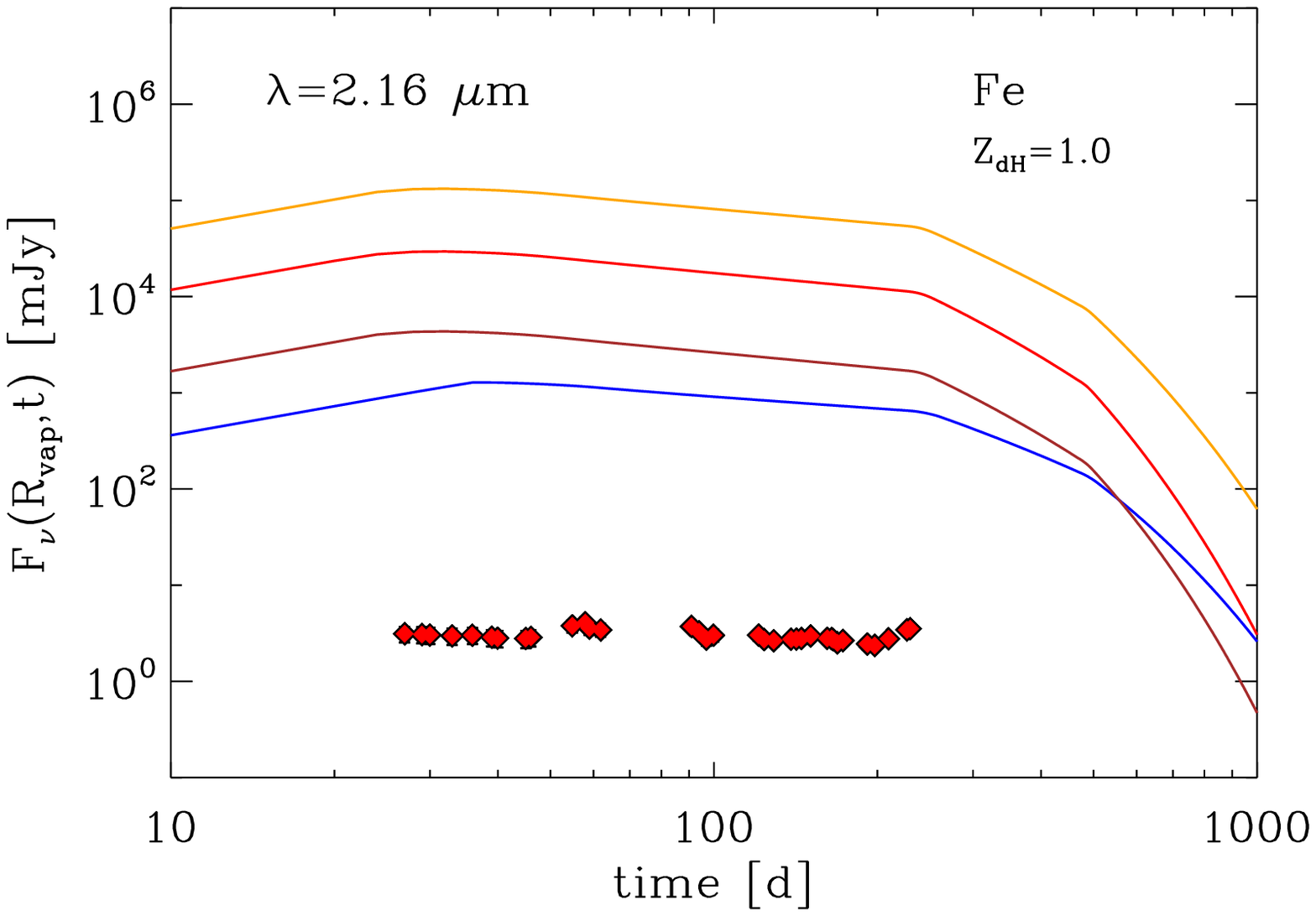}
\includegraphics[width=3.3in]{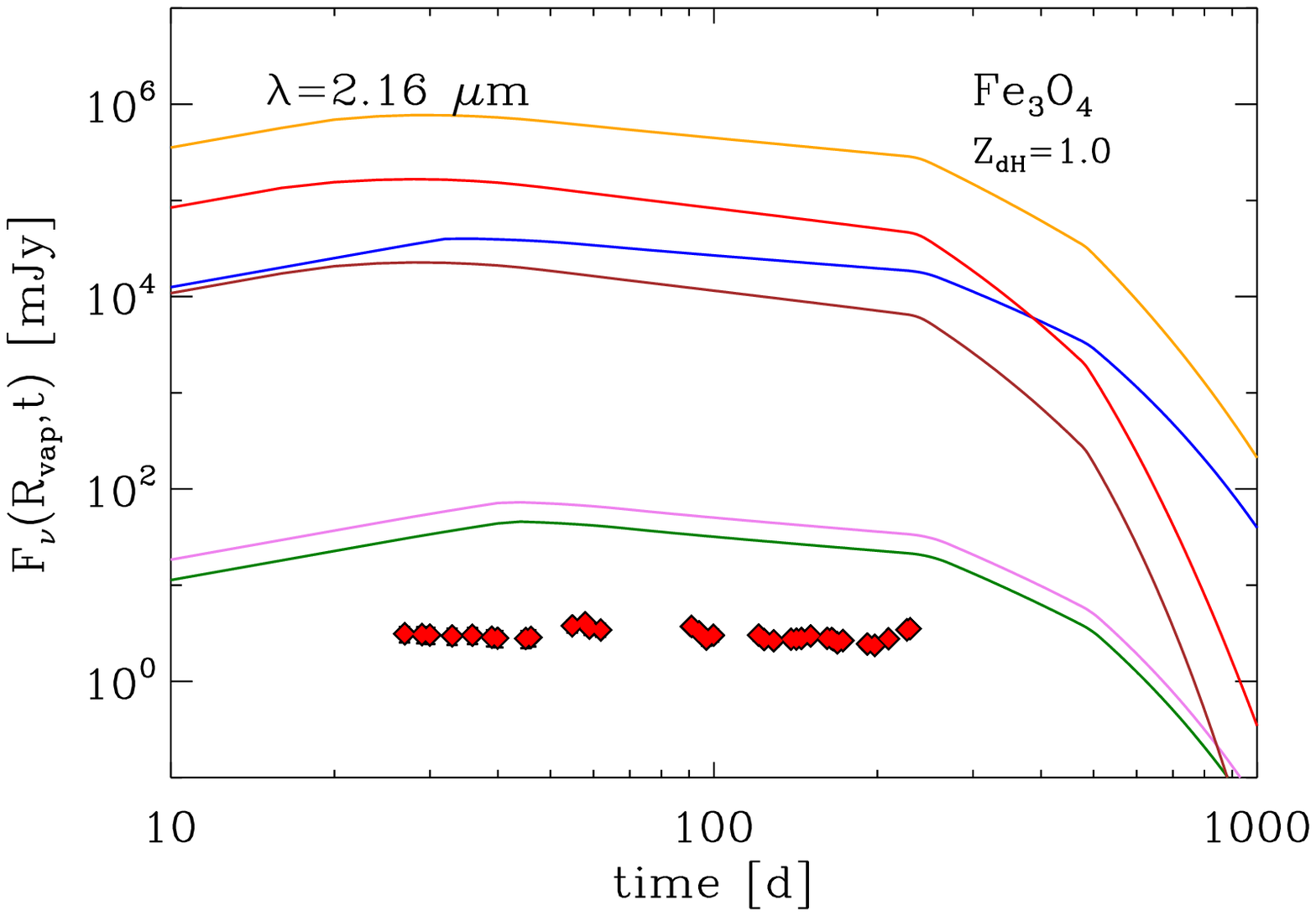}\\
\includegraphics[width=3.3in]{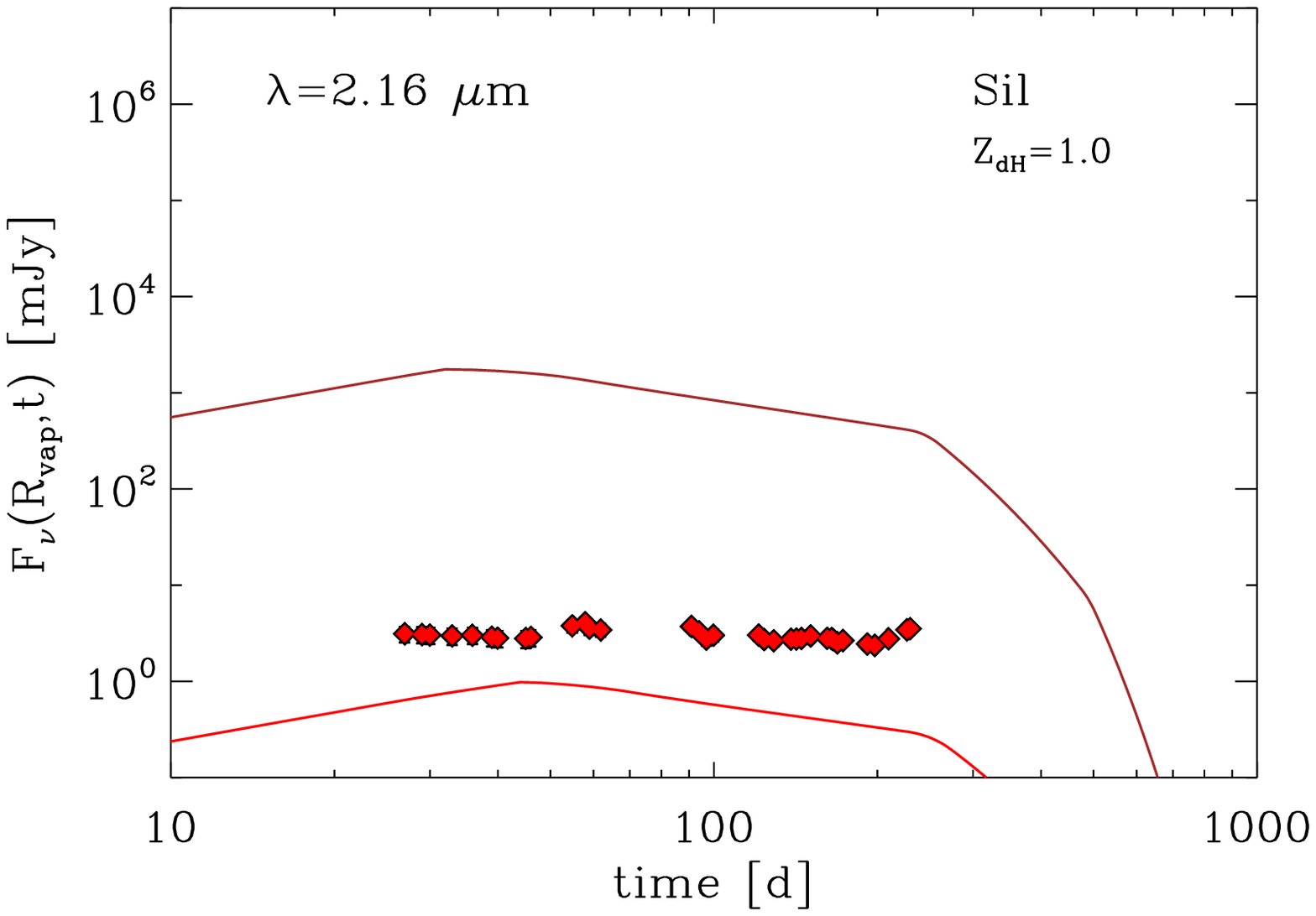}
\includegraphics[width=3.3in]{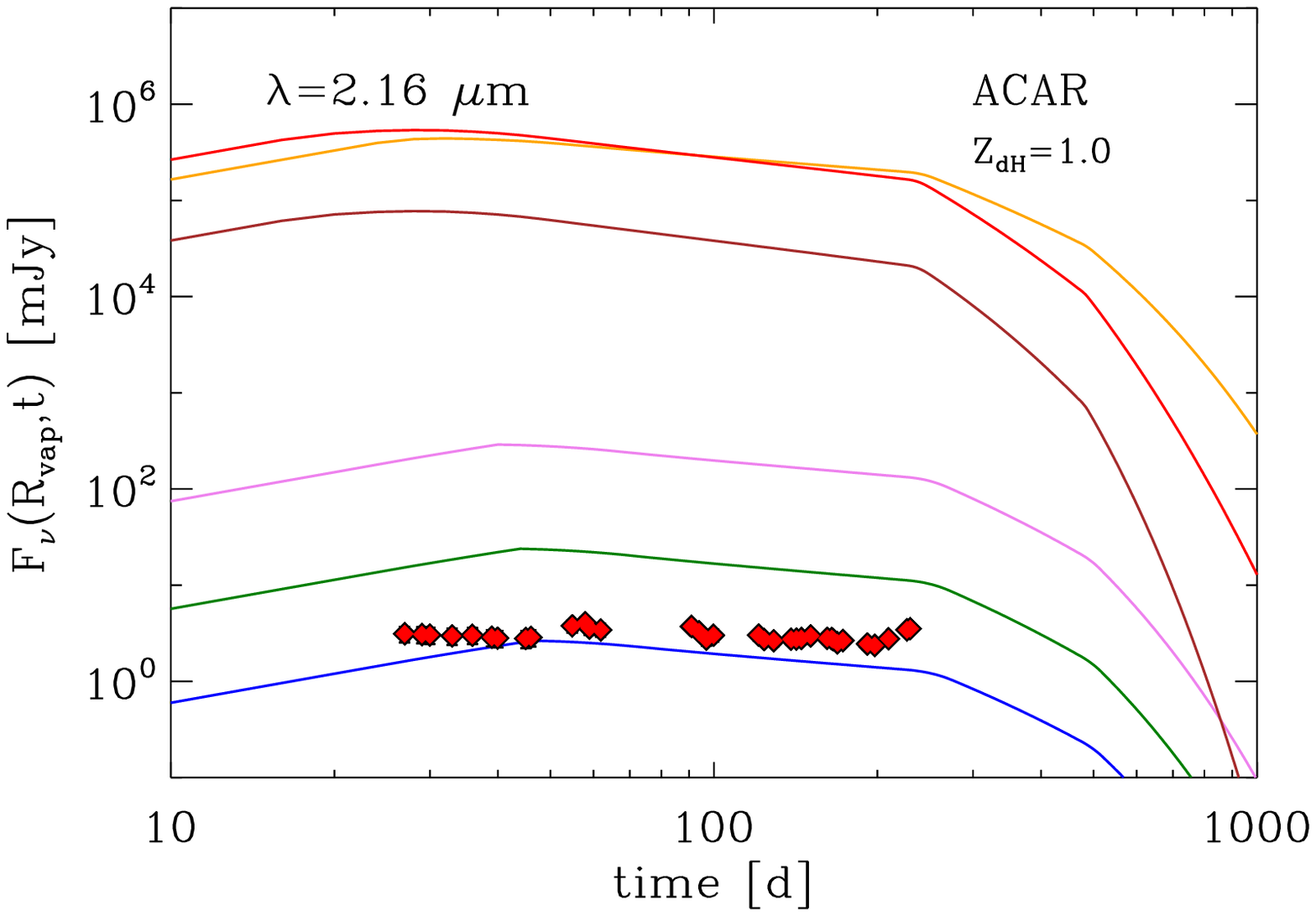}
\caption{\label{FshellsK} Same as figure~\ref{FshellsH} for the $K$ band.}
\end{center}
\end{figure*}

Any viable dust model must reproduce the slope of the observed light curves with a value of $Z_{dH}$, that is $\lesssim 3\times 10^{-3}$, the average solar dust-to-H mass ratio of the dust compositions considered in this paper. We derived the best fit to the evolution of the observed echo by simultaneously fitting both, the $H$ and $K$ light curves for all dust compositions and grain radii. For each dust composition and grain radius, the best fits were obtained when the inner radius of the dust shell was equal to $R_{evap}$, the dust evaporation radius. The results are shown in  figures~\ref{FshellsH} and \ref{FshellsK}, which  show the evolution of the echo in the $H$ and $K$ bands. The rise in the NIR echo reflects here the time for the SN light to sweep across the various shells which sizes are determined by the different dust vaporization radii. For example, for a CSM composed of ACAR dust, the shortest rise in the echo corresponds to 0.3-1.0~\mic\ radii grains which have the smallest evaporation radii (see Figure~\ref{Revap}). In these figures, we purposely plotted the models for a value of $Z_{dH}=1$, to illustrate the abundance constraints on viable echo models 

\begin{figure*}[t]
\begin{center}
\includegraphics[width=3.3in]{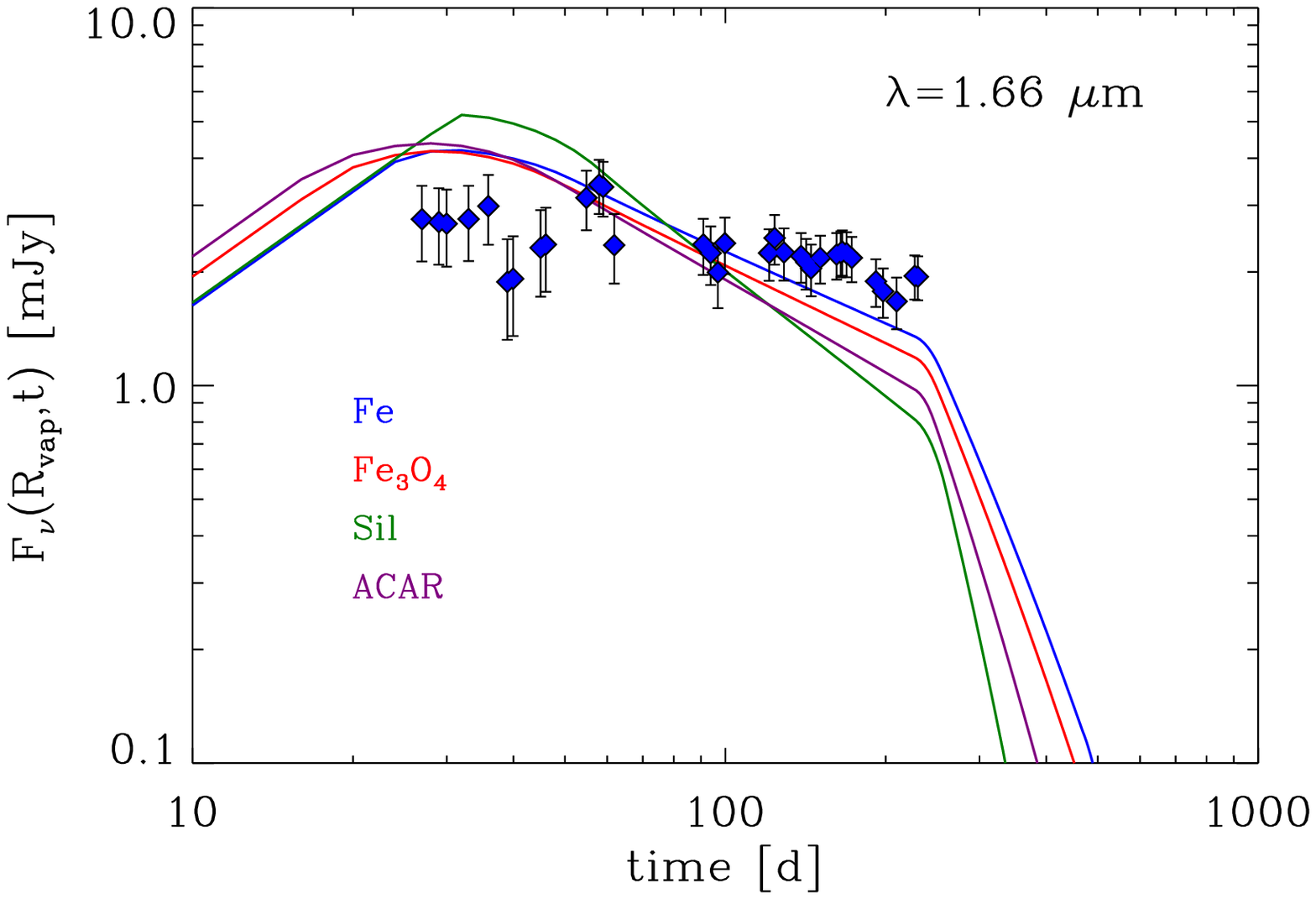}
\includegraphics[width=3.3in]{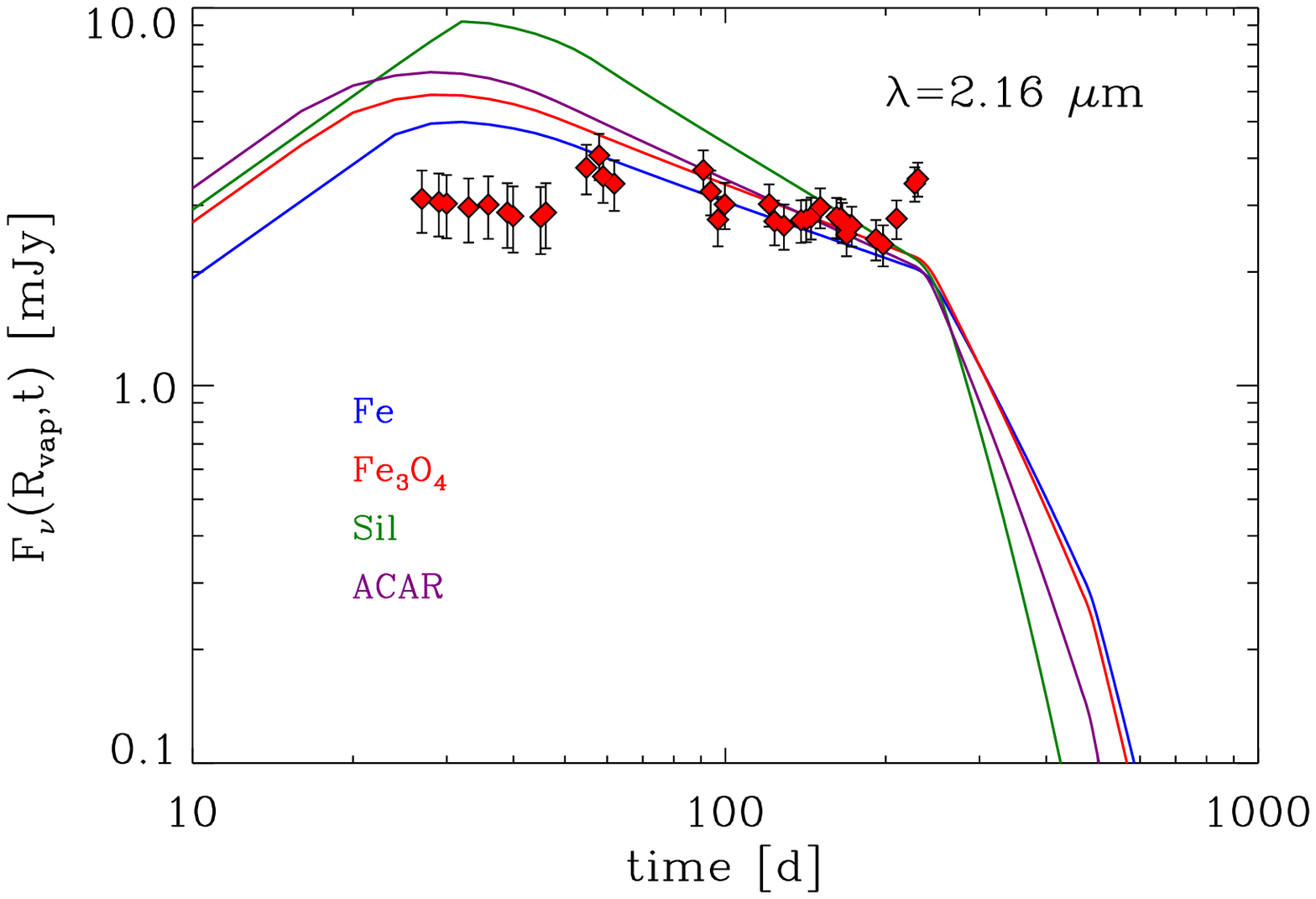}
\caption{\label{echofit} The best fitting dust models to the $H$ (left panel) and $K$ (right panel) band light curves.The fit parameters are listed in Table~1. }
\end{center}
\end{figure*}

 The subsequent evolution of the echo in the $H$ and $K$ bands depends on the evolution of the UVO luminosity of the SN and on the temperature of the radiating dust.  
The echo from all dust constituents arises from a  range of temperatures below the evaporation temperature. 
As seen by the central source, dust temperatures vary with distance. As seen by the observer, they also depend on the temporal evolution of the SN luminosity. This range of temperatures evolves with time, but from the two component fit to the UVO-NIR spectrum we have shown that it can be approximately fit with an effective temperature of about 1250~K, lower that the highest attainable value of the evaporation temperature. 

The SN luminosity decreases by a factor of $\sim 3$ during the first 230~days of evolution,  causing  the dust temperatures to decrease by a factor of $\sim 0.8$ over this period. The resulting decrease in the $H$ and $K$ band fluxes is about a factor of $\sim 3-5$ for Fe, Fe$3$O$_4$, and ACAR grains since these wavelengths are close the peak of their emission spectrum. Because of their lower evaporation temperature, the emission from silicate grains declines more steeply, by a factor of $7-8$ over this period, since the $H$ and $K$ bands are on the Wien side of their emission spectrum. This decline does not match the observed evolution of the IR echo. The sharper decline of the silicate produced echo, compared to the other echoes, is more pronounced in Figure~\ref{echofit}, in which the observed and calculated light curves are plotted with a narrower range of flux intensities.

 Figures~\ref{FshellsH} and \ref{FshellsK} illustrate which grain sizes require a normalization factor that is larger than that allowed from cosmic abundance constraints. 
The figures show that silicate grains are also ruled out as CSM dust constituents by abundance constraints. Only large $\sim 5$~\mic\ radii silicate grains can survive close enough to the source of radiation. The mass absorption coefficient of these large grains is significantly lower compared to smaller radii silicates and other dust candidates for the echo emission (Figure~\ref{kappa}). Consequently a significantly larger mass of silicates is required to produce the echo, compared to other dust species. Furthermore, the silicate emission arises from a low-density region of the CSM, so that the resulting dust-to-H mass ratio exceeds that expected from a CSM containing a solar abundance of silicon (Table~1).    
At first glance, 0.1~\mic\ radii ACAR grains seem to provide a good fit to the slopes of the $H$ and $K$ band light curves. However, the fit requires a value of $Z_{dH}$ of unity, which violates the abundance constrains by a few orders of magnitude.

Figure~\ref{echofit}, shows the best fitting models to the $H$ (left panel) and $K$ (right panel) band light curves.  The figure shows that the calculated smooth light curves are not able to reproduce the scatter in the observed light curves. These may be a consequence of the simple blackbody approximation used to represent the SN spectrum in deriving the residual emission from the echo. The scatter may also reflect the presence of real inhomogeneities in the CSM, which cannot be reproduced with the smooth spherical CSM model adopted in our calculations. In spite of these discrepancies, the model illustrates the constraints on any given dust model, given CSM and dust properties that are required to produce the observed SN echo. Given the time dependence of the SN luminosity and the density profile of the CSM, the observed echo provides strong constraints on the composition and size of the dust that can give rose to the echo.

Table~1 summarizes the parameters of the best fitting dust models considered in the paper: the grain radius, the dust evaporation radius and its corresponding delay time, the total dust mass giving rise to the echo, the required dust-to-H mass ratio, the maximum allowed value for a CSM of solar composition, and the {\it pre-explosion} visual optical depth. The latter was derived assuming the dust existed throughout the CSM, $R_d=R_{min}=R_0$ in eq.~(\ref{tauW}), before it was partially evaporated by the burst of radiation from the shock breakout or the subsequent shock-CSM interaction.
A CSM consisting of ACAR dust provides the best fit to the observed echo. The model shows that about $3.5\times 10^{-5}$~\msun\ of ACAR dust located at a  distance of $\sim 2.2\times 10^{16}$~cm from the center of explosion is required to produce the NIR echo.
The dust mass is significantly smaller than that derived by \cite{andrews11}, a difference we attribute to the significantly larger dust temperature derived in our analysis. The pre-explosion optical depth of the CSM is 0.035, consistent with the observed upper limit of $\tau(V) = 0.15\pm0.07$ \citep[][and references therein]{dwek17}.

\begin{figure*}[t]
\begin{center}
\includegraphics[width=3.3in]{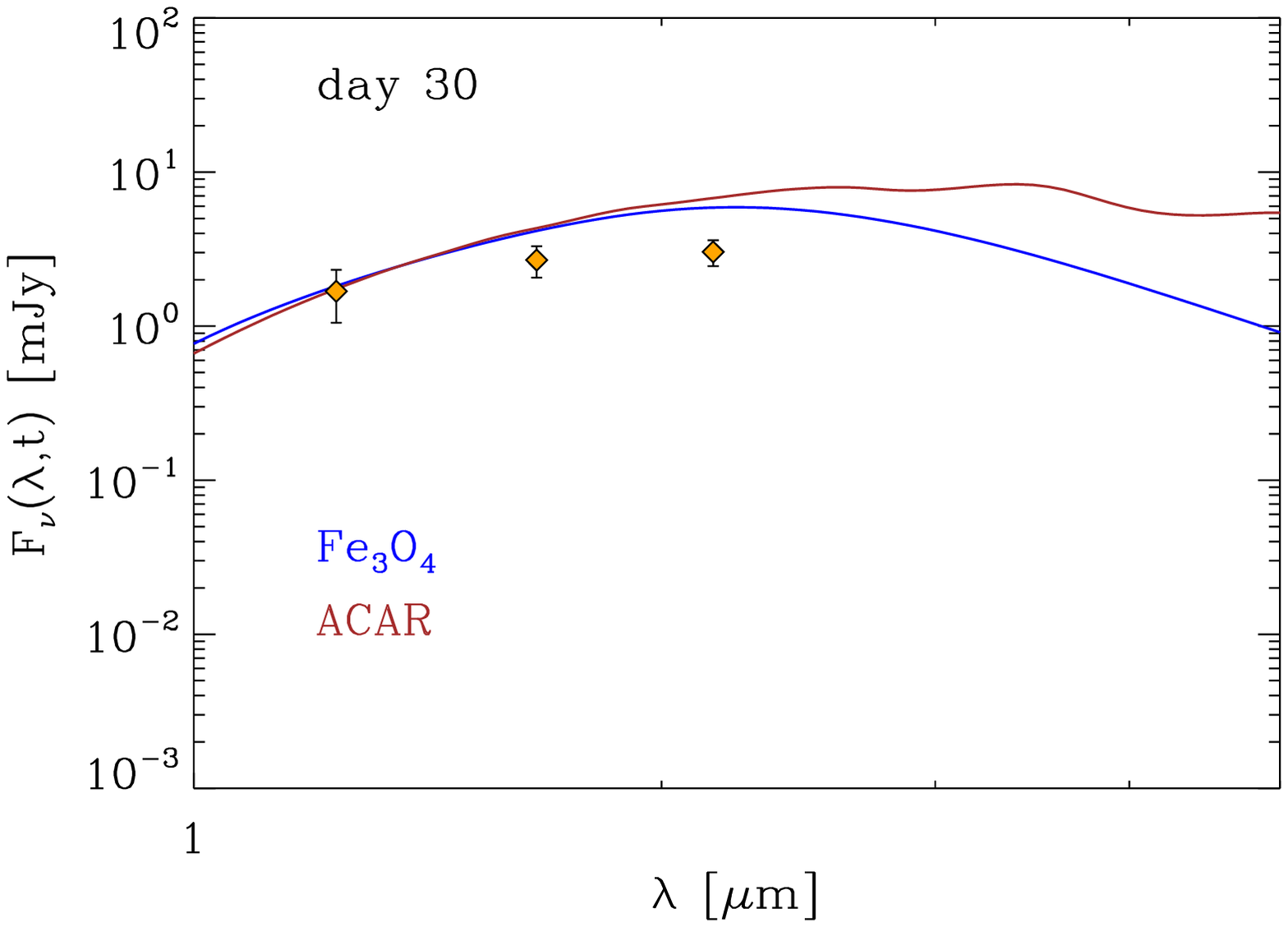}
\includegraphics[width=3.3in]{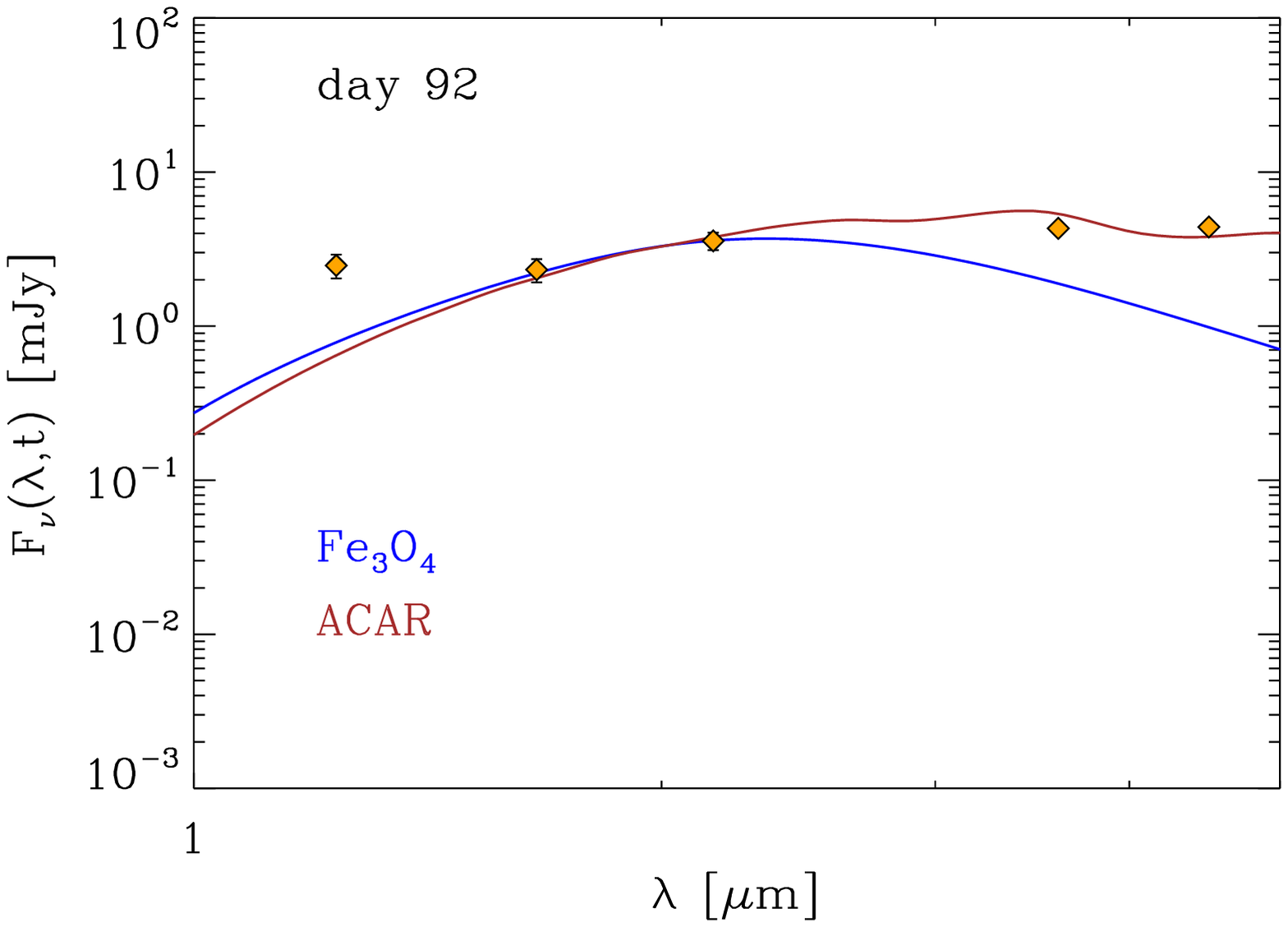}\\ 
\includegraphics[width=3.3in]{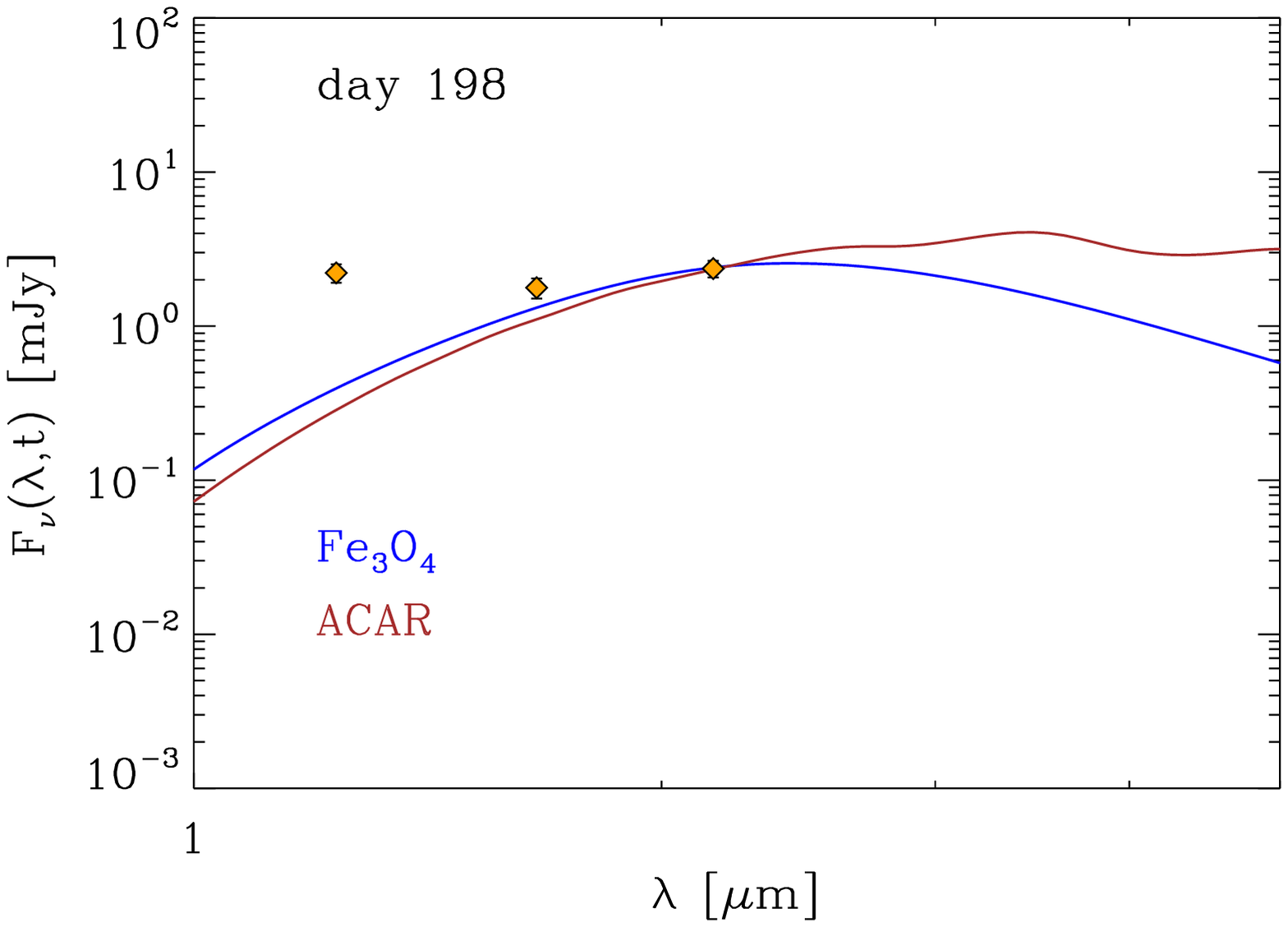}
\includegraphics[width=3.3in]{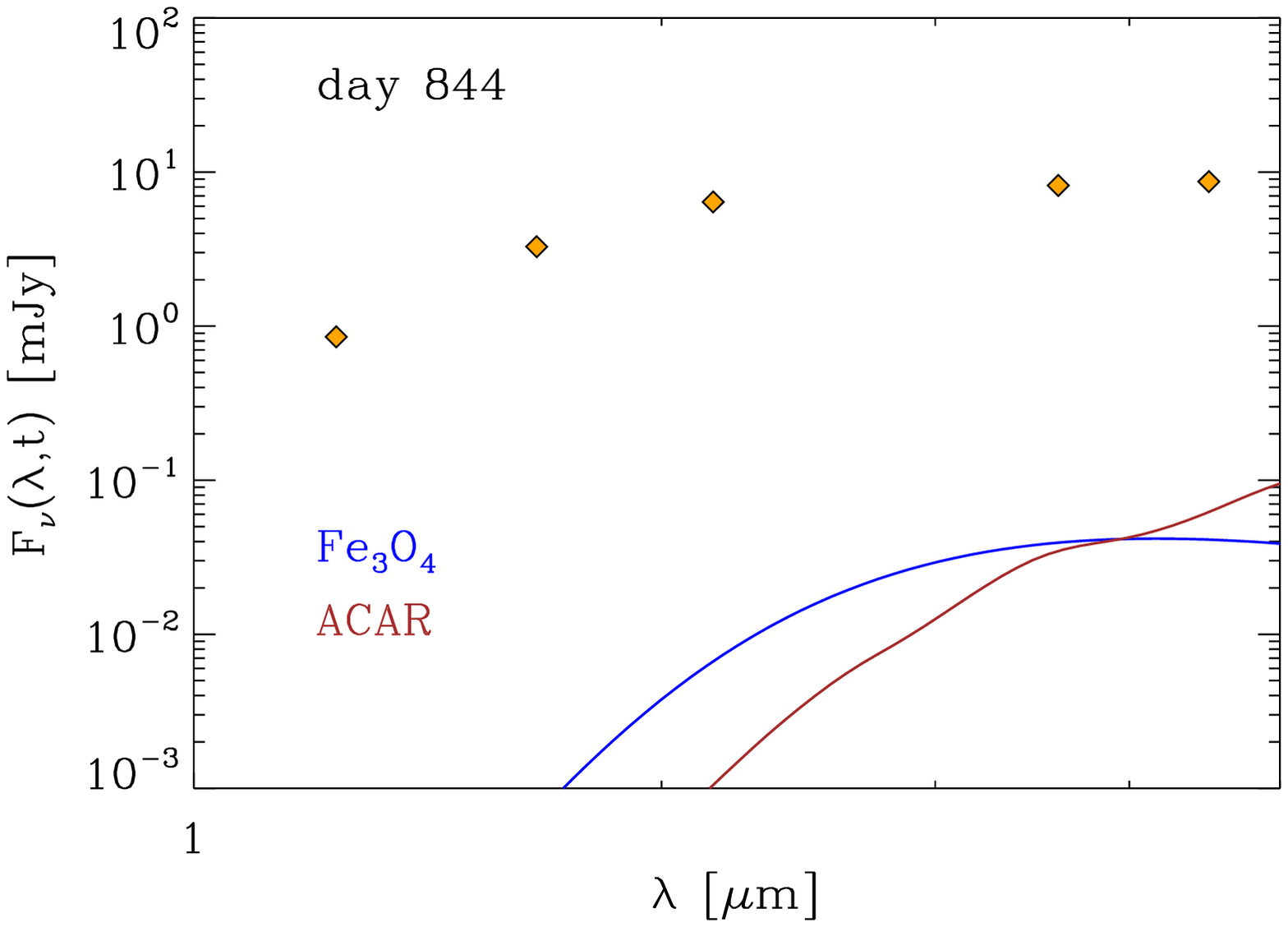}
\caption{\label{specfit} The IR emission from the echo is compared to the observed fluxes at select epochs. The echo provides a good fit to the early IR spectrum on days $\lesssim 210$, but makes a negligible contribution to the IR spectrum at later times, when it is generated by the emission from newly-formed CSM dust. }
\end{center}
\end{figure*}

Figure~\ref{specfit} shows the spectrum of the echo for two different epochs for which IRAC data were available. On day~92 the echo provides a good fit to the IR excess emission from the SN. On day 844 the echo contribution to the IR light curve is negligible. This result is not surprising because of the rapid decline in the SN light curve after day $\sim 230$ which is needed to power the echo. This comparison provides conclusive support to our assumption that the echo does not contribute significantly to the late time NIR emission.  As shown by \cite{sarangi18}, the IR emission at epochs $\gtrsim 300$~d is generated by dust that is newly formed in the postshock region of the shocked CSM.  
 
\begin{deluxetable*}{lcccccccc}
\tablecaption{Dust fits to the NIR echo}
\tablehead{
\colhead{Dust composition} &
\colhead{$a(\mu$m)} &
\colhead{$R_{evap}$(cm)} &
\colhead{$2R_{evap}/c$\ (d)} &
\colhead{$M_d$(\msun)}   &
\colhead{$Z_{dH}$}   &
\colhead{$Z_{dH}(\odot)$\tablenotemark{a}}   &
\colhead{$\tau$(V)\tablenotemark{b}}  
}
\startdata
  Fe                  &   0.30             &   $3.14\times10^{16}$  & 24.2    & 2.65 $\times10^{-4}$    & 3.76$\times10^{-5}$   & $1.77\times10^{-3}$   & 0.31    \\
 Fe$_3$O$_4$          &  0.30              &   $2.48\times10^{16}$  & 19.2    & 8.94 $\times10^{-5}$    & 7.62$\times10^{-6}$   & $2.44\times10^{-3}$   & 0.14    \\
  Sil                 &  5.0               &   $4.24\times10^{16}$  & 32.7    & 5.06 $\times10^{-3}$    & 5.25$\times10^{-3}$   & $2.68\times10^{-3}$   & 6.66    \\
   ACAR               &  1.0               &   $2.22\times10^{16}$  & 17.2    & 1.64 $\times10^{-4}$    & 1.25$\times10^{-5}$   & $3.22\times10^{-3}$   & 0.16    \\
 \enddata
 \label{Table1}
\tablenotetext{a}{Solar abundances were taken from \cite{asplund09}.}
\tablenotetext{b}{The pre-explosion optical depth in the V-band.}
\end{deluxetable*}


 \section{Echo Constraints on the Breakout Luminosity and Temperature}    

In the previous section we derived the composition, typical grain radii, and proximity to the SN of the dust that is required to produce the NIR echo. This scenario may not reflect the pre-SN conditions of the CSM, since the radiative energy that powers the observed echo is preceded by an intense flash of radiation that is produced by the breakout of the shock through the stellar surface.  If the conditions of the CSM prior to the SN event were known, then any changes in the dust morphology and grain sizes could be used to constrain the characteristic of the shock breakout. 
In the following we will assume that dust formation was a continuous process in the wind that generated the CSM, so that the preexisting dust extended to radius $R_0$, the radius of the dust-free cavity. We found that the best fitting models to the IR echo require the the inner radius of the dust shell to be equal to the dust evaporation radius, which in all cases is larger than $R_0$. The echo requires the survival of all the dust beyond the distance of $R_{evap}$. Consequently, the radiation from the shock breakout must not evaporate the dust to larger distances.  Constraining the evaporation radius of the shock breakout to be $\lesssim R_{evap}$ allows us to derive limits on the intensity, temperature, and duration of the shock breakout.
\subsection{Shock Breakout Characteristics} 
The collapse of the iron core in CCSN and the subsequent bounce generates an outward moving shock that breaks out through the stellar surface. This shock breakout produces an intense burst of radiation that precedes the radioactively-powered luminosity of the SN, or the shock-CSM powered luminosity in SN~IIn. The  burst characteristics, its luminosity, effective temperature, and duration, are determined by the energy of the explosion, the stellar mass and radius, and the radiative transfer of the photons through the optically thick stellar surface \citep[e.g.][]{blinnikov00}.

The best fitting echo models requires the presence of ACAR dust at a distance of $\sim 2.2\times 10^{16}$~cm from the explosion, and that it not be evaporated by the shock breakout radiation. This requirement provides important constraints on the characteristics of the burst.  Similar considerations were used to characterize the shock breakout radiation that followed the collapse of the progenitor of the Cas~A SNR \citep{dwek08b}. 

Numerical simulation \citep{klein78,ensman92,blinnikov00,blinnikov11a,blinnikov11b} suggest a range of possible burst luminosities, effective temperatures, and burst duration. An EUV spectrum characterized by an effective blackbody temperature of $6\times 10^5$~K was  inferred from the need to reproduce the UV line ratio from the circumstellar ring around SN1987A \citep{lundqvist96}. A burst luminosity of $\sim 2\times 10^{11}$~\lsun with an effective blackbody temperature of $2\times 10^4$~K, was needed to produce the  thermal IR echo from the interstellar dust around Cas~A \citep{dwek08b}. To bracket all possible burst luminosities and temperatures, we considered bursts with luminosities ranging from $10^9$ to $10^{12}$~\lsun, and  blackbody spectra with temperatures  ranging from 8000 to $10^6$~K. The lowest temperature and luminosity were chosen to bracket the parameters of the SN light curve generated by the CSM shock.
Burst duration times, derived from numerical simulations and corrected for light travel time delays across the stellar disk, are around 1000~s \citep{ensman92, blinnikov11a, fryer20}.

In calculating the dust evaporation radius, one has to take two effects that modify the burst spectrum into account: the effect of Thomson scattering off CSM electrons en route to the dust, and the absorption and reradiation of ionizing photons giving rise to a nebular spectrum.  

 \subsection{The Effect of Thomson Scattering}

A burst of photons emitted over a time interval $\Delta t_b$ will be spread out in time as a result of Thomson scattering with the electrons in the ambient plasma. For a source embedded in an electron cloud with an $r^{-2}$ density distribution the time spread of the emerging signal is given by a distribution function $P(u)$  \citep[][eq. (10)]{sunyaev80}

\begin{equation}
\label{pu}
P(u) = {3\over2}\, {ln(\tau_0)\over \tau_0 \sqrt{\pi}}\, \left({3\tau_0^2\over u}\right)^{3/2}\ \exp\left[-{3 u\over 4\tau_0^2} - {3 \tau^2\, ln^2(\tau_0)\over 4u}\right]
\end{equation} 
where   $u=\sigma_0\, n_e\, ct$, $\sigma_0=6.65\times 10^{-25}$~cm$^2$ is the Thomson cross section for electrons, and $\tau_0$ is the Thomson optical depth to the location of the dust in the CSM.   

\begin{figure}[t]
\includegraphics[width=3.3in]{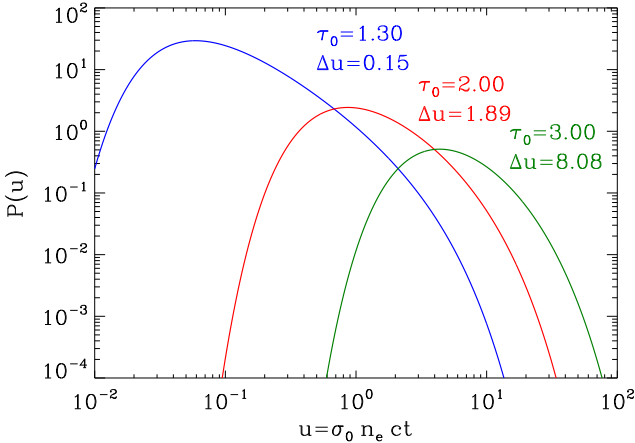}
\caption{\label{Pu} The distribution of a $\delta-$function pulse of photons escaping from a shell as a function of the dimensionless time parameter $u=\sigma_0\, n_e\, ct$. The curves show the spread in time of the burst after traveling a Thomson optical depth, $\tau_0$. The time spread is proportional $\Delta u$, calculated at the FWHM of $P(u)$.}
\end{figure}

Figure~\ref{Pu} depicts the function $P(u)$ for the derived optical depth of 1.30 to the dust shell (blue curve). The function $P(u)$ for values of $\tau_0=2.0$ (red curve) and $\tau_0=3.0$ (green curve) are shown for comparison.  Also listed in the figure is $\Delta u$, defined as the FWHM of $P(u)$ for a given value of $\tau_0$.  The figure shows that after traversing an optical depth of 1.30 the burst will be dispersed over a time interval $\Delta t$, given by 
 \begin{equation}
\label{deltau}
\Delta t ={ \Delta u \over \sigma_0 \, n_e(R_0)\, c } \approx  5\times10^4\ {\rm s} \qquad.
\end{equation}

 Since the total energy of the burst is conserved, the effective luminosity of a burst, $L_{eff}$ will be reduced by a factor $\Delta t/\Delta t_b$, so that
 \begin{equation}
\label{leff}
L_{eff} = {\Delta t_b \over \Delta t}\, L_b
\end{equation}
where $L_b$ and $\Delta t_b$ are the initial burst luminosity and duration, respectively. Photon energies are significantly smaller than the mass energy of the electrons, so that the burst spectrum is not affected by Thomson scattering in the CSM.

\subsection{The Effect of Nebular Interaction} 
Ionizing photons from the burst  will interact with the CSM ionize and excite its various atomic constituents. The absorbed energy will be reradiated in the form of line and continuum emission. We used the spectral synthesis code CLOUDY \citep{ferland98} to calculate the processed burst spectrum that the dust will be exposed to. Calculations were performed for a number of burst spectra  traversing an H-column density of $1\times 10^{24}$~cm$^{-2}$.   
The calculations used the CSM CNO abundances derived  by \cite{fransson15}, which differ from their solar values as the result of CNO processing by the progenitor star. In earlier considerations of dust content (Table~1), solar abundances were assumed.

The processed burst spectra are shown in Figure~\ref{spec_burst}. The three panels correspond to initial burst spectra characterized by temperatures of $10^4$, $10^5$, and $10^6$~K. Each panel depicts the spectra for burst luminosities of $10^9$, $10^{10}$, $10^{11}$, and $10^{12}$~\lsun. The initial burst spectra are shown as a dashed line.     

The processed spectra comprise of several emission components: (a) a transmitted component consisting of ionizing photons that were not absorbed within the intervening column density; (b) a nebular continuum and line emission component; and (c) a component of transmitted non-ionizing photons. Most of the luminosity in the $10^4$~K bursts spectra is emitted at UV-optical wavelengths. The number of ionizing photons, $N_{ion}$ in a $10^{10}$~\lsun\ burst is $1.7\times10^{50}$, $7.2\times10^{57}$, and $1.0\times10^{61}$, for the three temperatures, respectively, whereas the  number of intercepting H-atoms in the scattering region is $\sim 1\times 10^{24}$~cm$^{-2}$ column density is about $5\times 10^{57}$. 
The intensity of the nebular spectrum at wavelengths longer than the Lyman continuum is proportional to the number of ionizing photons in a inonization bounded CSM (left two panels of Figure~\ref{spec_burst}), a relation that breaks down in a matter-bounded CSM (right panel of the figure).   

All components are spread out in time as a result of Thomson scattering, so that the effective luminosity of these components is reduced as given by  eq. (\ref{leff}). 
The reradiated nebular emission is also spread in time because of the different recombination times caused by the $r^{-2}$ density gradient in the inner region of the CSM. The density changes by a factor of $\sim 2$ within the column density of  $10^{24}$~cm$^{-2}$, so that the spread in the reradiated nebular emission is approximately given by $\Delta t_{neb} \approx 0.5\times (\alpha\, n_0)^{-1} \approx 8\times 10^3$~s, where $\alpha=4.18\times 10^{-13}$~cm$^3$~s$^{-1}$ is the case~A recombination coefficient for H \citep{osterbrock06}. The time dispersion of the burst is therefore dominated by Thomson scattering and given by eq. (\ref{deltau}). 
The reprocessed spectra shown in the figure~\ref{spec_burst} were {\it not} corrected for the dilution effects of caused by Thomson scattering.

\subsection{Constraints on burst characteristics} 

\begin{figure*}[t]
\begin{center}
\includegraphics[width=2.3in]{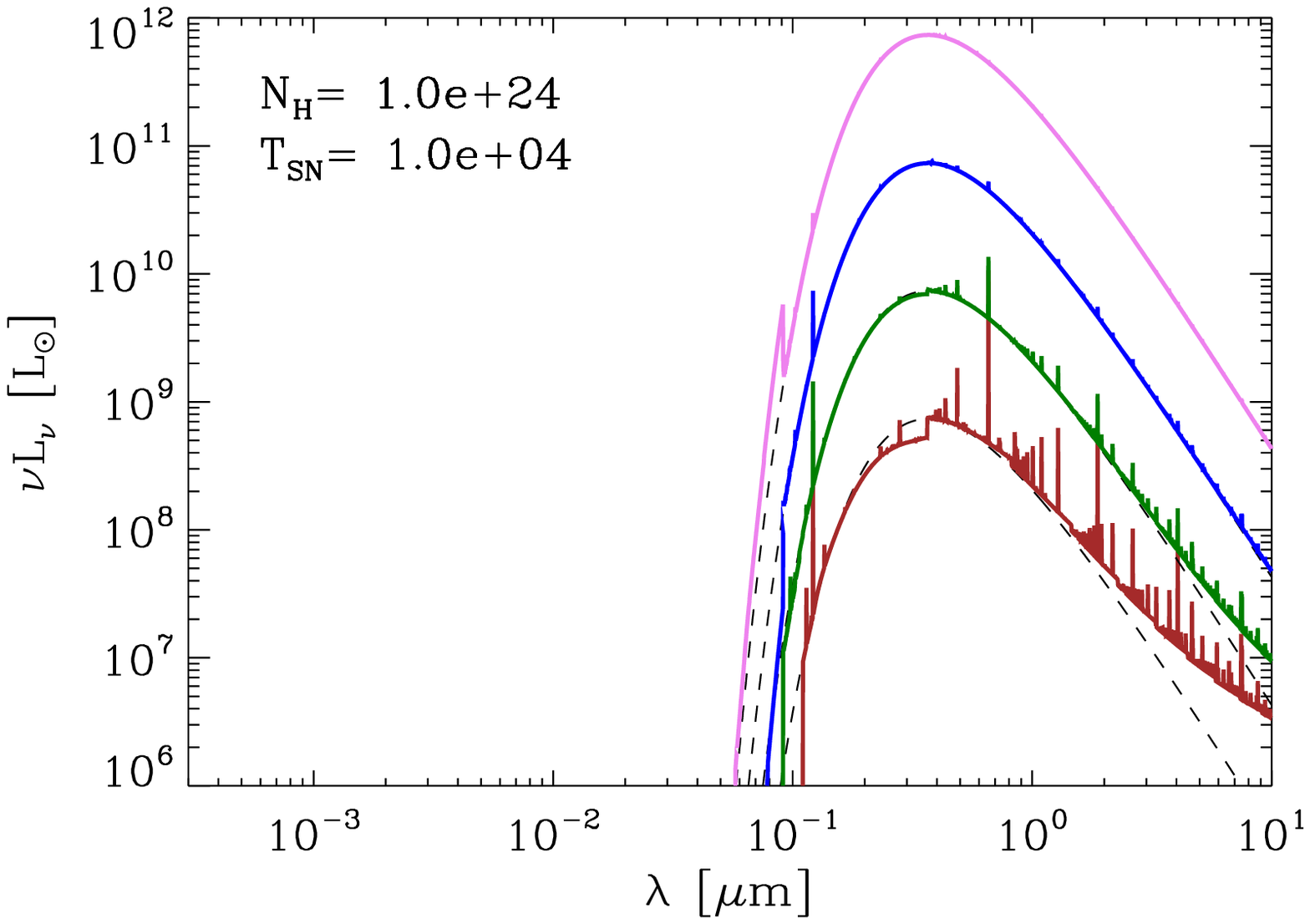} 
\includegraphics[width=2.3in]{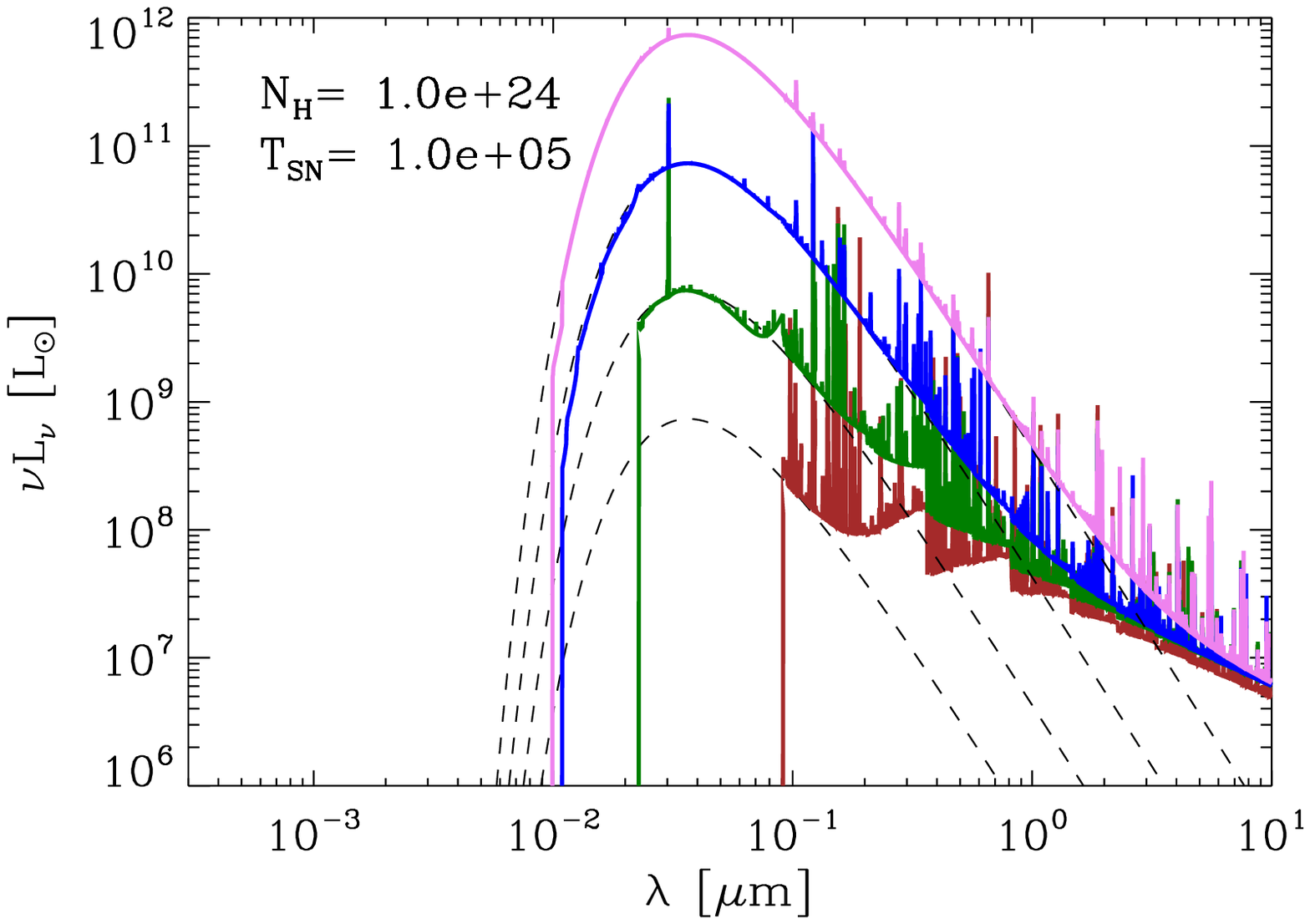} 
\includegraphics[width=2.3in]{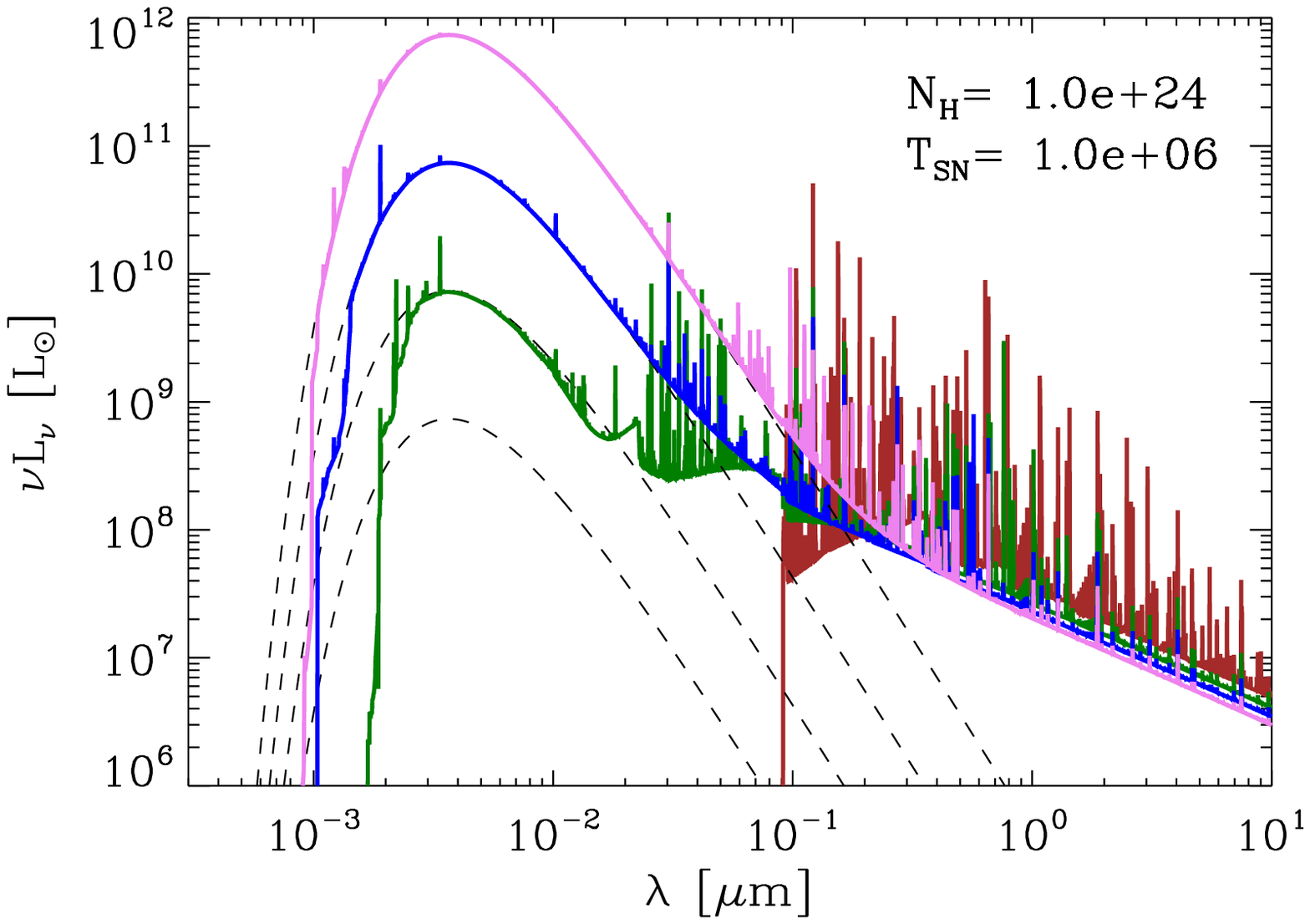} 
\caption{\label{spec_burst} Shock breakout spectra after propagating through a H-column density of $1\times10^{24}$~cm$^{-2}$ through the CSM. The dashed lines represent the initial burst spectra, and the colored ones represent the transmitted and reprocessed spectra after propagating through the CSM. The different colored curves represent burst spectra with luminosities (bottom to top) of $10^9, 10^{10}, 10^{11}$ and $10^{12}$~\lsun. The different panels correspond to different burst temperatures of (from left to right) of $10^4, 10^5$ and $10^6$~K, respectively. }
\end{center}
\end{figure*}

\begin{figure*}[t]
\begin{center}
\includegraphics[width=3.3in]{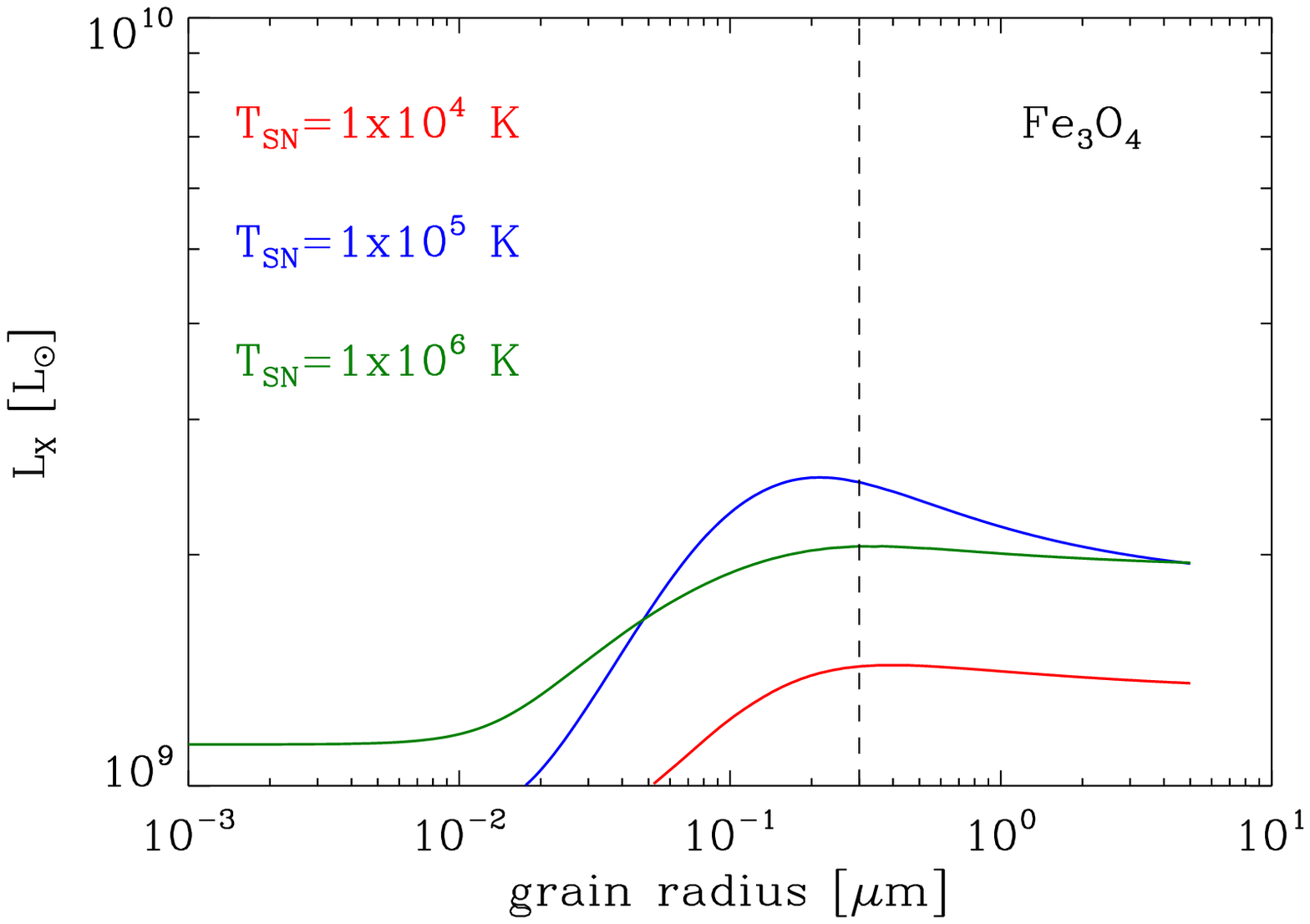}
\includegraphics[width=3.3in]{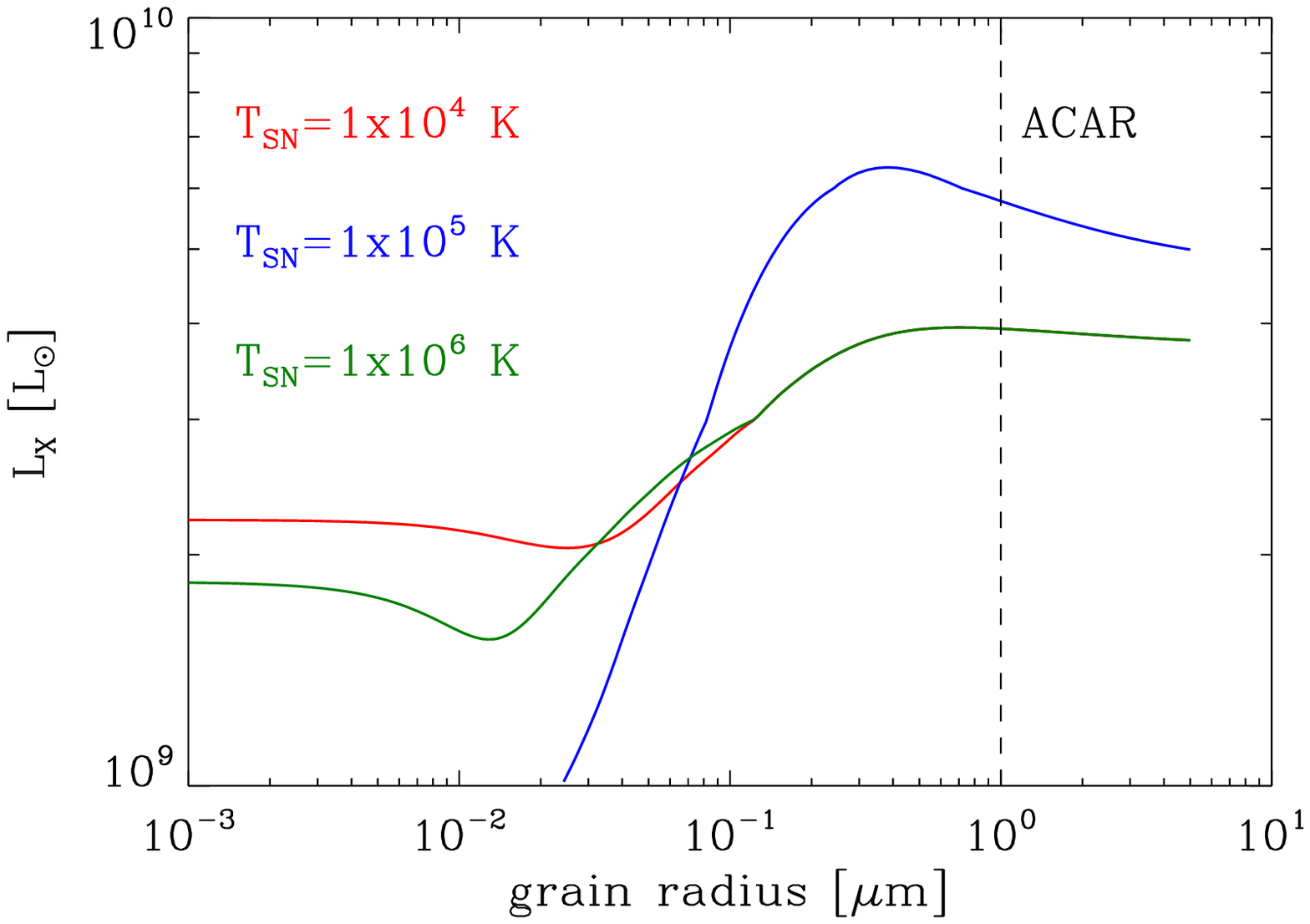}
\caption{\label{RevapX} Burst luminosities, $L_X$, that will evaporate Fe$_3$O$_4$ and ACAR grains out to their evaporation radius as a function of grain size. The different curves correspond to the different burst temperatures. The dashed vertical lines indicate the grains radii that give the best fit to the observed echo.}
\end{center}
\end{figure*}

Using the processed burst spectra presented in Figure~\ref{spec_burst}, we depict in Figure~\ref{RevapX} the burst luminosity and temperature that will evaporate ACAR or Fe$_3$O$_4$ grains of different radii out to a distance $R_{evap}$ from the center of the explosion, where $R_{evap}$ is the radius out to which the dust is evaporated by the radiation from the shock-CSM interaction (Figure~\ref{Revap}). The curves in the figure represent an upper limit on the burst luminosities for the different burst temperatures. Higher luminosities would have vaporized the dust that is observed to exist and produce the echo.

For example, the best fitting echo model is that generated by 1.0~\mic\ ACAR dust located at $R_{evap}=2.2\times 10^{16}$~cm. The curves in the right panel of the figure represent the burst luminosities and temperatures that will vaporize the dust out to that radius. The vertical dashed line corresponds to the  1.0~\mic\ grain radius that provides the best fit to the observed echo. The figure shows that burst luminosities above $\sim 4\times10^9$~\lsun\ will vaporize the dust needed to generate the echo. 

However, these consideration do not take into account the effect of Thomson scattering on the burst intensity as experienced by the dust.
Taking the value of $4\times10^9$~\lsun\ as the maximum effective luminosity the dust can be exposed to, and the dispersion time from eq.~(\ref{deltau}) we get that the burst characteristics are then constrained to be 
\begin{equation}
\label{burst}
\Delta t_b(s)\, L_b(L_\odot) \lesssim 2\times 10^{15}
\end{equation} 	   
So burst luminosities of $10^{12}$~\lsun are viable, as long as they do not last for more than $2\times 10^3$~s. Figure~\ref{RevapX} provides therefore a valuable aid for determining viable burst models that will preserve the CSM dust needed to produce the IR echo.

\section{Summary}

In this paper we studied the nature and composition of the dust that gives rise to the IR echo from SN2010jl, and introduced the connection between the echo  and the characteristics of the shock breakout from the stellar surface.
  
Our analysis of the origin of the IR echo from SN2010jl reveals a significantly different scenario than that presented by \cite{andrews11} and \cite{bevan20}. Firstly, we show that there is no need for adopting a toroidal structure for the emitting dust. 
No progenitor star was present in pre-SN images of SN2010jl suggesting that a visual extinction was needed to hide the most likely types of progenitor of the SN \citep{fox17,dwek17}. The requirement that a small amount of dust be present along the line-of-sight to the SN alleviates the need for the tilted toroidal configuration in the Andrews-Bevan models. 

Secondly, we rule out the flash of radiation from the shock breakout as the heating source of the echo. The flash duration and luminosity needed to sustain the temperature and duration of the epoch require a shock breakout energy that exceeds all theoretical models.

In our model we have derived a CSM density profile from X-ray observations of evolution of the intervening H-column density to the SN shock  \citep{chandra15}. This density profile determines the spatial distribution of the dust around the SN, the amount of CSM extinction and IR emission. We found that the visual optical depth of the pre-existing dust to the progenitor star is only 0.16, consistent with observational constraints \citep{dwek17}.

The echo arises from a dusty, spherically symmetric CSM that is heated by the radiation from the shocked CSM. 
The dust is heated by the radiation that is emitted by the SN shock-CSM interaction. We derive effective dust temperatures of $\sim 1250$~K,  higher than those found by \cite{andrews11}, and lower than those derived  by \cite{fransson14} and \cite{gall14}. The higher dust temperatures and closer proximity of the echoing dust to the radiation source have significantly lowered the  mass of dust required to generate the echo. The mass of ACAR dust in our model is about $1.6\times 10^{-4}$~\msun, compared to the values of (0.03-0.35)~\msun\ and $1.5\times 10^{-2}$~\msun\ derived by \cite{andrews11} and \cite{bevan20}, respectively. 

The derived dust mass is significantly lower that the mass of dust that could have condensed in the CSM prior to the SN explosion. However, the echo samples only the hottest dust in the CSM, since the observations only cover the NIR emission. It is conceivable that a significant amount of colder dust could be hidden, and only detected by  mid- to far-IR observations.

We have also shown that the temporal behavior of the observed near-IR echo is primarily determined by the dust evaporation temperature. The luminosity of the shock breakout and the subsequent SN luminosity will evaporate and clear the CSM of any dust in close proximity to the source of radiation.  This evaporation radius is a function of the dust composition and grain radius. In our model, the echo is generated by large, $\sim1$~\mic, ACAR grains located at a distance of $\sim 2.2\times10^{16}$~cm from the center of the explosion. This location is close to the edge of the CSM cavity, where its density peaks. If the CSM were dominated by silicate dust, they would be evaporated out to a distance of $\sim 4.2\times 10^{17}$~cm, close to the edge of the CSM as marked by the precipitous decline in its density. 

Silicate grains are definitely ruled out as the source of the near-IR echo. The near-IR emission from the silicate grains is on the Wien side of the emission spectrum, and is therefore more sensitive to the decrease in dust temperature resulting from the declining SN luminosity. An echo generated by silicate grains will therefore decline more rapidly than the observed echo light curves. Furthermore, a silicate origin will require a large mass of dust and a dust-to-H mass ratio that will exceed the mass of CSM silicon allowed from solar abundance constraints. 

Finally, we introduced a new relation between the characteristic of the shock breakout parameters and the characteristics of the dust that generates the IR echo.  The creation of an echo requires any pre-existing CSM dust to survive the initial burst of radiation generated by the shock breakout. The requirement that the dust evaporation radius of the burst be smaller than the radius of the dust giving rise to the echo provides strong constraints on the shock breakout characteristics. 

The observed light curve exhibits temporal fluctuations which cannot be replicated with the smooth, spherically-symmetric CSM adopted in our model.
Several reasons may be responsible for the behavior of the observed NIR echo:
\begin{enumerate}
  \item The CSM is not spherically symmetric or entirely homogenous. Asymmetries are suggested by spectropolarimetry observations \citep{patat11} and by the analysis of different velocity components of the CSM \citep{smith12}.
  \item The intensity of the echo was derived by subtracting the contribution of the radiation from the shock-CSM interaction from the observations. We fit the UVO emission with a constant temperature blackbody. The same approach was adopted in previous studies \citep[e.g.][]{gall14, fransson14}. Unlike other echoes, this radiation is not generated by the decay of radioactivities in the SN ejecta, and is not emerging from a stellar surface. It is generated by the cooling of the shocked CSM, and consists of nebular continuum and line emission. A constant temperature blackbody may not be a reliable representation of its spectrum. A time variant SN spectrum will affect the intensity of the echo at different epochs.  
  \item We have assumed that the radiation arises from the center of the explosion and also neglected the diffusion time of the shock radiation or the IR echo through the CSM. These simplifications may introduce temporal changes in the SN spectrum which may effect the derived intensity of the near-IR fluxes that were attributed to the echo. Such detailed analysis is beyond the scope of this paper. 
\end{enumerate}

Future observation will detect the IR echoes from numerous Type IIn CCSNe. We hope that the analysis we have presented this and in a series of papers \citep[][and references therein]{dwek17, sarangi18}, will provide a roadmap for the infrared analysis of these observation.

\acknowledgements
We thank the referee for his/her useful comments that significantly improved the manuscript. This work was supported by NASA's 16-ADP16-0004 research grant. RGA was supported by NASA under award number 80GSFC17M0002.

\clearpage
\bibliographystyle{aasjournal}
\bibliography{sn2010jl_echo_arXiv.bbl}
 \end{document}